\documentclass{pasj00} 

\newcommand{\cnfw}{c_{\rm nfw}}
\newcommand{\rvir}{r_{\rm vir}}

\begin{document}
\SetRunningHead{T. Hamana et al.}{Subaru Weak Lensing survey -- II: 
Multi-object Spectroscopy and Cluster Masses}
\Received{2008/8/28}
\Accepted{2001/01/01}

\title{Subaru Weak Lensing survey -- II: \\
Multi-object Spectroscopy and Cluster Masses}

\author{Takashi \textsc{Hamana}\altaffilmark{1}, 
Satoshi \textsc{Miyazaki}\altaffilmark{1},
Nobunari \textsc{Kashikawa}\altaffilmark{1},
Richard S. Ellis\altaffilmark{2},
Richard J. Massey\altaffilmark{3},\\
Alexandre Refregier\altaffilmark{4}, 
James E. Taylor\altaffilmark{5}}
\altaffiltext{1}{National Astronomical Observatory of Japan, Mitaka, 
Tokyo 181-8588, Japan}
\altaffiltext{2}{California Institute of Technology, 105-24 Astronomy,
Pasadena CA 91125, USA}
\altaffiltext{3}{Institute for Astronomy, Royal Observatory, Blackford Hill,
Edinburgh EH9 3HJ, UK}
\altaffiltext{4}{Service d'Astrophysique CEA Saclay, Bat. 709 F-91191 Gif sur
 Yvette, France}
\altaffiltext{5}{University of Waterloo, Department of Physics \& Astronomy,
 Waterloo, Ontario N2L 3G1, Canada}

\maketitle

\begin{abstract}
We present the first results of a multi-object spectroscopic (MOS)
campaign to follow up cluster candidates located via weak lensing.
Our main goals are to search for
spatial concentrations of galaxies that are plausible optical 
counterparts of the weak lensing signals, and to determine the cluster
redshifts from those of member galaxies.
Around each of 36 targeted cluster candidates,
we obtain $15-32$ galaxy redshifts.
For 28 of these targets, we confirm a secure cluster identification, with 
more than five spectroscopic galaxies within a velocity of $\pm 3000$km/s.
This includes three cases where two clusters at different redshifts are 
projected along the same line-of-sight. 
In 6 of the 8 unconfirmed targets, we find 
multiple small galaxy concentrations at different redshifts, each containing
at least three spectroscopic galaxies.
The weak lensing signal around those systems is thus probably created by the
projection of groups or small clusters along the same line-of-sight. 
In both the remaining two targets, a single small galaxy concentration
is found.

We evaluate the weak lensing mass of confirmed clusters 
via two methods: aperture densitometry and by fitting 
to an NFW model.
In most cases, these two mass estimates agree well.
In some candidate super-cluster systems, we find additional evidence of 
filaments connecting the main  density peak to additional nearby structure.
For a subsample of our most cleanly measured clusters,
we investigate the statistical relation between their weak lensing mass 
($M_{\rm NFW}$, $\sigma_{\rm sis}$) and the velocity dispersion of 
their member galaxies ($\sigma_{\rm v}$), comparing our 
sample with optically and X-ray selected samples from the literature.
Our lensing-selected clusters are consistent with 
$\sigma_v = \sigma_{\rm sis}$, with a similar scatter to
the optically and X-ray selected clusters.
We thus find no evidence of selection bias compared to these other
techniques.
We also derive an empirical relation between the cluster mass 
and the galaxy velocity dispersion, 
$M_{200}=9.6\times 10^{14}\times (\sigma_v/1000$km/s$)^{2.7}/E(z) 
h^{-1}M_\odot$, which is in reasonable agreement with the prediction
of $N$-body simulations in the $\Lambda$CDM cosmology.
\end{abstract}

\section{Introduction}
\label{sec:intro}

The development of weak lensing techniques, coupled with deep panoramic
imaging surveys, has enabled us to locate clusters of
galaxies via the gravitational distortion of background galaxies' shapes.
Since the first, spectroscopically confirmed discovery of a shear-selected 
cluster by Wittman et al. (2001), there has been rapid progress toward 
a large, weak-lensing selected cluster catalogue. 
Miyazaki et al. (2003) first reported the 
detection of several significant shear-selected cluster candidates in an 
untargeted 2.1 deg$^2$ field. 
Hetterscheidt et al. (2005) found 5 cluster candidates in 
50 randomly selected VLT FORS1 fields (0.64 deg$^2$ in total), all of which 
are associated with an overdensity of galaxies.
Wittman et al. (2006) reported 8 candidates in the 8.6 deg$^2$ Deep Lens Survey.
Gavazzi \& Soucail (2007) found 14 cluster candidates in the 4
deg$^2$ CFHT Legacy Survey (Deep), of which nine have optical or X-ray 
counterparts and are thus secure clusters.

The first sizable sample of weak lensing shear-selected cluster
candidates was presented by Miyazaki et al. (2007; hereafter P1). 
Their sample was obtained solely via peak finding in weak lensing 
density maps, and includes 100 significant peaks 
in a 16.7 deg$^2$ survey area.

Before such a sample is used for statistical cosmological or astronomical 
analyses, two additional follow-up observations are required. Firstly, each
cluster candidate should be confirmed by independent observations, since a
fraction of lensing peaks could be false positives from  e.g.\ the
chance tangential alignment of galaxies' intrinsic ellipticities (White, van
Waerbeke \&  Mackey 2002; Hamana, Takada \& Yoshida, 2004; Hennawi \& Spergel
2005). Secondly, the redshifts of confirmed clusters need to be determined in
order to derive their physical quantities, including mass. 

We have conducted a multi-object spectroscopic (MOS) campaign that
accomplishes both goals. We have measured the redshifts of a few tens of
galaxies within an expected cluster scale radius
(or core radius, typically a few arcmins), and searched for
spatial concentrations that are
plausible optical counterparts of the weak lensing signals. Once a
galaxy overdensity is found, it is easy to determine the cluster
redshift from the redshifts of member galaxies.
It is important to note that cluster confirmation based on prominent galaxy
concentrations would not be very effective for very high mass-to-light
ratio ($M/L$), galaxy-poor systems.
Although our methodology can {\it confirm} normal or galaxy-rich clusters,
the absence of a galaxy concentration in our fairly sparsely-sampled data
therefore does not necessarily prove that a weak lensing signal is false.

In addition to our primary goals, multi-object
spectroscopic observations provide several useful by-products. 
If redshifts can be obtained for sufficient galaxies in a cluster, their
line-of-sight velocity dispersion provides an estimate of 
the cluster's dynamical mass.
MOS observations can also detect multiple structures along the same line of sight.
Because of the relatively broad redshift window function of
gravitational lensing,
physically unrelated structures in the same line of sight may contribute
to a single peak in a weak lensing density map, resulting in 
an overestimation of the cluster mass (White et al. 2002). 
It will therefore be important to quantify and properly account for such projections
when computing statistics of cluster masses from weak lensing observations.

In this paper, we present results of cluster confirmations and cluster 
redshifts. We discuss the detailed weak lensing properties of each system and,
for a clean subset of our clusters, examine statistical relations between the 
weak lensing masses and dynamical masses. 
A statistical analysis of the {\it entire} sample, taking into account additional 
selection effects, will be presented in Green et al. (in preparation). 

This paper is organized as follows.
In \S2, we discuss our selection of cluster candidates.
In \S3, we describe our new observations, data reduction and measurements of galaxy redshifts.
In \S4, we identify optical counterparts to cluster candidates, and measure their velocity dispersions and dynamical masses.
In \S5, we analyze the weak lensing signal of confirmed clusters.
In \S6, we investigate cluster scaling relations within our sample.
In \S7, we summarize our results.
In Appendix~\ref{appendix:nfw}, we calculate the 
gravitational lensing shear profile of a truncated NFW model.
Detailed discussions of each system, including comparisons of the dynamic and
lensing masses, follow in Appendices~\ref{appendix:prop} (for clusters we have 
observed) and \ref{appendix:xmmlss} (for observations taken from the
literature).

Throughout this paper, we adopt a flat $\Lambda$CDM cosmology with 
the matter density $\Omega_{\rm m}=0.3$, the cosmological constant
$\Omega_\Lambda=0.7$, the Hubble constant 
$H_0=100 h$ km~s$^{-1}$~Mpc$^{-1}$ with  $h=0.7$.

\begin{table*}
\caption{Summary of spectroscopic observations. 
(a) Cluster name in the IAU convention.
(b) Field and catalogue number given in P1 (if listed).
(c) The peak $\kappa$ value. 
(d) The number density of galaxies with $18<mag<23$ (Vega mag) within 1 
arcmin of the $\kappa$ peak, and the mean 
value over the same Suprime-Cam FOV in parentheses. 
Here the luminosity is defined by the {\it SExtractor} {\tt MAG\_AUTO}
in the $R_C$-band, except for the COSMOS field where 
$i'$-band data is used.
(e) Exposure time.
(f) The date of the observation.
(g) The number of galaxies for which a spectroscopic redshift was obtained.
(h) The number of clusters identified (see \S \ref{sec:dynamics} for our cluster
identification criteria) and, when no clusters were identified,
the number of small galaxy concentrations in parentheses.}
\label{table:obs}
\begin{tabular}{llccllcc}
\hline
Name$^{(a)}$ & Field-No.$^{(b)}$ & $\kappa_{peak}$$^{(c)}$ &
$n_g$$^{(d)}$ & $t_{\rm exp}$$^{(e)}$ &obs date$^{(f)}$ &
$N_{spec}$$^{(g)}$ & $N_{c}$$^{(h)}$ \\ 
{} & {} & {} & [arcmin$^{-2}$]& [sec] & {} & {}\\ 
\hline
SL~J0217.3$-$0524 & SXDS & 0.041 & 12(7.2) & $2 \times 1200+900$ & 2005/12/23 & 24 & 1\\
SL~J0217.6$-$0530 & SXDS & 0.041 & 11(7.2) & $3 \times 900$ & 2005/12/23 & 19 & 0(1)\\
SL~J0217.9$-$0452 & SXDS & 0.061 & 12(6.2) & $2 \times 1200 + 900 $ & 2005/12/9 & 21 & 0(1)\\
SL~J0218.0$-$0444 & SXDS & 0.047 & 13(6.6) & $3 \times 1200 $ & 2005/12/9 & 19 & 0(2)\\
SL~J0219.6$-$0453 & SXDS & 0.060 & 12(6.9) & $2 \times 1200 + 900$ & 2005/12/23 & 25 & 1\\
SL~J0222.8$-$0416 & XMM-LSS-23 & 0.060 & 12(6.7) & $2 \times 1500 + 1200$ & 2005/12/24 & 27 & 1\\
SL~J0224.4$-$0449 & XMM-LSS-02 & 0.065 & 14(6.0) & $2 \times 1200+900 $ & 2005/12/23 & 25 & 1\\
SL~J0224.5$-$0414 & XMM-LSS-12 & 0.047 & 15(6.7) & $3 \times 900 $ & 2004/12/9 & 25 & 1\\
SL~J0225.3$-$0441 & XMM-LSS-15 & 0.086 & 7.1(6.0) & $3 \times 1200 $ & 2005/12/24 & 26 & 1\\
SL~J0225.4$-$0414 & XMM-LSS-22 & 0.083 & 14(6.9) & $3 \times 1200 $ & 2004/12/9 & 20 & 1\\
SL~J0225.7$-$0312 & XMM-LSS-01 & 0.114 & 8.1(7.1) & $3 \times 1200$ & 2005/12/24 & 25 & 1\\
SL~J0228.1$-$0450 & XMM-LSS-16 & 0.082 & 13(6.1) & $3 \times 1200 $ & 2005/12/24 & 25 & 1\\
SL~J0850.5$+$4512 & Lynx-08 & 0.099 & 11(6.2) & $3 \times 900$ & 2005/12/17 & 26 & 1\\
SL~J1000.7$+$0137 & COSMOS-00 & 0.092 & 14(7.0) & $3 \times 900$ & 2004/12/9 & 32 & 1\\
SL~J1001.2$+$0135 & COSMOS & 0.081 & 13(8.8) & $2 \times 900$ & 2004/12/9 & 32 & 2\\
SL~J1002.9$+$0131 & COSMOS & 0.047 & 18(8.8) & $2 \times 1800$ & 2004/5/16 & 21 & 0(2)\\
SL~J1047.3$+$5700 & Lockman Hole-05 & 0.101 & 18(6.9) & $3 \times 1200 $ & 2005/12/23 & 27 & 2\\
SL~J1048.1$+$5730 & Lockman Hole-15 & 0.077 & 10(7.8)  & $3 \times 1200 $ & 2005/12/23 & 27 & 1\\
SL~J1049.4$+$5655 & Lockman Hole-06 & 0.087 & 14(6.9) & $3 \times 1200 $ & 2005/12/23 & 26 & 1\\
SL~J1051.5$+$5646 & Lockman Hole-03 & 0.082 & 8.4(8.6)  & $2 \times 1200 + 600 $ & 2005/6/2 & 15 & 0(2)\\
SL~J1052.0$+$5659 & Lockman Hole & 0.085 & 7.1(8.6) & $2 \times 1200 $ & 2005/6/2 & 24 & 0(2)\\
SL~J1052.5$+$5731 & Lockman Hole & 0.075 & 11(8.6)  & $2 \times 1200 + 600 $ & 2005/6/2 & 17 & 0(2)\\
SL~J1057.5$+$5759 & Lockman Hole-00 & 0.091 & 23(6.3) & $3 \times 1200 $ & 2004/12/9 & 32 & 1\\
SL~J1135.6$+$3009 & GD140-00 & 0.117 & 10(6.7) & $2 \times 900 $ & 2004/12/17 & 17 & 1\\
SL~J1201.7$-$0331 & PG1159-035-05 & 0.070 & 16(6.6) & $2 \times 1800 $ & 2004/5/16 & 18 & 1\\
SL~J1204.4$-$0351 & PG1159-035  & 0.079 & 14(6.4) & $3 \times 900 $ & 2004/12/17 & 20 & 1\\
SL~J1334.3$+$3728 & 13hr field-00 & 0.097 & 24(7.1) & $2 \times 1800 + 420$ & 2004/5/16 & 31 & 1\\
SL~J1335.7$+$3731 & 13hr field-01 & 0.074  & 16(7.1) & $3 \times 900$ & 2005/6/1 & 23 & 1\\
SL~J1337.7$+$3800 & 13hr field-04  & 0.082 & 14(7.3) & $3 \times 1200$ & 2005/6/1 & 27 & 1\\
SL~J1601.6$+$4245 & GTO 2deg$^2$ & 0.107 & 14(6.2) & $3 \times 1200 $ & 2005/7/31  & 29 &2\\
SL~J1602.8$+$4335 & GTO 2deg$^2$-00 & 0.099 & 22(7.0) & $3 \times 900 $ & 2005/6/1 & 22 & 1\\
SL~J1605.4$+$4244 & GTO 2deg$^2$-09 & 0.067 & 11(7.8) & $3 \times 1200 $ & 2005/7/31 & 20 & 1\\
SL~J1607.9$+$4338 & GTO 2deg$^2$ & 0.079 & 11(6.5) & $3 \times 1200 $ & 2005/8/1 & 21 & 1\\
SL~J1634.1$+$5639 & CM DRA-06 & 0.104 & 5.1(6.3) & $3 \times 900$ & 2005/6/1 & 16 & 1\\
SL~J1639.9$+$5708 & CM DRA-04 & 0.044 & 12(7.3)  & $3 \times 1200$ & 2005/8/1 & 24 & 0(2)\\
SL~J1647.7$+$3455 & DEEP16-00 & 0.070 & 9.7(5.0) & $3 \times 1200$ & 2004/12/9 & 32 & 1\\
\hline
\end{tabular}
\end{table*}

\section{Cluster candidates}
\label{sec:target}

Targets for our spectroscopic follow-up campaign were selected from amongst the 
Subaru weak lensing survey's shear-selected cluster candidates.
We refer the reader to P1 for details of the Subaru weak lensing survey, 
including the image
acquisition and analysis, the creation of weak lensing density maps, and
the selection of cluster candidates.
Briefly, $R_C$-band imaging data were obtained in thirteen fields 
(except for the COSMOS field where $i'$-band data was used).
Each survey area is $1.07-2.8$ degree$^2$ and the total area is 
21.82 degree$^2$. 
Weak lensing density maps were computed with a pixel size of
$15\times 15$ arcsec, using the Kaiser \& Squires (1993)
inversion algorithm, with a Gaussian smoothing filter of $\theta_G=1$
arcmin. Noise maps were also created using the same algorithm, 
but from the root-mean-square (RMS) of 100 realizations in which the
orientations of the measured shears were randomized.
Maps of the signal-to-noise ratio ($S/N$) were created by dividing the 
density map by the noise RMS map. Positive peaks were searched for in the
$S/N$ maps, and peaks with $S/N$ greater than a threshold value were defined 
as candidates of massive halos.
P1 took a threshold value of $S/N=3.69$.
After carefully avoiding areas close to bright stars or field
boundaries, where the weak lensing density map could be affected by a lack
of background lensed galaxies, 100 significant detections remained in a 
16.7 deg$^2$ area (see Table 2 of P1 for their weak lensing and optical
properties).

We have carried out multi-object spectroscopic follow-up observations of 36
cluster candidates (see Table \ref{table:obs} for our targets' optical and weak lensing 
properties, galaxy number densities, peak $\kappa$ values, and the amplitudes
of their radial shear profiles, $\gamma_{\rm sis}$, which is defined in
\S \ref{sec:sis}).
Our target selection differs slightly from that of P1 because, in the
planning stages of this follow-up program, we were still evaluating the 
optimal criteria for reliable cluster selection.
The current selection was based on both the peak $\kappa$ values and a 
visual inspection of the optical images.
The peak $\kappa$ value was evaluated with two filter scales (1 and
2 arcmin). Since the Gaussian filter acts as a matched filter, large nearby
systems can be detected with a higher $S/N$ in the $2\arcmin$ filter.
Due to the visual inspection, our target could be biased toward
optically rich clusters.
Of our 36 targets, 24 are listed in the P1 catalog.
The remaining 12 targets are either below the P1 $S/N$ threshold with a single
filter scale, or in discarded survey areas (close to bright stars or field 
boundaries).

\section{Spectroscopic observations and data reduction}
\label{sec:specobs}

\subsection{Spectroscopic observations}
We used the Subaru telescope's Faint Object Camera and Spectrograph (FOCAS; Kashikawa et
al. 2002) in Multi-Object Spectroscopy (MOS) mode. 
Each cluster candidate was observed with one $6\arcmin$ diameter slit mask,
on which we placed $25-38$ slits.
We used the 150/mm grating and the SY47 order sorting filter, resulting in 
a wavelength coverage between 4700 and 9400{\AA}, with a pixel
resolution of 2.8\AA~pixel$^{-1}$.
The slit width was 0.8 arcsec for all cases, which corresponds to a
spectral resolution power of $R\sim250$, or $\Delta \lambda \sim
30${\AA} at 8000{\AA}. 

We conducted FOCAS Observations in 2004 May 16, December 9, 17, 
2005 June 1-2, July 31-August 1, December 9 and 23-24.
Observing conditions were not always good: some targets were observed
under a cloudy/cirrus sky.
We took two or three exposures per mask. The total exposure times
are listed in Table \ref{table:obs} and are
$30-70$ minutes depending on the apparent magnitude 
of targeted galaxies and the observing conditions.

Within each field, target galaxies were selected by their apparent $R_C$ magnitude 
(with higher priorities for brighter galaxies) 
and color information when it was available.
To obtain multicolor imaging,
we searched the Subaru archive or took pre-imaging data for 
the MOS mask design with the $I_C$-band filter.
When color information was available, we gave a higher priority to galaxies of a color
consistent with any early type red-sequences evident in the color-magnitude diagram.
To avoid possible selection effects, we set the
color range for this increased priority relatively broad ($\pm 0.5$ mag in 
color from a possible red-sequence).
Finally, we visually inspected the selected galaxies to avoid obviously
nearby galaxies.

\subsection{Data reduction}

The data were reduced with the standard FOCAS data reduction package 
{\it FOCASRED}\footnote{The Subaru data reduction manual is available from\\ 
{\tt http://www.naoj.org/Observing/DataReduction/}}, which operates in
IDL and IRAF.
After bias subtraction and flat-fielding, each slit-let spectrum was
extracted, then wavelength- and flux-calibrated.
Night sky lines within the spectra themselves were used to define
the wavelength scale.
We then carried out skysubtraction using a 2nd order Chebyshev function.
Individual spectra of 420$-$1800 sec exposure time were combined using the 
{\it imcombine} IRAF task. 
The final wavelength determination accuracy was a few angstroms.

\begin{table*}
\caption{Summary of dynamical analysis of galaxy groups/clusters.
(a) Number of spectroscopic galaxies within a velocity
space of $\pm 3000$km/s.
(b) The cluster redshift.
(c) The velocity dispersion.
(d) The angular harmonic radius.
(e) The harmonic radius estimated by eq (\ref{Rh}).
(f) The dynamical mass estimated by eq (\ref{mdyn}).
(g) Note: PS indicates poor statistics in the estimation of velocity dispersion
($N<12$), NS the existence of a neighbor system, 
FB proximity ($<3$ arcmin) to the Suprime-Cam field boundary,  
and CS membership of the clean sample (see \S \ref{sec:scaling}).}
\label{table:mdym}
\begin{tabular}{lccccccc}
\hline
Name & $N$$^{(a)}$ & $z_c$$^{(b)}$ & $\sigma_v$$^{(c)}$ & $\theta_h$$^{(d)}$ & $R_h$$^{(e)}$ & $M_{dyn}$$^{(f)}$ &Note$^{(g)}$\\ 
{} & {} & {} & [km/s] & [arcmin] & [$h^{-1}$Mpc] & [$\times 10^{14}h^{-1}M_\odot] $ {}& \\ 
\hline
SL~J0217.3$-$0524 & 8 & $ 0.4339_{-0.0041}^{+0.0010} $ & $ 958_{-224}^{+350} $ & $ 2.06\pm 0.78 $ & $ 0.487\pm 0.217 $ & $ 6.25_{-3.51}^{+5.89} $ & PS \\
SL~J0219.6$-$0453  & 11 & $ 0.3322_{-0.0005}^{+0.0008} $ & $ 404_{-78}^{+203} $ & $ 2.08\pm 0.58 $ & $ 0.416\pm 0.160 $ & $ 0.95_{-0.42}^{+1.22} $ & PS \\
SL~J0222.8$-$0416 & 6 & $ 0.3219_{-0.0035}^{+0.0036} $ & $ 970_{-9}^{+164} $ & $ 2.89\pm 1.62 $ & $ 0.566\pm 0.328 $ & $ 7.43_{-4.18}^{+4.99} $ & PS \\
SL~J0224.4$-$0449 & 10 & $ 0.4945_{-0.0002}^{+0.0006} $ & $ 271_{-202}^{+355} $ & $ 1.52\pm 0.26 $ & $ 0.388\pm 0.101 $ & $ 0.40_{-0.38}^{+1.73} $ & PS \\
SL~J0224.5$-$0414 & 12 & $ 0.2627_{-0.0004}^{+0.0003} $ & $ 268_{-51}^{+223} $ & $ 1.87\pm 0.36 $ & $ 0.319\pm 0.089 $ & $ 0.32_{-0.13}^{+0.76} $ & FB \\
SL~J0225.3$-$0441 & 7 & $ 0.2642_{-0.0006}^{+0.0013} $ & $ 530_{-148}^{+622} $ & $ 2.22\pm 0.84 $ & $ 0.379\pm 0.226 $ & $ 1.48_{-0.91}^{+5.55} $ & PS \\
SL~J0225.4$-$0414 & 8 & $ 0.1419_{-0.0007}^{+0.0003} $ & $ 400_{-113}^{+411} $ & $ 1.43\pm 0.79 $ & $ 0.150\pm 0.257 $ & $ 0.34_{-0.25}^{+1.06} $ & PS \\
SL~J0225.7$-$0312 & 15 & $ 0.1395_{-0.0010}^{+0.0006} $ & $ 739_{-86}^{+150} $ & $ 1.91\pm 0.31 $ & $ 0.197\pm 0.068 $ & $ 1.50_{-0.41}^{+0.71} $ & CS \\
SL~J0228.1$-$0450 & 13 & $ 0.2948_{-0.0006}^{+0.0006} $ & $ 447_{-52}^{+82} $ & $ 2.42\pm 0.33 $ & $ 0.447\pm 0.049 $ & $ 1.25_{-0.32}^{+0.53} $ & NS \\
SL~J0850.5+4512 & 15 & $ 0.1935_{-0.0009}^{+0.0007} $ & $ 650_{-53}^{+115} $ & $ 2.57\pm 0.37 $ & $ 0.346\pm 0.069 $ & $ 2.05_{-0.43}^{+0.84} $ & CS \\
SL~J1000.7+0137 & 14 & $ 0.2166_{-0.0006}^{+0.0002} $ & $ 729_{-439}^{+526} $ & $ 2.39\pm 0.50 $ & $ 0.352\pm 0.096 $ & $ 2.61_{-2.27}^{+5.17} $ & CS \\
SL~J1001.2+0135A & 11 & $ 0.2205_{-0.0026}^{+0.0024} $ & $ 1382_{-160}^{+337} $ & $ 2.15\pm 0.60 $ & $ 0.322\pm 0.156 $ & $ 8.58_{-3.02}^{+5.25} $ & PS \\
SL~J1001.2+0135B & 11 & $ 0.3657_{-0.0013}^{+0.0012} $ & $ 931_{-170}^{+315} $ & $ 2.16\pm 0.46 $ & $ 0.460\pm 0.110 $ & $ 5.57_{-2.20}^{+4.57} $ & PS \\
SL~J1047.3+5700A & 6 & $ 0.2427_{-0.0002}^{+0.0013} $ & $ 412_{-133}^{+205} $ & $ 2.54\pm 0.67 $ & $ 0.408\pm 0.081 $ & $ 0.97_{-0.58}^{+1.23} $ & PS \\
SL~J1047.3+5700B & 10 & $ 0.3045_{-0.0016}^{+0.0007} $ & $ 691_{-95}^{+143} $ & $ 2.88\pm 0.53 $ & $ 0.544\pm 0.088 $ & $ 3.62_{-1.15}^{+1.78} $ & PS \\
SL~J1048.1+5730 & 9 & $ 0.3173_{-0.0008}^{+0.0008} $ & $ 448_{-59}^{+138} $ & $ 1.69\pm 0.36 $ & $ 0.328\pm 0.071 $ & $ 0.92_{-0.30}^{+0.68} $ & PS \\
SL~J1049.4+5655 & 6 & $ 0.4210_{-0.0004}^{+0.0004} $ & $ 276_{-71}^{+112} $ & $ 2.23\pm 1.05 $ & $ 0.520\pm 0.282 $ & $ 0.55_{-0.36}^{+0.60} $ & PS \\
SL~J1057.5+5759 & 18 & $ 0.6011_{-0.0021}^{+0.0021} $ & $ 1552_{-204}^{+320} $ & $ 2.52\pm 0.44 $ & $ 0.709\pm 0.087 $ & $ 23.83_{-7.20}^{+11.62} $ & FB \\
SL~J1135.6+3009 & 15 & $ 0.2078_{-0.0012}^{+0.0008} $ & $ 893_{-145}^{+365} $ & $ 2.30\pm 0.63 $ & $ 0.329\pm 0.168 $ & $ 3.65_{-1.48}^{+3.73} $ & CS \\
SL~J1201.7$-$0331 & 8 & $ 0.5218_{-0.0019}^{+0.0042} $ & $ 1221_{-279}^{+425} $ & $ 1.38\pm 0.65 $ & $ 0.361\pm 0.269 $ & $ 7.51_{-4.66}^{+7.08} $ & PS \\
SL~J1204.4$-$0351 & 14 & $ 0.2609_{-0.0005}^{+0.0008} $ & $ 568_{-94}^{+374} $ & $ 2.70\pm 0.54 $ & $ 0.457\pm 0.083 $ & $ 2.06_{-0.75}^{+3.63} $ & CS \\
SL~J1334.3+3728 & 21 & $ 0.3012_{-0.0017}^{+0.0015} $ & $ 1443_{-164}^{+199} $ & $ 2.42\pm 0.21 $ & $ 0.454\pm 0.024 $ & $ 13.20_{-3.04}^{+4.04} $ & NS \\
SL~J1335.7+3731 & 14 & $ 0.4070_{-0.0017}^{+0.0040} $ & $ 1064_{-88}^{+166} $ & $ 2.22\pm 0.38 $ & $ 0.506\pm 0.087 $ & $ 8.00_{-1.86}^{+3.02} $ & CS \\
SL~J1337.7+3800 & 16 & $ 0.1798_{-0.0004}^{+0.0009} $ & $ 783_{-248}^{+285} $ & $ 3.04\pm 0.34 $ & $ 0.387\pm 0.045 $ & $ 3.31_{-1.80}^{+2.87} $ & CS \\
SL~J1601.6+4245A & 7 & $ 0.2075_{-0.0007}^{+0.0004} $ & $ 285_{-36}^{+97} $ & $ 3.07\pm 0.68 $ & $ 0.437\pm 0.086 $ & $ 0.50_{-0.16}^{+0.41} $ & PS \\
SL~J1601.6+4245B & 8 & $ 0.4702_{-0.0014}^{+0.0015} $ & $ 965_{-285}^{+694} $ & $ 3.12\pm 0.54 $ & $ 0.774\pm 0.086 $ & $ 10.05_{-5.35}^{+19.73} $ & PS \\
SL~J1602.8+4335 & 15 & $ 0.4156_{-0.0018}^{+0.0005} $ & $ 675_{-265}^{+589} $ & $ 2.74\pm 0.37 $ & $ 0.632\pm 0.043 $ & $ 4.03_{-2.59}^{+10.11} $ & CS \\
SL~J1605.4+4244 & 6 & $ 0.2233_{-0.0068}^{+0.0014} $ & $ 1525_{-188}^{+382} $ & $ 2.03\pm 0.80 $ & $ 0.306\pm 0.200 $ & $ 9.94_{-4.56}^{+6.85} $ & PS \\
SL~J1607.9+4338 & 9 & $ 0.3109_{-0.0004}^{+0.0004} $ & $ 273_{-40}^{+66} $ & $ 1.52\pm 0.44 $ & $ 0.291\pm 0.173 $ & $ 0.30_{-0.12}^{+0.19} $ & PS \\
SL~J1634.1+5639 & 13 & $ 0.2377_{-0.0021}^{+0.0017} $ & $ 1402_{-121}^{+334} $ & $ 2.28\pm 0.66 $ & $ 0.360\pm 0.126 $ & $ 9.87_{-3.29}^{+5.99} $ & CS \\
SL~J1647.7+3455 & 12 & $ 0.2592_{-0.0008}^{+0.0010} $ & $ 673_{-99}^{+158} $ & $ 2.09\pm 0.26 $ & $ 0.352\pm 0.053 $ & $ 2.23_{-0.67}^{+1.20} $ & CS \\
\hline
\end{tabular}
\end{table*}

\subsection{Redshift measurement}
\label{sec:zmesurement}

Redshifts were determined by centroiding multiple 
emission and/or absorption lines.
The statistical error of the redshift measurement is less than $2\times
10^{-4}$.
In our samples, there are 18 galaxies whose redshifts were also
obtained by SDSS. The differences between SDSS and our measurements
are less than $4\times 10^{-3}$. Thus we conclude that any
systematic errors in our redshift measurement are very small.

We adopted a simple spectral classification of galaxies, following 
Cohen et al.\ (1999), and refer the reader to their paper for details.
Briefly, {\it ``E''} (emission) denotes a galaxy in which emission
lines dominate the spectrum; {\it ``A''} (absorption) denotes a galaxy
where no emission lines are detected;
and {\it ``C''} (composite) is an intermediate case where both emission
and absorption (usually [O{\sc ii}]~$\lambda 3727$ and H + K) are seen.
Note that the [O{\sc ii}]~$\lambda 3727$ emission line 
is blueward of our wavelength coverage for galaxies
at $z\lesssim 0.2$, so the discrimination between {\it C} and
{\it A} was somewhat ambiguous for such nearby galaxies.

We successfully obtained the redshifts of $15-32$ galaxies near each cluster candidate.
All the results\footnote{Machine-readable tables are available from\\
{\tt http://th.nao.ac.jp/\~{}hamanatk/SLmosz.dat}} 
(the spatial and redshift distributions) are presented in Figures
\ref{fig:sxds_6}--\ref{fig:deep16_m1_15}.

\begin{figure*}
\includegraphics[width=180mm,clip]{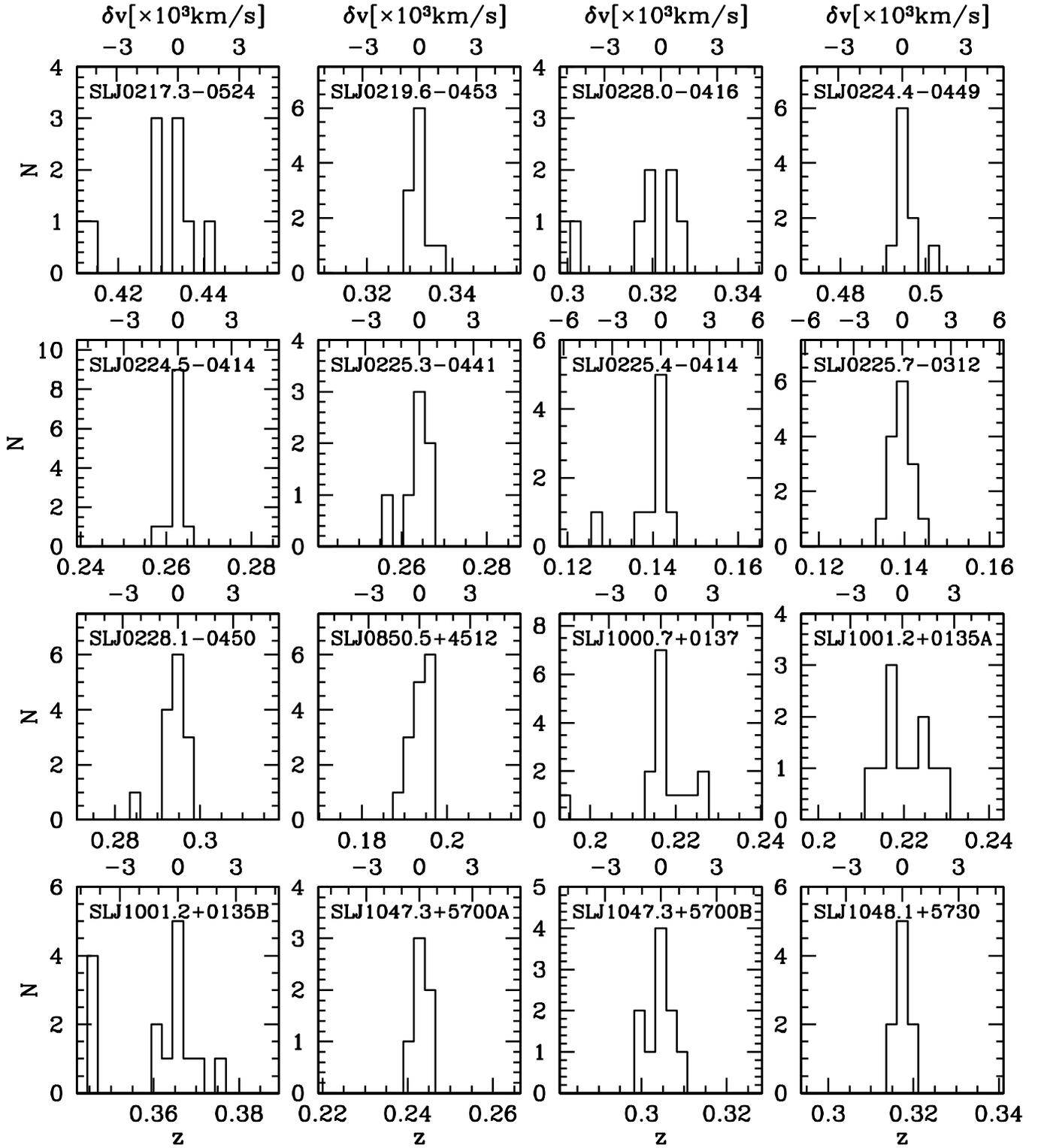}
\caption{The distribution of spectroscopically confirmed member
galaxies (located within $\pm 3000$km/s), in 
redshift (lower scale) and velocity (upper scale) space.}
\label{fig:vzhist1}
\end{figure*}

\begin{figure*}
\includegraphics[width=180mm,clip]{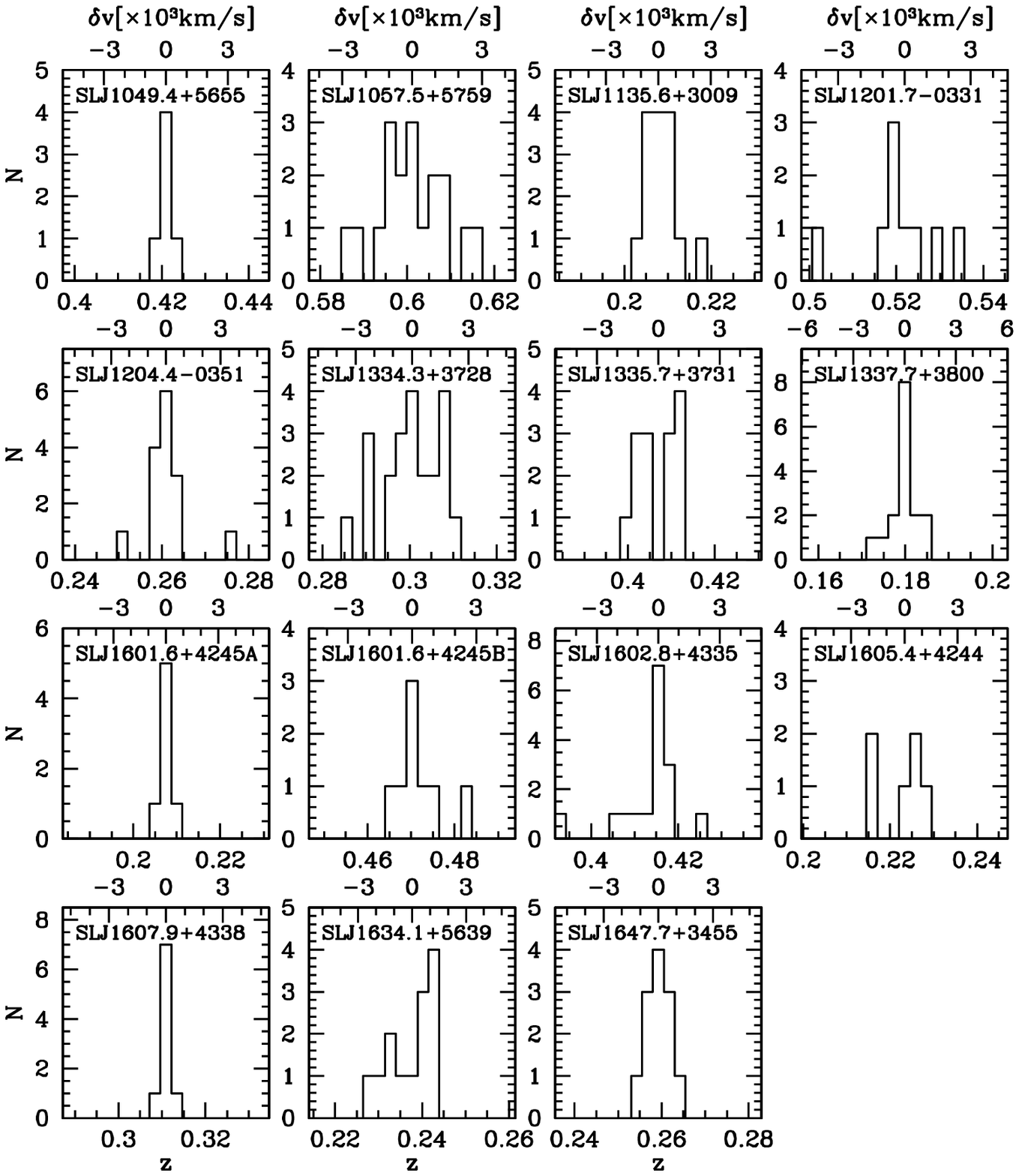}
\begin{footnotesize}
{\bf Fig.~\ref{fig:vzhist1}.} Continued from previous page.
\end{footnotesize}
\end{figure*}

\section{Dynamical masses}
\label{sec:dynamics}

We next searched for galaxy concentrations in the redshift data.
With only information on the line-of-sight velocity and the sky
coordinates of relatively sparse samples, it was often difficult to judge 
whether galaxies that appear clustered were really gravitationally bound.
We adopted a quantitative criterion for galaxy cluster identification 
that there be {\it more than five galaxies within $\pm 3000$km/s.}
As shown in Table \ref{table:mdym}, 31 galaxy concentrations satisfied this condition.
Three cluster candidates (SL~J1001.2+0135, SL~J1047.3+5700 and SL~J1601.6+4245) 
contained two galaxy concentrations.
The velocity distribution of galaxies within or near each concentration are 
presented in Figure \ref{fig:vzhist1}.
Note that there may be additional galaxy clusters that remain unconfirmed by
our MOS observation because of our relatively sparse sampling. 

We then computed the velocity dispersion of galaxies within the 31 detected clusters.
To do this, we used the {\it ROSTAT} routine, an
implementation of the robust bi-weight algorithm by Beers, Flynn \&
Gebhardt (1990).
We recorded
the bi-weight estimators for the cluster redshift
($z_c$) and velocity dispersion ($\sigma_v$), 
computing their errors by a bootstrap technique.
The results are summarized in Table \ref{table:mdym}.

The virial mass of a well relaxed cluster can be estimated from
its measured velocity dispersion 
(see, e.g., Cohen \& Kneib 2002) via
\begin{equation}
\label{mdyn}
M_{dyn} = {{6 \sigma_v^2 R_h} \over G},
\end{equation}
where $R_h$ is the harmonic radius, related to the virial radius by 
$r_{\rm vir} =\pi R_h /2$ (Limber \& Mathews 1960),  
but computed from angular positions of the 
member galaxies (see e.g., Saslow 1985; Nolthenius \& White 1987),
\begin{equation}
\label{Rh}
R_h = D_A(z_c) {\pi \over 2} {{N_m (N_m-1)} \over 2}
(\Sigma_i \Sigma_{j>i} \theta_{ij}^{-1})^{-1},
\end{equation}
where $\theta_{ij}$ is the angular
distance between galaxies $i$ and $j$, $N_m$ is the number of cluster
members, and $D_A$ is the angular diameter distance.
We estimated the error on $R_h$ by a bootstrap technique. 

The values of $M_{dyn}$ and $R_h$ that we measured for the 31 clusters are 
summarized in Table \ref{table:mdym}.
Note that in deriving eq. (\ref{Rh}), it was assumed
that member galaxies are homogeneously sampled.
In our case, the selection of target galaxies was restricted 
by the constraints of MOS mask design. 
Accordingly, our sampling fraction (the ratio of targeted to member galaxies) 
could be smaller in the densely populated cores of clusters than in their outer parts. 
It is thus possible that our estimates of $R_h$ were slightly overestimated. 
We did not have enough information to propagate the potential systematic bias 
caused by this inhomogeneous sampling. Therefore, our estimates of dynamical
masses should be regarded with some caution.

\section{Weak lensing analysis}
\label{sec:wlmass}

\subsection{Basic equations}
Let us first summarize several expressions useful for weak lensing analysis.
We closely follow the notation of Bartelmann \& Schneider (2001).
For more details, see that excellent review and references therein.

The aim of weak lensing mass reconstruction is to measure the
dimensionless surface mass density, frequently called the 
gravitational lensing density, 
\begin{equation}
\label{kappa}
\kappa = {{\Sigma}\over{\tilde\Sigma_{cr}(z_l)}},
\end{equation}
where $\Sigma$ is the two-dimensional projected mass density, and
$\tilde\Sigma_{cr}(z_l)$ is the source weighted critical density
\begin{equation}
\label{Sigmqcr}
\tilde\Sigma_{cr} (z_l) = {{c^2}\over{4 \pi G}} {1\over{D_l}}
{{\int_{z_l}^\infty dz_s~n_s(z_s) {{D_s}/ {D_{ls}}}} 
\over
{\int_{0}^\infty dz_s~n_s(z_s)}},
\end{equation}
where $n_s(z)$ is the redshift distribution of source galaxies, and 
$D_l$, $D_s$, $D_{ls}$ are the angular diameter distances from 
the observer to a lens, the observer to a source and the lens and a 
source, respectively.

Our weak lensing mass reconstruction used the Fourier space relation 
between $\kappa$ and the gravitational lensing shear $\gamma$ 
(Kaiser \& Squires 1993). 
Note that the directly observable quantity is not $\gamma$ but 
the reduced shear
\begin{equation}
\label{g}
g={{\gamma}\over{1-\kappa}}.
\end{equation}
We adopt the weak lensing approximation, $g\simeq\gamma$, because most of 
our signal lies in the wings of clusters, where $\kappa \lesssim 0.1$ (see
Figures \ref{fig:sxds_6}-\ref{fig:deep16_m1_15}).

\subsection{Source redshift distribution}
\label{sec:nz}

For the redshift distribution of source galaxies, we adopted the conventional
parametric model,
\begin{equation}
\label{ns}
n_s(z)={\beta \over {z_\ast \Gamma[(1+\alpha)/\beta]}}
\left({z\over {z_\ast}}\right)^\alpha
\exp\left[-\left({z\over {z_\ast}}\right)^\beta \right].
\end{equation}
The mean redshift is given by $\langle z_s \rangle = z_\ast
\Gamma[(2+\alpha)/\beta]/ \Gamma[(1+\alpha)/\beta]$. 
Since our galaxy catalog ($R_C < 25.5$ mag [Vega]) was much deeper than 
any sizable galaxy catalog with spectroscopic redshifts, 
the redshift distribution of our galaxies is uncertain.
We took a fiducial model of $\langle z_s \rangle = 1$, $\alpha=1.5$
and $\beta=1$, and evaluated two errors arising from uncertainty
in the model parameters.
The first error, due to uncertainly in $\langle z_s
\rangle$, we denote as $\sigma_{zs}$.
We evaluate this in the standard manner as 
$\sigma_{zs} = d\tilde{\Sigma}_{cr}/d z_s \delta z_s$,
and consider an uncertainty of $\delta z_s = 0.2$.
We found  $\sigma_{zs} \simeq 0.2 \times z_l$ for $z_l < 0.35$ and 
$\sigma_{zs} \simeq 0.07$ for $z_l > 0.35$.
The second error, due to uncertainly in $\alpha$ and
$\beta$, we denote as $\sigma_{ns}$.
In order to evaluate this, we considered two models of 
($\alpha$, $\beta$)=(2, 1.5) and (3, 2).
The relative differences in $\tilde{\Sigma}_{cr}$ from the fiducial
model were found to be about 10 percent or less at the
redshifts of our clusters.
We therefore decided to take  $\sigma_{ns}=0.1\times \tilde{\Sigma}_{cr}$.
These errors were properly propagated to the errors in weak lensing mass
and the SIS velocity dispersion (see below).

\subsection{Aperture densitometry}
\label{sec:zeta}

One convenient way to compute the mass of galaxy clusters exploits the
relation between $\kappa$ and the tangential shear, $\gamma_t$ (Squires
\& Kaiser 1996)
\begin{equation}
\label{eq:gamma-t}
\langle \gamma_t \rangle (\theta) 
= -{1\over 2} {{d\bar{\kappa} \over {d\ln \theta}}},
\end{equation}
where $\langle \gamma_t \rangle $ is the azimuthal average of the
tangential shear, and $\bar{\kappa}$ is the averaged $\kappa$ over a circular
aperture with radius $\theta$. From this, one can obtain an
expression for the average projected density within a radius $\theta_1$,
$\bar{\kappa}(\theta<\theta_1)$, subtracted from that within an annulus $\theta_2<\theta<\theta_3$, 
$\bar{\kappa}(\theta_2<\theta<\theta_3)$, 
\begin{eqnarray}
\label{eq:eta}
\zeta(\theta_1,\theta_2,\theta_3) &=& \bar{\kappa}(\theta<\theta_1) - \bar{\kappa}(\theta_2<\theta<\theta_3)\nonumber\\
&=& 2 \int_{\theta_1}^{\theta_2} d\ln \theta \langle \gamma_t \rangle\nonumber\\
&& + {2\over {1-\theta_2^2/\theta_3^2}} \int_{\theta_2}^{\theta_3} d\ln \theta \langle \gamma_t
\rangle.
\end{eqnarray}
This is the so-called $\zeta$-statistic.
In our computation of $\zeta$, we adopted the weak lensing
approximation $\gamma \simeq g$, which should be
valid in all but the very inner part of clusters 
($r \lesssim 100$kpc or $\theta \lesssim 0.5$ arcmin). 
It is then straightforward to relate $\zeta$ with the aperture mass, 
\begin{equation}
\label{eq:Map}
M(<\theta) = \pi \theta^2 \zeta(\theta,\theta_2,\theta_3) \tilde\Sigma_{cr}.
\end{equation}
This is the mass within an aperture $\theta$ minus an unknown mass.
While the value of this additional mass is difficult to evaluate, 
it presumably tends to zero if $\theta_2$ is sufficiently large to be
hardly affected by the cluster mass.
We took $\theta_2=10$ arcmin and $\theta_3=20$ arcmin.
The statistical error in $\zeta$ 
was estimated from the rms of 100 recalculations of $\zeta$, each time
randomizing the orientation of galaxy ellipticities.
This error was propagated to $M(<\theta)$ in the standard manner.
The results are presented in Figures
\ref{fig:sxds_6}-\ref{fig:deep16_m1_15}.

\begin{table*}
\caption{Summary of weak lensing analyses: 
(a) the amplitude of tangential shear profile at 1 arcmin, when fitted
  with an {\it SIS} model (see \S \ref{sec:sis}).
(b) the SIS velocity dispersion parameter.
(c) the virial mass estimated by fitting the radial shear profile with 
an NFW model.
(d) the $M_{200}$ estimated adopting the NFW model.
(e) the $M_{500}$ estimated adopting the NFW model.
(f) the virial radius computed from the NFW mass using the
relation eq (\ref{stm}).}
\label{table:wlmass1}
\begin{tabular}{lcccccccc}
\hline
Name & $z_{c}$ & $\gamma_{\rm sis}$$^{(a)}$ & $\sigma_{\rm SIS}$$^{(b)}$ & $M_{\rm
  NFW}$$^{(c)}$ & $M_{200}$$^{(d)}$ & $M_{500}$$^{(e)}$ & $r_{\rm vir}$$^{(f)}$ \\ 
{} & {} & {} & [km/s] &  
\multicolumn{3}{c}{[$\times 10^{14}h^{-1}M_\odot$]} & [comoving Mpc$h^{-1}$] \\ 
\hline
SL~J0217.3$-$0524 & 0.4339 & 0.048 & $ 723_{-451}^{+ 439} $ & $ 1.95_{-0.95}^{+ 1.44} $ & $ 1.71_{-0.83}^{+ 1.26} $ & $ 1.20_{-0.58}^{+ 0.88} $ & 1.3 \\
SL~J0219.6$-$0453 & 0.3322 & 0.072 & $ 789_{-409}^{+ 383} $ & $ 2.75_{-1.02}^{+ 1.25} $ & $ 2.37_{-0.88}^{+ 1.08} $ & $ 1.67_{-0.62}^{+ 0.76} $ & 1.5 \\
SL~J0222.8$-$0416 & 0.3219 & 0.058 & $ 695_{-408}^{+ 388} $ & $ 2.09_{-0.40}^{+ 0.09} $ & $ 1.81_{-0.35}^{+ 0.08} $ & $ 1.28_{-0.24}^{+ 0.06} $ & 1.3 \\
SL~J0224.4$-$0449 & 0.4945 & 0.071 & $ 946_{-477}^{+ 437} $ & $ 4.79_{-1.42}^{+ 0.83} $ & $ 4.19_{-1.24}^{+ 0.73} $ & $ 2.85_{-0.84}^{+ 0.49} $ & 1.8 \\
SL~J0224.5$-$0414 & 0.2627 & 0.066 & $ 679_{-340}^{+ 310} $ & $ 1.78_{-0.59}^{+ 0.73} $ & $ 1.53_{-0.51}^{+ 0.63} $ & $ 1.09_{-0.36}^{+ 0.45} $ & 1.3 \\
SL~J0225.3$-$0441 & 0.2642 & 0.087 & $ 800_{-425}^{+ 405} $ & $ 3.02_{-1.04}^{+ 1.32} $ & $ 2.58_{-0.89}^{+ 1.13} $ & $ 1.82_{-0.63}^{+ 0.80} $ & 1.5 \\
SL~J0225.4$-$0414 & 0.1415 & 0.059 & $ 573_{-293}^{+ 273} $ & $ 1.86_{-0.45}^{+ 0.45} $ & $ 1.57_{-0.38}^{+ 0.38} $ & $ 1.14_{-0.27}^{+ 0.27} $ & 1.2 \\
SL~J0225.7$-$0312 & 0.1395 & 0.112 & $ 790_{-379}^{+ 356} $ & $ 2.95_{-0.79}^{+ 0.83} $ & $ 2.48_{-0.66}^{+ 0.70} $ & $ 1.77_{-0.47}^{+ 0.50} $ & 1.4 \\
SL~J0228.1$-$0450 & 0.2948 & 0.102 & $ 898_{-527}^{+ 523} $ & $ 4.37_{-1.86}^{+ 1.67} $ & $ 3.73_{-1.59}^{+ 1.43} $ & $ 2.60_{-1.11}^{+ 0.99} $ & 1.7 \\
SL~J0850.5+4512 & 0.1935 & 0.068 & $ 650_{-371}^{+ 361} $ & $ 1.95_{-0.77}^{+ 1.17} $ & $ 1.66_{-0.65}^{+ 0.99} $ & $ 1.19_{-0.47}^{+ 0.71} $ & 1.3 \\
SL~J1000.7+0137 & 0.2166 & 0.091 & $ 775_{-347}^{+ 332} $ & $ 2.69_{-0.32}^{+ 1.03} $ & $ 2.29_{-0.27}^{+ 0.88} $ & $ 1.63_{-0.19}^{+ 0.62} $ & 1.4 \\
SL~J1048.1+5730 & 0.3173 & 0.069 & $ 758_{-435}^{+ 441} $ & $ 2.51_{-1.03}^{+ 1.13} $ & $ 2.16_{-0.89}^{+ 0.97} $ & $ 1.53_{-0.63}^{+ 0.69} $ & 1.4 \\
SL~J1049.4+5655 & 0.4210 & 0.091 & $ 983_{-525}^{+ 502} $ & $ 3.72_{-0.70}^{+ 3.28} $ & $ 3.23_{-0.61}^{+ 2.85} $ & $ 2.23_{-0.42}^{+ 1.97} $ & 1.6 \\
SL~J1057.5+5759 & 0.6011 & 0.087 & $ 1194_{-544}^{+ 523} $ & $ 8.71_{-1.56}^{+ 2.58} $ & $ 7.65_{-1.37}^{+ 2.27} $ & $ 5.05_{-0.90}^{+ 1.50} $ & 2.2 \\
SL~J1135.6+3009 & 0.2078 & 0.100 & $ 804_{-373}^{+ 344} $ & $ 4.17_{-1.06}^{+ 0.69} $ & $ 3.52_{-0.90}^{+ 0.58} $ & $ 2.48_{-0.63}^{+ 0.41} $ & 1.6 \\
SL~J1201.7$-$0331 & 0.5219 & 0.087 & $ 1085_{-530}^{+ 527} $ & $ 7.24_{-2.46}^{+ 2.97} $ & $ 6.32_{-2.15}^{+ 2.59} $ & $ 4.23_{-1.44}^{+ 1.74} $ & 2.1 \\
SL~J1204.4$-$0351 & 0.2609 & 0.098 & $ 844_{-460}^{+ 427} $ & $ 3.55_{-1.28}^{+ 1.64} $ & $ 3.03_{-1.09}^{+ 1.40} $ & $ 2.13_{-0.77}^{+ 0.98} $ & 1.6 \\
SL~J1334.3+3728 & 0.3006 & 0.123 & $ 991_{-482}^{+ 454} $ & $ 4.79_{-1.35}^{+ 1.67} $ & $ 4.09_{-1.15}^{+ 1.43} $ & $ 2.84_{-0.80}^{+ 0.99} $ & 1.8 \\
SL~J1335.7+3731 & 0.4070 & 0.093 & $ 978_{-493}^{+ 452} $ & $ 6.61_{-1.92}^{+ 2.18} $ & $ 5.70_{-1.66}^{+ 1.88} $ & $ 3.88_{-1.13}^{+ 1.28} $ & 2.0 \\
SL~J1337.7+3800 & 0.1798 & 0.071 & $ 655_{-346}^{+ 313} $ & $ 1.62_{-0.42}^{+ 0.69} $ & $ 1.38_{-0.36}^{+ 0.59} $ & $ 1.00_{-0.26}^{+ 0.42} $ & 1.2 \\
SL~J1602.8+4335 & 0.4155 & 0.087  & $ 954_{-418}^{+ 418} $ & $ 4.79_{-1.05}^{+ 1.20} $ & $ 4.15_{-0.91}^{+ 1.04} $ & $ 2.85_{-0.62}^{+ 0.71} $ & 1.8 \\
SL~J1605.4+4244 & 0.2233 & 0.066 & $ 665_{-342}^{+ 319} $ & $ 1.82_{-0.63}^{+ 0.79} $ & $ 1.55_{-0.54}^{+ 0.67} $ & $ 1.12_{-0.39}^{+ 0.48} $ & 1.2 \\
SL~J1607.9+4338 & 0.3109 & 0.063 & $ 718_{-447}^{+ 418} $ & $ 1.74_{-0.74}^{+ 0.04} $ & $ 1.50_{-0.64}^{+ 0.03} $ & $ 1.07_{-0.46}^{+ 0.02} $ & 1.3 \\
SL~J1634.1+5639 & 0.2377 & 0.076 & $ 724_{-404}^{+ 402} $ & $ 2.09_{-0.87}^{+ 1.01} $ & $ 1.79_{-0.74}^{+ 0.86} $ & $ 1.28_{-0.53}^{+ 0.62} $ & 1.3 \\
SL~J1647.7+3455 & 0.2592 & 0.079 & $ 759_{-560}^{+ 551} $ & $ 4.90_{-1.89}^{+ 2.48} $ & $ 4.16_{-1.60}^{+ 2.11} $ & $ 2.90_{-1.12}^{+ 1.47} $ & 1.8 \\
\hline
\end{tabular}
\end{table*}

\subsection{Lens models of clusters}
An alternative way to estimate the cluster mass from weak lensing
data is to fit the shear profile to an analytical model of a
galaxy cluster.
One merit of this method over the $\zeta$-statistic is that this
is free from the uncertainty in the additive mass (see \S
\ref{sec:zeta}). 
It is also of fundamental importance to check that a model prediction 
agrees with the observed shear profile, to provide a direct
observational test on the theoretical model.

\subsubsection{NFW model}
\label{sec:nfw}
Currently, the most successful model is that proposed by 
Navarro, Frenk \& White (1996; 1997; NFW hereafter).
We adopted a truncated NFW model, in which the density profile is truncated at
the virial radius
\begin{equation}
\label{eq:rho-nfw}
\rho_{\rm nfw}(x)=
\left\{
\begin{array}{ll}
{{\rho_s} \over {r/r_s (1+r/r_s)^2}},\hspace{1em}
&\mbox{for $r<\rvir$}\\
0 & \mbox{otherwise,}
\end{array}
\right.
\end{equation}
where $r_s$ and $\rvir$ are the scale radius and virial radius respectively.
It is convenient to introduce the concentration parameter 
$\cnfw=\rvir/r_s$.
Bullock et al. (2001) found from $N$-body simulations that the concentration
parameter is related to halo mass as
\begin{equation}
\label{cnfw}
\cnfw(M,z)={{c_\ast}\over {1+z}}
\left( {{M}\over {10^{14}h^{-1}M_\odot}}\right)^{-0.13}, 
\end{equation}
where $c_\ast \simeq 8$ for the $\Lambda$CDM model.
By definition, the mass enclosed within a sphere of radius $\rvir$ gives the virial
mass
\begin{equation}
\label{nfw:Mvir}
M_{\rm vir} = 4\pi \rho_s r_s^3 \left[\log(1+\cnfw)-{\cnfw\over{1+\cnfw}}\right].
\end{equation}
The virial mass is also defined from the spherical top-hat collapse model
as
\begin{equation}
\label{stm}
M_{\rm vir}={{4\pi}\over 3} \delta_{\rm vir}(z) \bar{\rho}(z) \rvir^3,
\end{equation}
where $\delta_{\rm vir}$ is the threshold
over-density for spherical collapse (see Nakamura \& Suto 1997 and Henry
2000 for useful fitting functions).
Using equations (\ref{nfw:Mvir}) and (\ref{stm}),
one can express 
$\rho_s$ in terms of $\delta_{\rm vir}(z)$ and $\cnfw$.
Introducing $\delta_s=\rho_s/\bar{\rho}-1$, one finds
\begin{equation}
\label{delta_s}
\delta_s={{\delta_{\rm vir}} \over 3} 
{{\cnfw^3} \over {\log(1+\cnfw)-\cnfw/(1+\cnfw)}}.
\end{equation}
Thus the density profile of the NFW halo can be characterized by two
parameters: the virial mass $M_{\rm vir}$ and the concentration parameter 
$\cnfw$. In the following analyses we fixed the 
mass-concentration 
relation to the empirical result, eq.~(\ref{cnfw}). This is necessary because,
with high $S/N$ measurements of the tangential shear in only a limited angular 
range (typically $1<\theta<4$ arcmin), only a poor
constraint on the concentration parameter is obtained. See Appendix
\ref{appendix:nfw} for details.

Cluster masses are also often defined by $M_{200}$, which is the mass
contained within the radius $r_{200}$, where the mean mass density of
the halo is equal to 200 times the critical density at the redshift of
the cluster. In order to allow the direct comparison with other works,
we compute $M_{200}$ assuming the NFW profile.

Analytical expressions for weak lensing convergence and shear behind an NFW profile 
are given in Takada \& Jain (2003a; b).
Note that those expressions differ from those by Bartelmann (1996) 
and Wright \& Brainerd (2000). The latter are for a non-truncated NFW halo
and give a larger surface mass density than ours due to the infinite
extend of the mass.
In Appendix \ref{appendix:nfw}, we illustrate how the shear and
projected mass profile depend on the virial mass and scale
radius.

\subsubsection{SIS model}
\label{sec:sis}

The singular isothermal sphere (SIS) model has frequently been
adopted as a lens model, because of its simplicity.
One convenient feature of this model is that it provides an estimate of
the velocity dispersion, which is a useful measure of the gravitational
potential and allows a direct comparison with the observed
velocity dispersion in galaxies.

The density profile of the SIS model is
\begin{equation}
\label{eq:rhosis}
\rho_{\rm sis}(r) = {{\sigma_{\rm sis}^2} \over {2 \pi G}} {1 \over {r^2}},
\end{equation}
where $\sigma_{\rm sis}$ is the SIS velocity dispersion.
The mass within a radius $r$ is
\begin{equation}
\label{eq:Msis}
M_{\rm SIS}(<r) = {{2 \sigma_{\rm sis}^2} \over {G}} r.
\end{equation}
The projected density is
\begin{equation}
\label{eq:Sigmasis}
\Sigma(y) = \int dz \rho_{\rm sis}(y,z) = {{\sigma_{\rm sis}^2} \over {2 G}} {1 \over {y}},
\end{equation}
where $y$ is a radial coordinate in the plane perpendicular to the
line-of-sight direction. The shear profile is
\begin{eqnarray}
\label{eq:gammasis}
\gamma(\theta) &=& {1 \over \tilde{\Sigma}_{cr}}
 {{\sigma_{\rm sis}^2} \over {2 G}} {1 \over {D_l \theta}}\nonumber\\
&\equiv & {{\gamma_{sis}} \over {(\theta/1{\rm arcmin}})},
\end{eqnarray}
where $\theta$ is the angular separation from the lens center,
and we define $\gamma_{sis}$ as the amplitude of shear profile at
$\theta=1$ arcmin.
We evaluated $\gamma_{sis}$ by fitting the measured shear 
over the range $1<\theta<4$ arcmin.
We used this range for three reasons: 
(1) for most clusters, the shear signal was measured with a
good $S/N$ throughout this interval, but degraded outside it.
(2) at larger radii, nearby structures (either
physically related or unrelated to the main cluster) could contribute to
the shear signal.
(3) at smaller radii, 
the reduced shear $g$, and 
the dilution effect due to the cluster member
galaxies could have non-negligible effect on the shear signal
(Broadhurst et al. 2004). 
The measured $\gamma_{\rm sis}$ values are summarised in Table
\ref{table:wlmass1}.

By combining equations for the definition of virial mass (\ref{stm}) and
the mass in a SIS (\ref{eq:Msis}), one finds
\begin{eqnarray}
\label{eq:Msisvir}
M_{\rm SIS}(<r_{\rm vir}) 
&=& \left[ {6 \over \pi} 
{{\sigma_{\rm sis}^6} \over  
{G^3 \delta_{\rm vir}(z_l) \bar{\rho} (z)}} \right]^{1/2} \nonumber\\
&\simeq& 6.6\times 10^{14} [h^{-1} M_\odot]\nonumber\\
&&\times
C
\times
\left(\sigma_{\rm sis} \over {1000{\rm km/s}} \right)^3,
\end{eqnarray}
where
\begin{equation}
\label{eq:C}
C \equiv \left[ {1 \over {\Omega_m}} {1 \over {(1+z_l)^3}} 
{{200} \over \delta_{\rm vir}(z_l)} \right]^{1 \over 2}.
\end{equation}
In the notation of Bryan \& Norman (1998) $C$ is written by 
$C = (200/\Delta_c)^{1/2}E(z)^{-1}$, where $\Delta_c$ is the critical
overdensity of the spherical collapse, and 
\begin{equation}
\label{eq:Ez}
E(z)=\frac{H(z)}{H_0}=\left[\Omega_m(1+z)^3+\Omega_\Lambda\right]^{1/2}
\end{equation}
for a
$\Lambda$-flat cosmology.
A reasonably accurate fitting function of $C$ for a $\Lambda$-cosmology 
($\Omega_m=0.3$, $\Omega_\Lambda=0.7$) over the redshift range $0<z_l<0.8$ is
found to be $C \simeq 1.44/(1+z_l)^{1.08}$, and in the case of Einstein-de
Sitter cosmology ($\Omega_m=1$, $\Omega_\Lambda=0$), it reduces to
$C = \sqrt{200/178}/(1+z_l)^{1.5} \simeq 1.06/(1+z_l)^{1.5}$.

\subsection{Results}
\label{sec:prop}

The results of our weak lensing analyses are compiled in Table~3. 
A detailed discussion of each system is presented in 
Appendix \ref{appendix:prop}, including a comparison of
weak lensing and optical properties, and the results of MOS observations.

\section{Cluster scaling relations}
\label{sec:scaling}

\subsection{Clean sample of clusters}
\label{sec:cleansample}

We shall now examine statistical relations between the dynamical mass
estimator ($\sigma_v$) and the weak lensing mass estimators 
($\sigma_{\rm sis}$ and $M_{\rm NFW}$).
In this paper, we shall restrict our analysis to clusters with the very cleanest measurements;
a statistical treatment of the entire sample, taking into account all selection effects,
will follow in Green et al. (in preparation).
We define a {\it clean (sub)-sample} of clusters (``CS'' in Table
\ref{table:mdym}) as those whose velocity dispersion was evaluated from 
at least 12 spectroscopic member galaxies, and whose weak lensing mass 
estimation could not have been affected by proximity to either a neighboring system (``NS'')
or a field boundary (``FB'').
Ten clusters satisfy these criteria.

We additionally include two clusters in our survey area whose velocity
dispersions have been measured by other authors. 
Observations by Willis et al.\ (2005)
of SL~J0221.7$-$0345 
($z=0.43$, $\sigma_v=821_{-74}^{+92}$km/s from 39 galaxy redshifts)
and SL~J0228.4$-$0425 
($z=0.43$, $\sigma_v=694_{-91}^{+204}$km/s from 13 redshifts)
satisfy the same stringent selection criteria as above. 
We examine the weak lensing properties of those clusters in the same
manner described in \S \ref{sec:wlmass}, and the results are described fully
in Appendix \ref{appendix:xmmlss}.

\subsection{Velocity dispersions}

\begin{figure}
\includegraphics[width=83mm,clip]{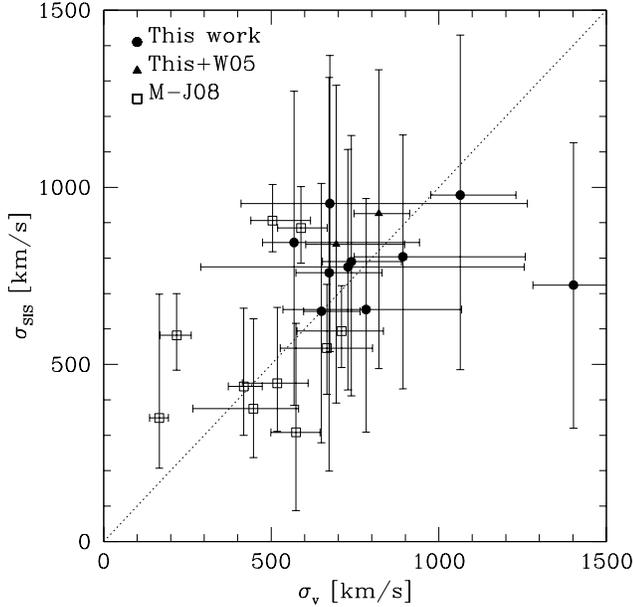}
\caption{Comparison between the velocity dispersion measured from 
galaxy redshifts ($\sigma_v$) and the velocity dispersion parameter 
of the weak lensing SIS model ($\sigma_{\rm SIS}$).
Filled circles show our {\it clean sample} (see text).
Filled triangles show two clusters whose $\sigma_v$ 
was measured by Wills et al.~(2005) (see Appendix
\ref{appendix:xmmlss}).
Open squares show optically selected clusters from
Milvang-Jensen al. (2008; note that clusters with structures
possibly affecting the velocity dispersion estimate are excluded, see \S
8 of their paper for details).} 
\label{fig:sigmav_sigmasis_EDisCS}
\end{figure}

\begin{figure}
\includegraphics[width=83mm,clip]{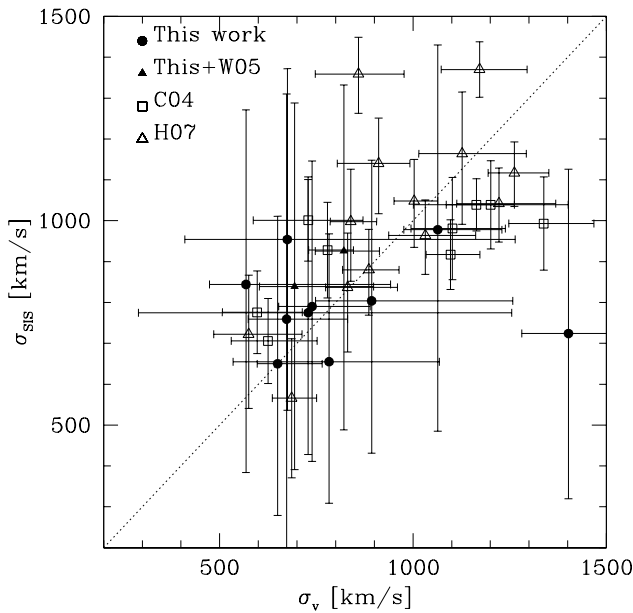}
\caption{Same as Fig \ref{fig:sigmav_sigmasis_EDisCS} but 
compared with X-ray selected clusters from
Cypriano et al. (2004; open squares) and Hoekstra 
(2007, open triangles).}
\label{fig:sigmav_sigmasis}
\end{figure}

The relation between cluster galaxies' velocity dispersion ($\sigma_v$) 
and the velocity dispersion parameter of the best-fit SIS model ($\sigma_{\rm sis}$)
has been measured for various clusters, in the context of their
dynamical status (Irgens et al. 2002; Hoekstra et al. 2002; Hoekstra
2007; Milvang-Jensen et al. 2008). Results for our sample are presented in 
figure~\ref{fig:sigmav_sigmasis_EDisCS} (with a comparison to optically selected clusters)
and figure~\ref{fig:sigmav_sigmasis} (with a comparison to X-ray selected clusters).
The optically selected catalog of Milvang-Jensen et al. (2008) 
contains less massive clusters, at higher redshift ($\bar{z}=0.58$)
than our sample ($\bar{z}=0.28$).
Conversely, the wide but shallow
Einstein Medium Sensitivity Survey (EMSS; Gioia et
al. 1990) and X-ray Brightest Abell Cluster Survey (XBACS; Ebeling et al
1996) include the most massive clusters at more modest redshift.
The mean redshift of the catalog by Cypriano et 
al.\ (2004) is $\bar{z}=0.13$ and by Hoekstra (2007) is $\bar{z}=0.31$.

Only a few outliers in Figures \ref{fig:sigmav_sigmasis_EDisCS}
and \ref{fig:sigmav_sigmasis} are inconsistent with 
$\sigma_v = \sigma_{\rm sis}$.
The main outlier from our sample, SL~J1634.1$+$5639, has 
$\sigma_v \sim 2\sigma_{\rm sis} \sim 1400$km/s, but the velocity distribution of
if spectroscopic member galaxies is also strongly skewed (see Figure~\ref{fig:vzhist1}).
The reason for this is currently not clear, but discussed further in 
Appendix \ref{sec:SLJ1634.1$+$5639}.
Overall, despite variations in samples' range of cluster masses and redshifts, 
the scatter in the 
$\sigma_v-\sigma_{\rm sis}$ relation is remarkably similar for all four catalogs.
Thus, as far as the relation between 
$\sigma_v$ and $\sigma_{\rm sis}$ is concerned, 
no strong selection bias is identified between the various 
cluster detection techniques.

It is worth noting that the density (and shear) profile of a real cluster
is typically {\it not} a single power law. The best-fit SIS model, and 
value of $\sigma_{\rm sis}$, may therefore depend on the specific
fitting method, and the range
over which data are fit.
Our above finding, that $\sigma_v \simeq \sigma_{\rm sis}$, may therefore 
be somewhat method-dependent. 
A corollary of this issue is that it might also be possible to minimize 
scatter in the $\sigma_v-\sigma_{\rm sis}$ relation
by optimizing the fitting method used to obtain $\sigma_{\rm sis}$.
We have not attempted to do this.

\subsection{Velocity dispersion versus mass}

\begin{figure}
\includegraphics[width=83mm,clip]{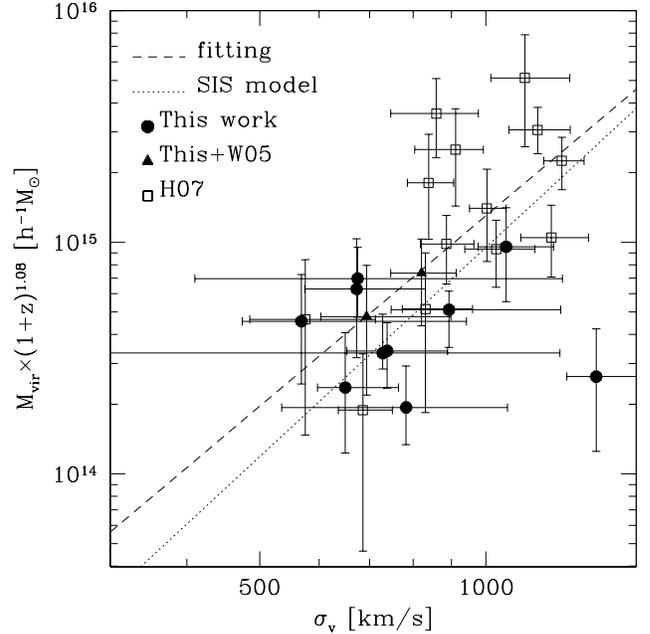}
\caption{Comparison between the velocity dispersion measured from 
galaxy redshifts ($\sigma_v$) and the virial mass measured from weak
lensing ($M_{\rm NFW}$).
In order to account for redshift evolution in the
relation, cluster masses were multiplied by a factor of $(1+z_c)^{1.08}$ 
(as motivated by eqs (\ref{eq:Msisvir}) and (\ref{eq:C})).
The filled circles show clusters in this study; the two filled triangles
 show clusters in this study whose $\sigma_v$ were measured by
Wills et al.~(2005) (see Appendix \ref{appendix:xmmlss}), and the open
circles show the sample of Hoekstra (2007).
The dotted line shows the prediction of a pure SIS model,
$M_{\rm vir}\times (1+z_c)^{1.08}
=9.5\times10^{14}\times(\sigma_v/1000{\rm km/s})^{3} h^{-1}M_\odot$.
The dashed line shows the best-fit  empirical relation,
$M_{\rm vir}\times (1+z_c)^{1.08}
=13\times10^{14}\times(\sigma_v/1000{\rm km/s})^{2.7} h^{-1}M_\odot$.}
\label{fig:sigmav_mnfw}
\end{figure}

We next examine the relation between the velocity dispersion of a cluster's 
galaxies ($\sigma_v$) and its weak lensing mass.
Since the NFW virial masses and aperture masses agree for all 12 
cleanly-measured clusters, we adopt $M_{\rm vir}=M_{\rm NFW}$ as our sole weak 
lensing mass estimate.
Figure \ref{fig:sigmav_mnfw} shows our results, and compares them to 
measurements of X-ray selected clusters by Hoekstra (2007).
It is important to note that weak lensing measurements do not depend upon 
the dynamical status of the clusters. 
Motivated by SIS model prediction (eqs (\ref{eq:Msisvir}) and
(\ref{eq:C})), we adopt a functional form of the scaling relation
$M_{\rm vir} \propto M_\ast {\sigma_v}^p/(1+z_c)^{1.08}$.
For pure SIS clusters in a $\Lambda$CDM cosmology, the parameters would be 
$p=3$ and $M_\ast = 9.5\times 10^{14}$ (see eq.\ (\ref{eq:Msisvir}) and
below). 
This simplistic toy model is shown as a dotted line, and already 
provides a reasonable approximation to the observed $M_{\rm vir}-\sigma_v$ data. 
A least-squares fit (excluding the outlier SL~J1634.1$+$5639) yields
$M_{\rm vir}(1+z_c)^{1.08}=(13\pm 2) \times 10^{14}\times
(\sigma_v/1000$km/s$)^{2.7\pm 0.6} h^{-1}M_\odot$.
The best-fit power-index is thus consistent with the SIS prediction of 3, but the
normalization is slightly higher.

\begin{figure}
\includegraphics[width=83mm,clip]{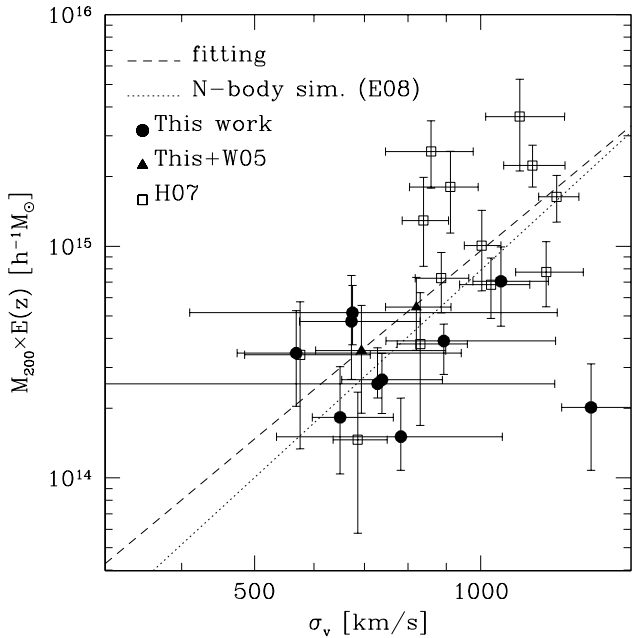}
\caption{Same as Figure \ref{fig:sigmav_mnfw} but with $M_{200}$ instead
of virial mass $M_{\rm NFW}$, for comparison with numerical simulations. 
The dotted line shows a prediction from $N$-body simulations
$M_{200}~E(z)=9.358\times10^{14}\times(\sigma_v/1000{\rm km/s})^{2.975}
h^{-1}M_\odot$ (Evrard et al.\ 2008).
The dashed line shows the best-fit empirical relation 
$M_{200}~E(z) = 9.6\times10^{14} \times (\sigma_v/1000{\rm km/s})^{2.7} h^{-1}M_\odot$.}
\label{fig:sigmav_m200}
\end{figure}

A more sophisticated prediction of the cluster $M-\sigma_v$ relation,
using $N$-body simulations, was obtained by Evrard et al.\ (2008). 
They find
$M_{200}~E(z)=9.358\times10^{14}\times(\sigma_v/1000{\rm km/s})^{2.975} 
h^{-1}M_\odot$, and argue that it is insensitive
to cosmological parameters in a variety of CDM models.
To aid in comparison, we estimate $M_{200}$ from our measurements of 
$M_{\rm NFW}$ by assuming every cluster has an NFW density profile.
These masses are listed in Table~\ref{table:wlmass1} and our 
results are shown in Figure~\ref{fig:sigmav_m200}, together with 
those of Hoekstra (2007).
Evrard et al.'s prediction is overlaid as a dotted line, and the 
best-fit power-law model (excluding the outlier SL~J1634.1$+$5639),
$M_{200}~E(z)
=(9.6\pm 1.6)\times10^{14}\times(\sigma_v/1000{\rm km/s})^{2.7\pm 0.6}
h^{-1}M_\odot$, as a dashed line.

Interpretation of this apparent consistency is not trivial because
Evrard et al.\ evaluated $\sigma_v$ from simulated dark matter particles
instead of galaxies. 
However, since both galaxies and cold dark matter particles may be
safely regarded as collisionless particles in the cluster potential,
it is reasonable to assume that they have
approximately the same velocity
dispersion outside a central region in which the effect of 
dynamical friction is important (Okamoto \& Habe 1999).
If this is the case, our findings indicate that the dynamical
structure of galaxy clusters is indeed consistent with that expected in the
standard CDM paradigm of structure formation.
It will be interesting to compare our observations with simulations of
cluster evolution that also incorporate mechanisms for galaxy formation.

\section{Summary and Discussions}
\label{sec:summary}

We have presented the results of a multi-object spectroscopic campaign 
to target 36 cluster candidates located by the Subaru weak 
lensing survey (Miyazaki et al. 2007).
We obtained the redshifts of $15-32$ galaxies within a few arcminutes
of each cluster candidate.
Our primary goals were to search for a spatial concentration of galaxies
as an optical counterpart of each 
weak lensing density peak, and to determine the cluster redshifts.
We found 31 galaxy concentrations containing more than five spectroscopic galaxies 
within a velocity of $\pm 3000$km/s, and determined their redshifts.
These included 25 detections of isolated clusters, and
three systems (SL~J1000.7.3+0137, SL~J1047.3+5700, SL~J1601.6+4245) 
in which two galaxy clusters are projected at different redshifts along the same line-of-sight.
This demonstrates that spectroscopic follow-up of weak lensing cluster
candidates is a reliable way not only to identify their optical
counterparts but also to distinguish superposed systems.

We have therefore identified secure optical counterparts of the
weak lensing signal in 28 out of 36 targets.
In 6 of the 8 unconfirmed cluster candidates, we found 
multiple small galaxy concentrations at different redshifts (each containing
at least 3 spectroscopic galaxies).
This suggests that the weak lensing signal in those cases may arise from 
the projection of small clusters along the same line-of-sight. 
However, it is also possible that a real, massive cluster
is responsible for the weak lensing density peak, but was missed by 
our relatively sparse MOS observations.
This is also the case for the final two unconfirmed candidates, 
where only a single
small galaxy concentration was identified. 
In order to obtain a firm confirmation of the optical counterpart of such
unconfirmed candidates, denser spectroscopic observations would be
required.  

We measured the mass of single cluster systems with known redshifts
using two weak lensing methods: aperture densitometry
and by fitting the shear profile to an NFW model.
In most cases, the two mass estimators agree well: 
providing observational support for the NFW model.
In the few clusters where the mass estimators did not agree, 
the weak lensing $\kappa$ signal clearly deviates from spherical symmetry.
This could account for the disagreement.
It was also found, by eye, that the aperture mass profile of some clusters
does not flatten even at a large radius of $\theta\sim 10$ arcmin.
This can be accounted for by the mass contribution from surrounding structures.
We found some candidates of super-cluster systems, whose weak lensing mass reconstructions
show evidence of filamentary structure connecting the main cluster to 
surrounding systems.

We investigated statistical relations between clusters' 
weak lensing properties ($\sigma_{\rm sis}$ and $M_{\rm vir}$ or $M_{200}$) 
and the velocity
dispersion of their member galaxies ($\sigma_{\rm v}$), comparing our results
to optically and X-ray selected cluster samples from the literature. 
Although our {\it clean sample} contained only 12 clusters, 
we found our clusters to be consistent with 
$\sigma_v = \sigma_{\rm sis}$, with 
a scatter as large as that of optically and X-ray selected samples.
Therefore, as far as the relation between 
$\sigma_v$ and $\sigma_{\rm sis}$ is concerned, no strong bias between the cluster
selection techniques was identified.
We also derived the empirical relation between the cluster virial mass 
and the galaxy velocity dispersion: 
$M_{\rm vir}(1+z_c)^{1.08}=(13\pm 2) \times 10^{14}\times
(\sigma_v/1000$km/s$)^{2.7\pm 0.6} h^{-1}M_\odot$.
The derived $M_{\rm vir}-\sigma_{\rm SIS}$ relation is similar to theoretical 
expectations from the SIS model, eq. (\ref{eq:Msisvir}).
It is important to note that, unlike the SIS model assumption, real
cluster shear profiles (and density profiles) are {\it not} single power-laws, 
so this result may depend upon details of the fitting technique. 
For comparison with numerical simulations, we also derived the 
$M_{200}-\sigma_v$ relation and found 
$M_{200}~E(z)
=(9.6\pm 1.6)\times10^{14}\times(\sigma_v/1000{\rm km/s})^{2.7\pm 0.6}
h^{-1}M_\odot$.
This is in good agreement with predictions by 
Evrard et al.\ (2008), demonstrating that 
the dynamical structure of galaxy clusters
is similar to that expected in the
standard CDM paradigm of structure formation.

\bigskip
We are very grateful to Subaru astronomers: Y. Ohyama, K. Aoki
and T. Hattori for their dedicated supports of the FOCAS observing.
Numerical computations presented in this paper were carried
out on computer system at CfCA 
(Center for Computational Astrophysics) of the National
Astronomical Observatory Japan.
Data reduction and analysis were in part carried out on 
general common use computer system at ADAC (Astronomical Data 
Analysis Center) of the National Astronomical Observatory of Japan.
This research was supported in part by the Grants--in--Aid from 
Monbu--Kagakusho and Japan Society of Promotion of Science: 
Project number 15340065 (TH\&SM) and 17740116 (TH).


\appendix

\section{Weak lensing properties of the truncated NFW model}
\label{appendix:nfw}

\begin{figure}
\includegraphics[width=83mm]{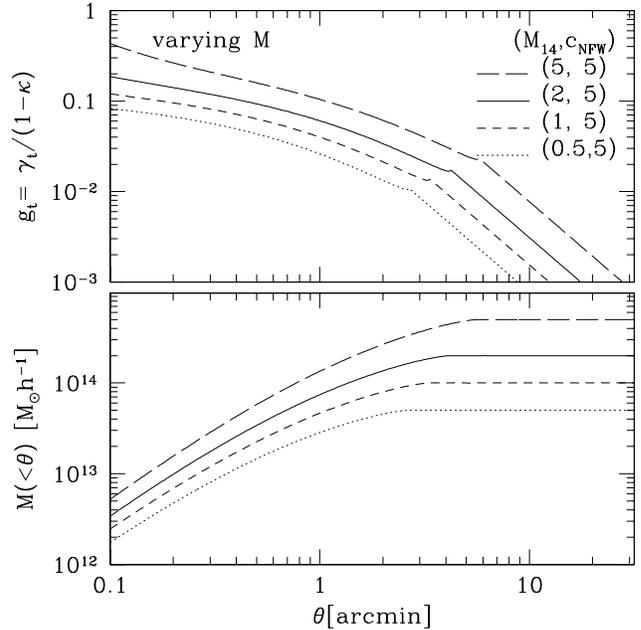}
\caption{The top panel shows the reduced shear profile of a 
truncated NFW model, for different virial masses 
($5\times$, $2\times$, $1\times$ and 
$0.5\times  10^{14}M_\odot h^{-1}$ from upper to lower).
The bottom panel shows 
the projected mass within an aperture $\theta$.
In all cases, the concentration parameter is $c_{\rm NFW}=5$, and
the lens and source redshifts are $z_l=0.4$ and $z_s=1$ (a
single source plane approximation is employed).}
\label{fig:nfw-m}
\end{figure}

\begin{figure}
\includegraphics[width=83mm]{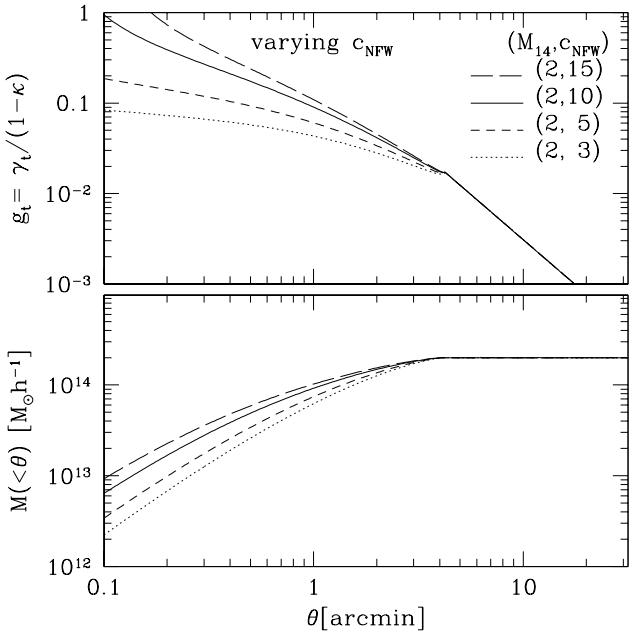}
\caption{Same as Fig \ref{fig:nfw-m} but for different concentration
parameters ($c_{\rm NFW} = 15$, 10, 5 and 3, from upper to lower).
In all cases, the virial mass $M_{\rm vir}=5\times 10^{14}M_\odot h^{-1}$.}
\label{fig:nfw-c}
\end{figure}

Here, we present the reduced shear profile of a truncated 
NFW model (\S \ref{sec:nfw}), to illustrate the dependence on model 
parameters.
In Figure \ref{fig:nfw-m} and \ref{fig:nfw-c}, we plot the reduced shear
(upper panel) and the projected mass within an aperture $\theta$
(lower panel) for various values of the model parameters $M_{\rm vir}$ and
$c_{\rm NFW}$.
Since we adopt the truncated model (eq. (\ref{eq:rho-nfw})), $\kappa$
becomes zero outside the virial radius, so the 
aperture mass flattens and the reduced shear
(where $g = \gamma$) scales as $\propto \theta^{-2}$. 

As shown in the Figure \ref{fig:nfw-m}, changes in $M_{\rm vir}$
alter the amplitude of the shear profile, but leave the overall shape almost 
unchanged. The slight discontinuity at the virial
radius in the reduced shear profile is an numerical artifact 
caused by the discontinuity there in $\kappa$.
Figure \ref{fig:nfw-c} illustrates that changes in the concentration parameter 
alter the slope of the shear profile, but not the amplitude at the virial radius.
The higher the concentration, the steeper the slope becomes, as
expected.

In principle, a measurement of the weak lensing shear profile over a
broad angular range therefore allows simultaneous constraints on both the 
cluster mass and the concentration parameter (i.e. the mass would be 
determined mainly by the shear amplitude near the virial radius, 
and the concentration parameter by the slope at $\theta<\theta_{\rm vir}$).
However, in our case, the angular range over which tangential shear is 
measured with a good $S/N$ is rather narrow (typically $1<\theta<4$ arcmin).
This especially prevents us from obtaining a tight constraint on the
concentration parameter, which requires measurements over a broad angular
range. Instead of treating both the virial mass and the concentration
parameter as free parameters, we therefore decided to adopt an empirically
observed 
relation between the concentration parameter and mass, eq. (\ref{cnfw}).

\section{Properties of individual targets}
\label{appendix:prop}

Here, we describe each target's weak lensing and optical properties, 
and discuss the full results of the MOS observations.
Quantitative summaries of these data are found in Tables 1, 2 and 3.

Figures~\ref{fig:sxds_6}-\ref{fig:deep16_m1_15} also present the
data, in a uniform format for easy visual comparison. Each figure is
laid out as follows. 
The top-left panel shows an optical image of the cluster core, 
overlaid with contours reproducing the weak lensing density. The contours start from
$\kappa=0.04$ and increase in increments of $\Delta \kappa = 0.01$.
Galaxies with successfully measured redshifts are marked with circles, their
target ID and their redshift (in parentheses).  The colors correspond to the
galaxies' observed spectral types.
The top-right panel shows the positions of those galaxies in cone diagrams.
The horizontal axis corresponds to radial comoving distance, and the vertical 
axes to a sky direction, with $x$ and $y$ standing for R.A.\ and Dec.\ respectively.
The bottom-left panel reproduces the weak lensing density map on a larger scale, 
to show any nearby structure. Overlaid on the gray scale map are red contours, 
starting from $\kappa=0.04$ and increasing in increments of $\Delta \kappa = 0.02$.
White contours show instead the smoothed number density of 
galaxies with $18<R_C<23$ ($18<i'<23$ for the COSMOS field), starting 
from $n_g=10$ arcmin$^{-2}$ and in increments of $\Delta n_g=2$ arcmin$^{-2}$.
For candidates where a single, well-defined concentration of galaxies was found,
the bottom-right panel shows the weak lensing tangential shear profile (upper
plot) and aperture mass profile (lower plot). 
Black points with error bars show measured data. 
The dashed line shows the best-fit SIS model, and the solid line
shows the best-fit NFW model, plotted up to the virial radius.
The red diamond shows the virial mass of this best-fit NFW model.

\subsection{SL~J0217.3$-$0524}
\begin{figure*}
\includegraphics[height=160mm,clip,angle=-90]{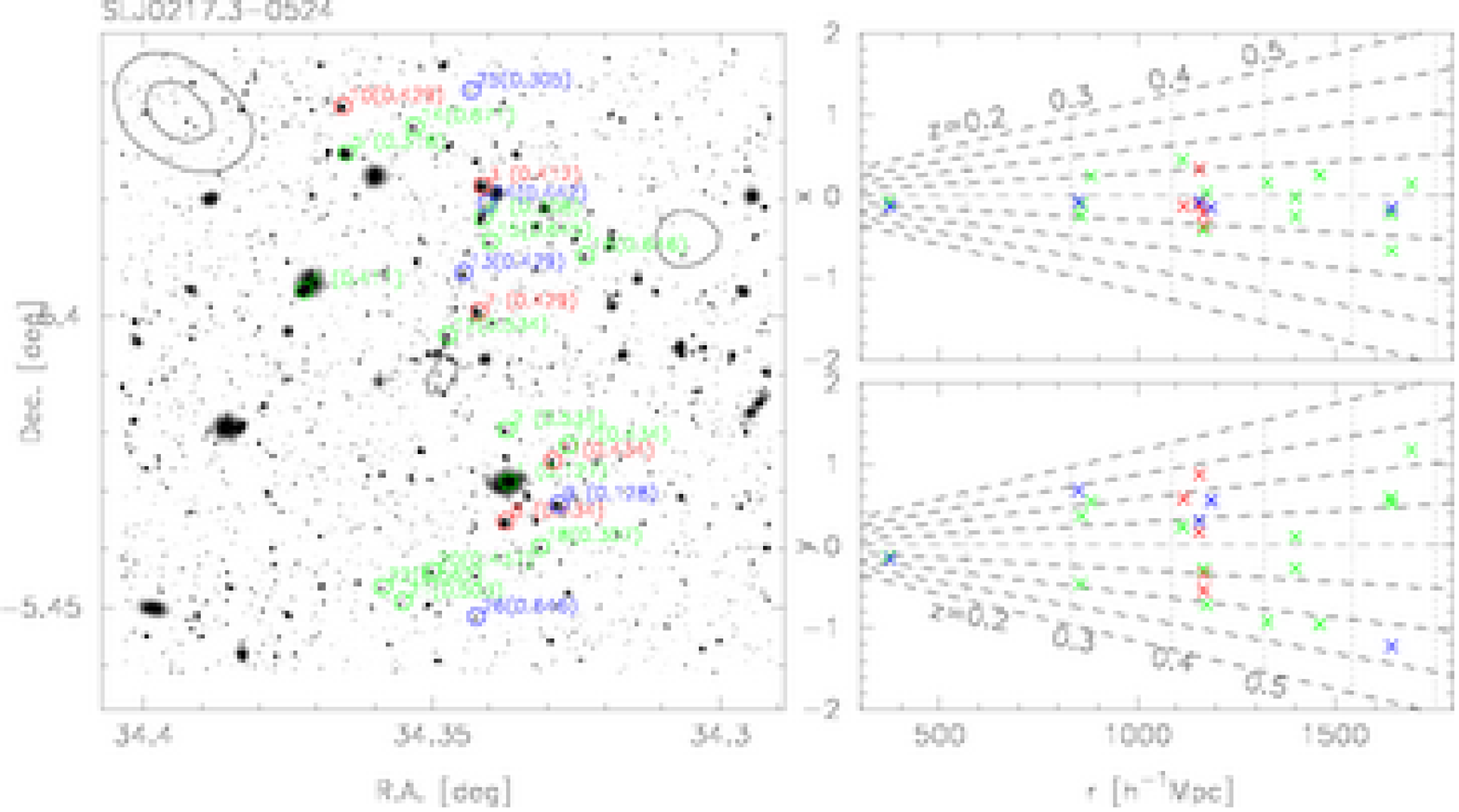}
\vspace{2mm}\\
\includegraphics[width=75mm,clip,angle=-90]{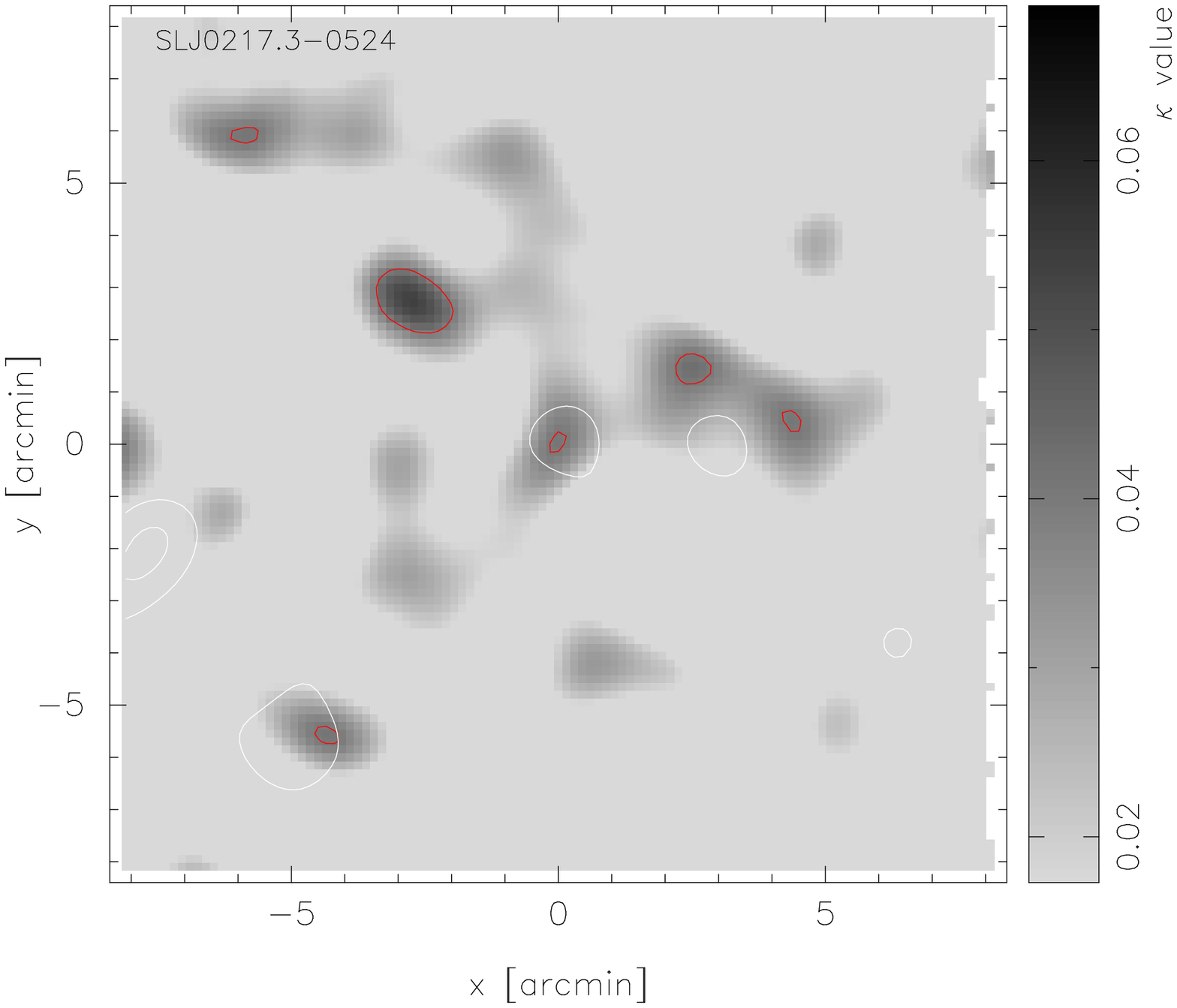}
\hspace{2mm}
\includegraphics[width=75mm,clip,angle=-90]{fig8c.ps}
\caption{SL~J0217.3$-$0524: 
{\it Top-left panel}: $R_C$-band image with the weak lensing density  
overplotted as contours (starting from $\kappa = 0.04$ and in increments of 
$\Delta \kappa = 0.01$).
Galaxies with measured redshifts are 
marked with circles, and labeled with their target ID and redshift 
(in parentheses). 
Red circles represent absorption galaxies, blue circles represent emission and 
green circles represent composite galaxies 
(see \S \ref{sec:zmesurement} and Cohen et al. 1999 for details).
{\it Top-right panels}: Cone diagrams showing the 3D locations of galaxies.
The horizontal axis shows the radial comoving distance. 
On the vertical axis, x and y correspond to the RA and Dec directions respectively.
{\it Bottom-left panel}: The gray scale shows the weak lensing $\kappa$
map (over an extended area), with red contours starting from 
$\kappa = 0.04$ and in increments of $\Delta \kappa = 0.02$.
White contours show the smoothed number density of galaxies 
($18 < mag < 23$), starting from $n_g=10/$arcmin$^2$ and in increments of 
$2/$arcmin$^2$.
{\it Bottom-right panel}: The measured weak lensing tangential shear profile
$g_t = \gamma_t/(1-\kappa)$, with the best-fit SIS model (dashed line)
and NFW model (solid line, plotted up to the virial radius). The aperture mass
profile $M(<\theta)$, computed from the tangential shear.
The red diamond shows the virial mass of the best-fit NFW model.
\label{fig:sxds_6}}
\end{figure*}

This is not listed in the P1 catalogue because the peak $\kappa$ $S/N$ does
not exceed P1's threshold.
As observed in  Figure \ref{fig:sxds_6}, the weak lensing $\kappa$ 
peak is well correlated with a galaxy over-density.
In the redshift data, there exists one galaxy concentration 
that passes our cluster criteria, with 8 members at $z_c=0.43$.
There is also is a small concentration at $z=0.31$. 
The velocity dispersion of the galaxy cluster is consistent with the SIS
velocity dispersion.
The NFW cluster mass agrees with the aperture mass at the corresponding
virial radius.
However, the aperture mass keeps increasing even outside of the
expected virial radius.
This is likely due to the mass associated with two $\kappa$ over-densities
located a few arcminutes to the west and the north-east of 
the cluster center.

\subsection{SL~J0217.6$-$0530}
This is not listed in the P1 catalogue because the peak $\kappa$ $S/N$ is below 
the threshold.
The weak lensing density peak appears isolated and is correlated with a 
galaxy over-density.
In the redshift data, there is no galaxy concentration passing 
our criterion of five galaxies within a velocity of $\pm 3000$km/s,
but there is a small group of three galaxies at $z=0.43$.
With only this information, it is currently not clear whether the weak 
lensing shear signal comes solely from the halo of the small galaxy 
concentration or whether there are other mass concentrations 
along the same line-of-sight. 
Since no galaxy concentration that passes our cluster criterion was found, 
no weak lensing mass estimation or radial profiles are displayed.

\subsection{SL~J0217.9$-$0452}
This is not listed in the P1 catalogue because the peak $\kappa$ $S/N$ is below 
the threshold.
The weak lensing $\kappa$ map shows a bimodal feature, with a second peak
located at 2 arcminutes to the west of the main peak.
There is a galaxy over-density near the main peak.
No galaxy concentration in the redshift data is sufficiently rich to qualify
as a cluster under our criterion, but there is a small group of four
galaxies at $z=0.19$.
With only the current information, it is not clear whether the weak 
lensing shear signal comes from the halo of the small galaxy 
concentration alone or from additional concentrations along the same or an 
adjacent line-of-sight. 
Since no galaxy
concentration passing our cluster criterion was found, 
we did not make the weak lensing mass estimation.

\subsection{SL~J0218.0$-$0444}
This is not listed in the P1 catalogue because the peak $\kappa$ $S/N$ does
not exceed P1's threshold.
The weak lensing density distribution is elongated along the south-east to 
north-west direction.
There is a galaxy over-density overlapping with the weak lensing $\kappa$
peak but elongated perpendicular to this.
In the redshift data, no galaxy concentration passes 
our cluster criterion, but there are two small concentrations at 
$z=0.37$ (five galaxies) and at $z=0.31$ (four galaxies).
Since no dominant galaxy concentration was
found, we have not estimated a weak lensing mass.
Note that in the vicinity of this target there is a known galaxy cluster,
identified in  optical-near infrared imaging and 
with an estimated photometric redshift of $z_p=0.71\pm 0.03$ (van
Breukelen et al.\ 2006, their ID 5, R.A.$=34.50$, Dec.$=-4.72$).
This cluster is located 2 arcminutes to the north-west and is 
within the elongated over-density region.
It is therefore likely that the $\kappa$ excess 
consists of the chance projection of several halos located at different
redshifts along adjacent lines-of-sight.

\subsection{SL~J0219.6$-$0453}
This is not listed in the P1 catalogue because the peak $\kappa$ $S/N$ is 
below the threshold.
As shown in Figure~\ref{fig:sxds_7}, the weak lensing $\kappa$ 
peak is elongated in the east-west direction, with a smaller, 
secondary peak 3 arcminutes east of the cluster center. 
An extended galaxy over-density overlaps with both $\kappa$ peaks.
In the redshift data, a galaxy concentration with 11 members at
$z_c=0.33$ passes our cluster criterion. 
There is also a small concentration of three galaxies at $z=0.3$. 
The measured velocity dispersion of the galaxy cluster is
smaller than the SIS velocity
dispersion, but within 1-$\sigma$ error.
The NFW cluster mass is slightly larger than the aperture mass at the
corresponding virial radius.
This could be due to a contaminated measurement of tangential shear.
Looking more closely at the aperture mass profile shows a flattening
at $\theta=1-2$ arcmin, followed by a second subsequent increase 
at larger radii up to $\sim 4$ arcminutes.
Mass associated with the secondary peak may account for that second
rise.

\subsection{SL~J0222.8$-$0416}
The weak lensing density distribution looks relaxed, and  correlates well
with the galaxy over-density.
In the redshift data, one concentration of six galaxies lies
$z_c=0.32$ with 6 members that passes our cluster criterion. 
In addition, there are small concentrations at $z=0.435$ (4 galaxies) 
and at $z=0.227$ (3 galaxies). 
The velocity dispersion of the galaxy cluster is larger than the SIS velocity
dispersion, though they are within
1-$\sigma$ error.
The NFW cluster mass agrees with the aperture mass at the corresponding
virial radius.
However, the aperture mass keeps increasing at outer radii of the
expected virial radius.
This is due to the mass associated with the structure
located at about $4-8$ arcminutes to the east$-$south from the cluster
center where no associated galaxy excess is observed.

\subsection{SL~J0224.4$-$0449}
This is a galaxy cluster previously identified by weak lensing shear
(Cl-02 of Gavazzi \& Soucail 2007).
The redshift was photometrically estimated to be $z=0.497$ (Gavazzi \& Soucail
2007) but had not been spectroscopically obtained.
The weak lensing density distribution is elongated in the north-south
direction.
The elongation of the galaxy distribution is less pronounced, but 
in the same direction.
In the redshift data, there one strong concentration of ten, mainly absorption 
galaxies at $z_c=0.49$. There may also be a small concentration at $z=0.32$.
The velocity dispersion of the galaxy cluster is found to be
smaller than the SIS velocity dispersion.
The reason for this discrepancy is not clear, but one possibility is
the small number of member galaxies used for 
the estimation of the velocity dispersion.
The NFW cluster mass agrees with the aperture mass at the corresponding
virial radius.
The aperture mass flattens at the scales larger than the virial radius
and no substructure is observed in the weak lensing density map.

\subsection{SL~J0224.5$-$0414}
The weak lensing density distribution shows irregular morphology,
with a second peak about 3 arcminutes east of the first peak.
There is an apparent galaxy over-density that largely overlaps with
the first peak.
In the redshift data, there is one strong galaxy concentration at
$z_c=0.26$ with 12 members, dominated by absorption galaxies.
In addition, there is a small concentration at $z=0.316$ (4 galaxies).
The velocity dispersion of the galaxy cluster is smaller than the SIS
velocity dispersion, though they are within 1-$\sigma$ error.
The NFW cluster mass agrees with the aperture mass at the corresponding
virial radius.
The aperture mass profile increases erratically at larger radii looks, 
probably reflecting contributions to the signal from nearby structures.

\subsection{SL~J0225.3$-$0441}
This is a galaxy cluster previously identified by weak lensing shear 
(Cl-05 of Gavazzi \& Soucail 2007).
The photometric redshift was estimated to be $z=0.269$ (Gavazzi \& Soucail
2007) but a spectroscopic redshift had not been obtained.
The weak lensing density distribution is slightly elongated in the
north-west to south-east direction, with an additional small extension 
to the north-east. Interestingly, the center of the galaxy over-density 
is at the position of the small extension, and a possible cD galaxy is also 
found there (target ID 1).
At $z_c=0.26$, there is a concentration of seven (mainly absorption) 
galaxies, including the possible cD galaxy.
In addition, there are nearby concentrations of three galaxies at both 
$z=0.21$ and $z=0.46$.
The velocity dispersion of the galaxy cluster is consistent with the 
best-fit SIS velocity dispersion parameter, and the NFW cluster mass
is consistent with the aperture mass at the virial radius.

\subsection{SL~J0225.4$-$0414}
This is a galaxy cluster previously identified by weak lensing shear  
(Cl-14 of Gavazzi \& Soucail 2007).
The photometric redshift was estimated to be $z=0.153$ (Gavazzi \& Soucail
2007) but a spectroscopic redshift had not been obtained.
The weak lensing density distribution is slightly elongated in the
north-south direction. Two additional peaks lie in the same direction:
one 4 arcminutes to the north and another 7 arcminutes to the 
south. There is a filamentary-like structure connecting the three
peaks.
The northern clump was previously identified by weak lensing shear 
(Cl-04 of Gavazzi \& Soucail 2007) but the spectroscopic redshift has not
been obtained. 

There are galaxy over-densities corresponding to (but slightly offset from)
all the three weak lensing peaks.
Interestingly, the galaxy over-density associated with the central peak 
is elongated in a perpendicular direction to the weak lensing density.
In the redshift data, there is a concentration of eight absorption galaxies at
$z_c=0.14$.
The velocity dispersion of the galaxy cluster is consistent with the SIS 
velocity dispersion.
The tangential shear profile is not well fit by an NFW model, with an 
excess at large radii due to the surrounding structures observed in the weak
lensing density map.
Similarly, the aperture mass profile does not flatten at scales even as large
as $\theta=10$ arcminutes.
The mass contribution from the surrounding structures may also account for
this.
It was therefore difficult to define the boundary of this galaxy cluster, 
in which to calculate the total mass.

\subsection{SL~J0225.7$-$0312}
This cluster has one of the strongest weak lensing signals 
($\kappa_{peak}=0.114$) of our catalog.
The weak lensing density distribution looks relaxed except for an
extension to the north-east. 
The galaxy distribution correlates with the weak lensing density but is
off-centered towards the extension.
In the redshift data, there is an apparent galaxy concentration at
$z_c=0.14$, with 15 members dominated by absorption galaxies.
The velocity dispersion of the galaxy cluster is in good agreement with
the SIS velocity dispersion.
The NFW cluster mass consistent with the aperture mass at the largest radius.

\subsection{SL~J0228.1$-$0450}
This is a galaxy cluster previously identified by weak lensing shear  
(Cl-14 of Gavazzi \& Soucail 2007).
The photometric redshift was estimated to be $z=0.292$ (Gavazzi \& Soucail
2007) but no spectroscopic redshift had been obtained.
We found a strong concentration of 13 absorption galaxies at $z=0.29$. 
The weak lensing density distribution shows irregular morphology.
The low-level feature to the south may be edge effects due to a bright star mask.
However, a second cluster, 7 arcminutes (16.5$h^{-1}$ comoving Mpc) 
to the west and at the same redshift is real. This
was listed as SL~J0227.7$-$0450 in P1 and was first identified from XMM-Newton
data by Pierre et al.\ (2006), where it was named XLSS~J022739.9$-$045129 [also XLSSC022].
The weak lensing map shows a filamentary structure connecting the two clusters,
which appear to form a super cluster system.

There is an apparent galaxy over-density that largely overlaps with 
the weak lensing high density region.
There is a possible cD galaxy (the target ID 3, $z=0.294$) slightly south-east 
of the weak lensing density peak.
The velocity dispersion of the galaxy cluster is found to be smaller
than the SIS velocity, though they are within 1-$\sigma$ error.
The NFW cluster mass agrees with the aperture mass. 
However, the boundary of the cluster is very uncertain because of the
filament.
The aperture mass profile does not show flattening at large scales 
of $\theta=10$ arcminutes.

\subsection{SL~J0850.5$+$4512}
This is a known cluster first identified from its galaxy overdensity 
(NSC~J85029+451141, Gal et al.\ 2003), but no spectroscopic redshift had
been obtained.
The weak lensing density distribution looks relaxed.
There is a possible associated substructure in the north-east from the
cluster.
The galaxy distribution agrees well with the weak lensing density map.
Two possible cD galaxies (target IDs 1 and 31) are located very 
close to the weak lensing density peak.
In the redshift data, there exists one strong galaxy concentration at
$z_c=0.19$ with 15 members dominated by the absorption galaxies.
The velocity dispersion of the galaxy cluster is found to be in a good
agreement with the SIS velocity
dispersion.
The NFW cluster mass agrees with the aperture mass.

\subsection{SL~J1000.7$+$0137}
\label{sec:cosmos-h1-143}
This is a known cluster first identified from its galaxy concentration 
(NSC~J100047+013912, Gal et al.\ 2003), and later via X-ray emission by Finoguenov
et al.\ (2006, their ID 67; the photometric redshift they obtained is  
$z=0.22$) but the spectroscopic redshift had not been determined.
The weak lensing density distribution shows irregular morphology.
There is an over-density of galaxies, but its peak is about 2
arcminutes north of the weak lensing density peak.
About 6 arcminutes east of the cluster center there is another weak lensing
density peak, which is our target SL~J1001.2$+$0135 described 
in \S \ref{sec:cosmos-h1-157}. 
A smaller, third peak lies a similar distance to the east.
In the redshift data, we find a strong galaxy concentration at
$z_c=0.22$, with 14 members dominated by absorption galaxies.
Note that we find galaxy cluster SL~J1001.2$+$0135 at the same redshift.
There are additional small concentrations of galaxies at $z=0.34$ and $z=0.52$.
The separation between the two main clusters is 11$h^{-1}$ comoving Mpc (at
$z=0.22$), and a filamentary structure connecting the two clusters is 
observed in the weak lensing density map.
Thus it is likely that they form a super cluster system.

The velocity dispersion of the galaxy cluster is in a good agreement
with the SIS velocity dispersion.
The NFW cluster mass is slightly larger than the aperture mass.
It is likely that the spherical NFW model does not give a good
description of this cluster, because of asymmetry in the density distribution.

\subsection{SL~J1001.2$+$0135}
\label{sec:cosmos-h1-157}
This is not listed in the P1 catalogue because the peak $\kappa$ $S/N$ does
not exceed the required threshold.
Note that an extended X-ray source discovered by XMM-Newton (Finoguenov
2006, their ID 54; RA.=150.33413, Dec.=1.60301) lies about 3 arcminutes east
of the weak lensing density peak, and a known
optically selected cluster 
(Gal et al.\ 2003, NSC~J100113+013335 R.A.=150.30812,
Dec.=1.55967, $z_phot=0.242$ ).
The weak lensing $\kappa$ map appears elongated, with filamentary 
structure connecting this cluster to SL~J0850.5$+$4512 (\S \ref{sec:cosmos-h1-143}).

In the redshift data, there are two strong galaxy concentrations at 
$z=0.22$ (11 members; we name it SL~J1001.2$+$0135A) and $z=0.37$ (11
members; we name it SL~J1001.2$+$0135B).
The velocity dispersions are $\sigma_{\rm sis}\sim 1380$km/s (A) and
930km/s (B).
Since we do not have enough information to de-project the weak lensing
density into two components (e.g., accurate photometric redshifts
of faint galaxies, e.g.\ Massey et al.\ 2007), 
we are unable to separately estimate the weak lensing mass of each cluster.

\subsection{SL~J1002.9$+$0131}
This is not listed in the P1 catalogue because the $\kappa$ peak is
located close to the field edge and thus outside the {\it secure survey area}.
There is a galaxy over-density correlated with the weak lensing $\kappa$
peak.
In the redshift data, there is no galaxy concentration passing 
our cluster criterion, but there are two small concentrations at 
$z=0.37$ (3 galaxies) and at $z=0.66$ (3 galaxies).
Since no galaxy concentration that passes our cluster criterion was
found, we did not estimate the weak lensing mass.

Note that very near this target is a known X-ray 
cluster with the estimated photometric redshift of $z_p=0.75$
(Finoguenov et al. 2006, their ID; R.A.=150.75121, Dec.=1.52793).
Since the lensing efficiency of such a high redshift cluster is low
(see Figure 3 of Hamana et al.\ 2004), it is unlikely to be solely responsible for
the $\kappa$ peak. Therefore one possible explanation of the observed 
$\kappa$ excess is a chance projection of several halos located at different
redshifts in adjacent lines-of-sight.

\subsection{SL~J1047.3$+$5700}
The weak lensing density distribution is elongated in the
north-west to south-east direction.
The galaxy over-density correlates with the $\kappa$ map very well.
However, the redshift information reveals that the weak lensing and 
galaxy over-densities arise from not one but two clusters, located at
different redshifts along the same line-of-sight.
The foreground cluster is at $z=0.24$ with 6 members ($\sigma_v=412$km/s); 
the background cluster is at $z=0.30$ with 10 members ($\sigma_v=619$km/s). 
We call these  SL~J1047.3$+$5700A and SL~J1047.3$+$5700B respectively.
Since we do not have enough information to de-project the weak lensing
density into two components (e.g., accurate photometric redshifts
of faint galaxies, e.g., Massey et al 2007), 
we can not make separate mass estimates.

\subsection{SL~J1048.1$+$5730}
The weak lensing density distribution is slightly elongated.
There is a separate weak lensing density peak about 3 arcminutes south-east 
of the cluster center, which was not listed in the P1 catalogue 
because its $\kappa$ $S/N$ (as opposed to the illustrated $\kappa$) 
is lower than the required threshold.
There is a galaxy over-density that largely overlaps with
the $\kappa$ peak.
The redshift data contains one strong galaxy concentration at
$z_c=0.31$, with 9 members of mainly absorption type.
There is also a small concentration at $z=0.36$.
The velocity dispersion of the galaxy cluster is found to be 
smaller than the SIS velocity
dispersion, but within
1-$\sigma$ error.
The aperture mass profile shows a jump at $\theta\simeq 3.3$ arcminutes,
probably due to the south-east peak.
The NFW cluster mass agrees with the aperture mass within $\theta<3$ arcmin.

\subsection{SL~J1049.4$+$5655}
The weak lensing density distribution is elongated 
from the north-east to the south-west, with a second peak about 3
arcminutes north-east of the cluster center.
The distribution of galaxies is more isotropic.
In the redshift data, there is one galaxy concentration at
$z_c=0.42$ (6 members) that passes our cluster criterion, plus
small groups at $z=0.24$ (4 galaxies), $z=0.31$ (4 galaxies) 
and possibly $z=0.59$ (3 galaxies).
The velocity dispersion of the galaxies is 
smaller than the best-fit SIS velocity dispersion.
This discrepancy is probably due to a combination of line-of-sight 
projections and the small number of redshifts used to compute $\sigma_v$.

\subsection{SL~J1051.5$+$5646}
The weak lensing density distribution is slightly
elongated, and there is a second peak (catalogued as SL~J1051.6+5647)
4 arcminutes to the north-east, for which we have not obtained galaxy spectra.
There is no galaxy concentration sufficiently rich to fulfill our cluster criterion in the 
redshift data, but two small groups lie at $z=0.33$ and $z=0.35$.
Note that galaxies with target IDs 2, 3 and 5 are found to be very nearby.
Also note that the bright galaxy located at the peak of the $\kappa$ map
(SDSS {\tt SpecObjID} 255522745545129984) is a nearby galaxy at $z=0.047$, 
and a second bright galaxy at R.A.=162.87, Dec.=56.82 (SDSS {\tt SpecObjID}
267344926176444416) is also at $z=0.46$.
Since no galaxy concentration passing our cluster criterion was
found, we did not make estimate a weak lensing mass.

\subsection{SL~J1052.0$+$5659}
This is not listed in the P1 catalogue because its peak $\kappa$ $S/N$ is
below the required threshold.
The weak lensing density distribution looks well relaxed.
A galaxy over-density is present but its peak is off-center, 
about 2 arcmin south of the $\kappa$ peak.
In the redshift data, there is no galaxy concentration passing 
our cluster criterion, but there are two small concentrations at 
$z=0.34$ (4 galaxies) and $z=0.52$ (4 galaxies).
Since no rich galaxy concentration was found, we did not
make the weak lensing mass estimation.

\subsection{SL~J1052.5$+$5731}
This is not listed in the P1 catalogue because its $\kappa$ $S/N$ 
is below the required threshold.
The weak lensing density distribution shows irregular morphology.
A projected galaxy over-density overlaps with the $\kappa$ peak but, 
in the redshift data, there is no galaxy concentration passing 
our cluster criterion. There are two small concentrations around 
$z=0.34$ (5 galaxies) and $z=0.61$ (3 galaxies).
Since no dominant galaxy concentration was
found, we did not calculate the weak lensing mass.

Note that this region also contains two X-ray cluster candidates
(Kolokotronis et al.\ 2006): 
SEXCLAS-12 (R.A.=163.159, Dec.=57.514, 
at photometric redshift $z_p=0.61$) 
and SEXCLAS-13 (R.A.=163.226, Dec.=57.536; 
$z_p=0.58$).
The sky position of SEXCLAS-12 is very close to the $\kappa$ peak ($\sim 1$
arcmin) and its estimated photometric redshift is very similar to the
redshift of our small galaxy concentration.
It is therefore likely that the galaxy concentration at $z=0.61$ is the
optical counter part of X-ray cluster candidate SEXCLAS-12
(Kolokotronis et al. 2006).
The observed $\kappa$ peak appears to consist of a
chance projection of halos of galaxy clusters at different redshifts.

\subsection{SL~J1057.5$+$5759}
The weak lensing density distribution looks relaxed, and is coincident with
a very prominent overdensity of galaxies. 18 (mainly absorption) galaxies
are found at $z_c=0.60$. Their velocity dispersion is larger than the velocity 
dispersion parameter of the best-fit SIS model, but within 1-$\sigma$ error.
The NFW cluster mass agrees with the aperture mass at the corresponding 
virial radius. The aperture mass does not flatten at large radii. The reason 
for this is not currently clear, although the measurements beyond 5~arcminutes 
are noisy because the cluster is near the edge of a field.

\subsection{SL~J1135.6$+$3009}
The weak lensing density distribution looks well isolated but with an 
extension to the north. 
A prominent over-density of bright galaxies includes a
cD galaxy precisely at the $\kappa$ peak position.
The redshift data reveals a concentration of 15 (mainly absorption)
galaxies at $z_c=0.21$. The velocity dispersion of the galaxy cluster 
is consistent with the best-fit SIS parameter and the NFW cluster 
mass is consistent with the aperture mass.

\subsection{SL~J1201.7$-$0331}
The weak lensing density distribution appears relaxed. 
There is a clear galaxy over-density which coincides with 
the $\kappa$ peak.
The redshift data contain a galaxy concentration at
$z_c=0.52$ with 8 members dominated by absorption galaxies.
The velocity dispersion of the galaxy cluster is consistent with 
the SIS velocity dispersion. 
The NFW cluster mass is larger than the aperture mass at the 
corresponding virial radius.
This small disagreement is likely due to poor measurements of the
shear near the edge of a Subaru field. High $S/N$ measurements are 
obtained only scales between $1 < \theta < 3.3$ arcmin.

\subsection{SL~J1204.4$-$0351}
This is not listed in the P1 catalogue because the $\kappa$ $S/N$ 
is below that threshold.
However, it is a known cluster, first identified by its extended X-ray emission 
(RX~J1204.3-0350, Vikhlinin et al.\ 1998) and later confirmed
with optical data (OC5 1204$-$0351, Donahue et al.\ 2002).
The spectroscopic redshift of this cluster is $z=0.261$ (Mullis et al.\ 2003).

The weak lensing density distribution shows irregular morphology.
The galaxy over-density is clearly observed and its peak position is
very close to the $\kappa$ peak. 
A concentration of 14 galaxies is indeed seen at
$z_c=0.261$, dominated the absorption spectral types.
Our measured velocity dispersion is consistent with 
the SIS velocity dispersion
The NFW cluster mass is slightly larger than the aperture mass at 
the corresponding virial radius.
This small disagreement is likely due to the deviations from 
spherical symmetry apparent in the $\kappa$ map.
In this case, the NFW model would not be a good description of the
cluster density distribution.

\subsection{SL~J1334.3$+$3728}
This is a known cluster, first identified by galaxy counts  
(NSC~J133424+372822, Gal et al.\ 2003), but a spectroscopic redshift has
only been obtained for the cD galaxy (R.A.=203.60, Dec.=37.48, $z=0.305$;
SDSS {\tt SpecObjID} 591610245776670720). This is located very close to the
$\kappa$ peak (R.A.=203.60, Dec.=37.48) whose redshift is $z=0.305$
(SDSS, {\tt SpecObjID} is 591610245776670720).

The weak lensing density distribution looks very irregular,
with elongations to the north-west and south-east, as well as 
a neighbouring structure about 2 arcminutes north-east of the $\kappa$ peak.
However, the distribution of galaxies is centered neatly on only the main 
$\kappa$ peak. The galaxy redshifts reveal 21 members of a cluster at
$z_c=0.30$, most of which are absorption galaxies.
Upon further inspection, there is also a significant trend for
southern (northern) galaxies to be at lower (higher) redshifts,
which may imply an ongoing merger of two clusters. 
The measured velocity dispersion is larger than the best-fit SIS velocity
dispersion parameter, but consistent within 1-$\sigma$ error.
The NFW cluster mass agrees with the aperture mass at the corresponding 
virial radius.
However, the aperture mass profile does not flatten even at $\theta=10$
arcminutes.
This is probably due to mass of neighbour structures, and it is very
difficult to define the boundary of the cluster.

\subsection{SL~J1335.7$+$3731}
The weak lensing density distribution looks very irregular, with 
two distinct main peaks aligned in the east-west direction, and an
additional structure between them extending towards the north.
Only the central structure was listed in P1, as the S/N in $\kappa$
(as opposed to the $\kappa$ values shown) is below the required
threshold.
The galaxy over-density closely follows this elongated structure.
A well-defined concentration of 14 galaxies is located at $z_c=0.41$. 
Interestingly, more than half of these have emission or composite type spectra.
There is also a small group of 3 galaxies at $z=0.20$.
The velocity dispersion of the galaxy cluster is in a reasonable
agreement with the SIS velocity dispersion parameter.
The NFW cluster mass is consistent with the aperture mass at the 
corresponding virial radius. However, it is unlikely that a 
spherical NFW model is a good description of this cluster, because 
of the asymmetry apparent in the weak lensing density map.

\subsection{SL~J1337.7$+$3800}
The weak lensing density distribution looks relaxed. 
There is a galaxy over-density that coincides with the $\kappa$ peak,
including a cD galaxy very close to the center.
In the redshift data, there is a prominent galaxy concentration at
$z_c=0.16$ with 16 members dominated by absorption galaxies.
The velocity dispersion of the galaxy cluster is consistent with 
the SIS velocity dispersion.
The NFW cluster mass is in a good agreement with the aperture 
mass at the corresponding virial radius.

\subsection{SL~J1601.6$+$4245}
This is not listed in the P1 catalogue because the $\kappa$ $S/N$ is
below the P1's threshold.
The weak lensing density distribution looks relaxed, except for an additional 
filament extending to the north-west.
A prominent over-density of galaxies coincides with the $\kappa$ peak.
This includes a cD galaxy at RA.=240.3934, 
Dec.=42.75902, whose spectrum was obtained by SDSS and
was found to be $z=0.208$ ({\tt SpecObjID} 375714238640947200).
Spectra of another two galaxies in this field were obtained by 
SDSS: {\tt SpecObjID}=375714238619975680 at 
RA.=240.34840, Dec.=43.73718, $z=0.208$ and 
{\tt SpecObjID}=375714238666113024 at 
R.A.=240.45011, Dec.=42.79000, $z=0.292$ .

Our MOS observations in fact reveal a projection of
several clusters at different redshifts along the same line-of-sight.
A foreground cluster (SL~J1601.6$+$4245A) is at $z=0.208$,
with 7 spectroscopically confirmed member galaxies, and
a background cluster (SL~J1601.6$+$4245B) at $z=0.47$ with 8 members. 
We also find a small group of 5 new galaxies at $z=0.29$.
Therefore, SL~J1601.6$+$4245A has $2+7$ spectroscopically confirmed
cluster members including a cD galaxy, and the small galaxy
concentration has $1+5$ spectroscopic members.
We therefore conclude that the observed $\kappa$ excess is caused by
the chance projection of at least three galaxy clusters located at
different redshifts. 
Since we do not have enough information to de-project the weak lensing
density into components, we can not compute their weak lensing masses.

\subsection{SL~J1602.8$+$4335}
The weak lensing density distribution looks relaxed, with an additional
low-level filamentary structure running from north to south.
There is a galaxy over-density, including a cD galaxies, that coincides
with the $\kappa$ peak.
In the redshift data, a strong concentration at
$z_c=0.42$ of 15 members is dominated by absorption galaxies.
The velocity dispersion of the galaxy cluster is consistent with the SIS
velocity dispersion.
The NFW cluster mass is in a good agreement with the aperture mass 
at the corresponding virial radius. 
No conclusive explanation is found for the north-south filament.

\subsection{SL~J1605.4$+$4244}
The weak lensing density distribution looks relaxed.
There is a galaxy over-density that largely overlaps with 
the $\kappa$ peak.
In the redshift data, there exists one galaxy concentration at
$z_c=0.22$ with 6 members.
The velocity dispersion is significantly larger than the SIS 
velocity dispersion, probably on account of the low number of 
observed member galaxies.
The aperture mass profile behaves irregularly at larger radii.
This may be noise due to a shortage of source galaxies, which 
are hidden by masks around nearby bright stars.

\subsection{SL~J1607.9$+$4338}
This is not listed in the P1 catalogue because the $\kappa$ $S/N$ is
below the P1 threshold.
This is a known, optically selected cluster (GHO 1606+4346; Gunn,
Hoessel, Oke, 1986), but no spectroscopic redshift had been obtained.
The weak lensing density distribution appears elongated towards the north
and south-west, with local $\kappa$ maxima about 3 arcminutes from the
target center in both directions.
The south-west peak has the highest $\kappa$ and is listed in the P1
catalogue (GTO 2deg$^2$ \#04).
A galaxy over-density overlaps the central and northern $\kappa$ peaks.
Our redshift data reveals a galaxy concentration at $z_c=0.31$, passing our
cluster criteria with 9 members. This is dominated by absorption galaxies. 
In addition, there is small concentration of five galaxies at $z=0.25$.
The velocity dispersion of the galaxy cluster is 
smaller than the SIS velocity dispersion.
The reason for this is not clear but is probably due to poor
statistics from the small number of spectroscopically confirmed cluster members. 
The NFW cluster mass is consistent with the aperture mass at the 
corresponding virial radius.

\subsection{SL~J1634.1$+$5639}
\label{sec:SLJ1634.1$+$5639}
The weak lensing density distribution looks relaxed.
The clustering of bright galaxies is apparent, but the number
density of galaxies with $18<R_C<23$ is lower than the surrounding 
mean density. The redshift data reveals a single concentration of
13 galaxies at $z_c=0.4$.
The northern galaxies tend to have absorption spectra
but, interestingly, the southern galaxies have emission spectra.
The velocity dispersion of the galaxy cluster is significantly 
larger than the SIS velocity dispersion.
As shown in Figure \ref{fig:vzhist1}, the velocity distribution of 
spectroscopic members appears strongly skewed toward the bluer side.
This skewness may account for the large measurement of dispersion.
The reason of the large skewness is currently not clear: dynamical
activity of the cluster may be involved, although there are poor 
statistics to constrain higher moments.
The NFW cluster mass is in a good agreement with the aperture mass at the
corresponding virial radius.

\subsection{SL~J1639.9$+$5708}
The weak lensing density distribution  is
elongated in the north-south direction.
The corresponding galaxy over-density is clearly found.
In the redshift data, there is no galaxy concentration passing 
our cluster criterion, but there are two small concentrations at 
$z=0.2$ and $z=0.63$.
Since no sufficiently rich concentration of galaxies was found,
we did not calculate the weak lensing mass.

\subsubsection{SL~J1647.7$+$3455}
The weak lensing density distribution is
elongated in the north-south direction.
A prominent overdensity of bright galaxies includes one cluster
of 12 galaxies at $z_c=0.26$. This is dominated by absorption galaxies.
There are small additional groups of galaxies at $z=0.41$ and $z=0.47$.
The velocity dispersion of the main galaxy cluster is in a reasonable
agreement with the SIS velocity dispersion.
The NFW cluster mass is consistent with the aperture mass.
However, the aperture mass profile does not flatten as expected at 
larger radii. This may be accounted for by the mass of a second cluster
located about 6 arcminutes south-west of the cluster. 
It is also likely that the spherical NFW model does not accurately
describe this cluster, as asymmetry is clearly visible in the weak lensing 
density map.

\clearpage

\clearpage
\begin{figure*}
\includegraphics[height=160mm,clip,angle=-90]{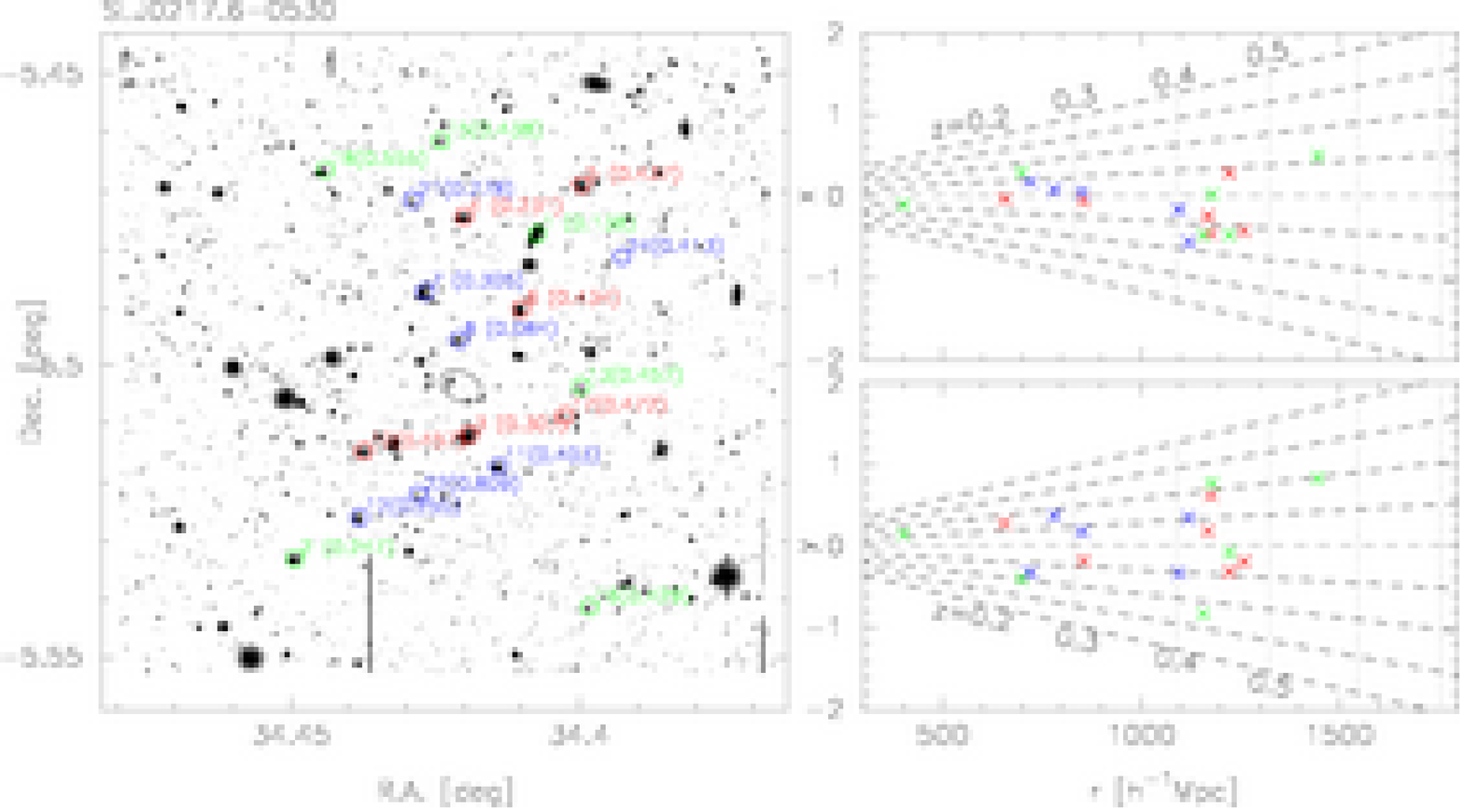}
\vspace{2mm}\\
\includegraphics[width=75mm,clip,angle=-90]{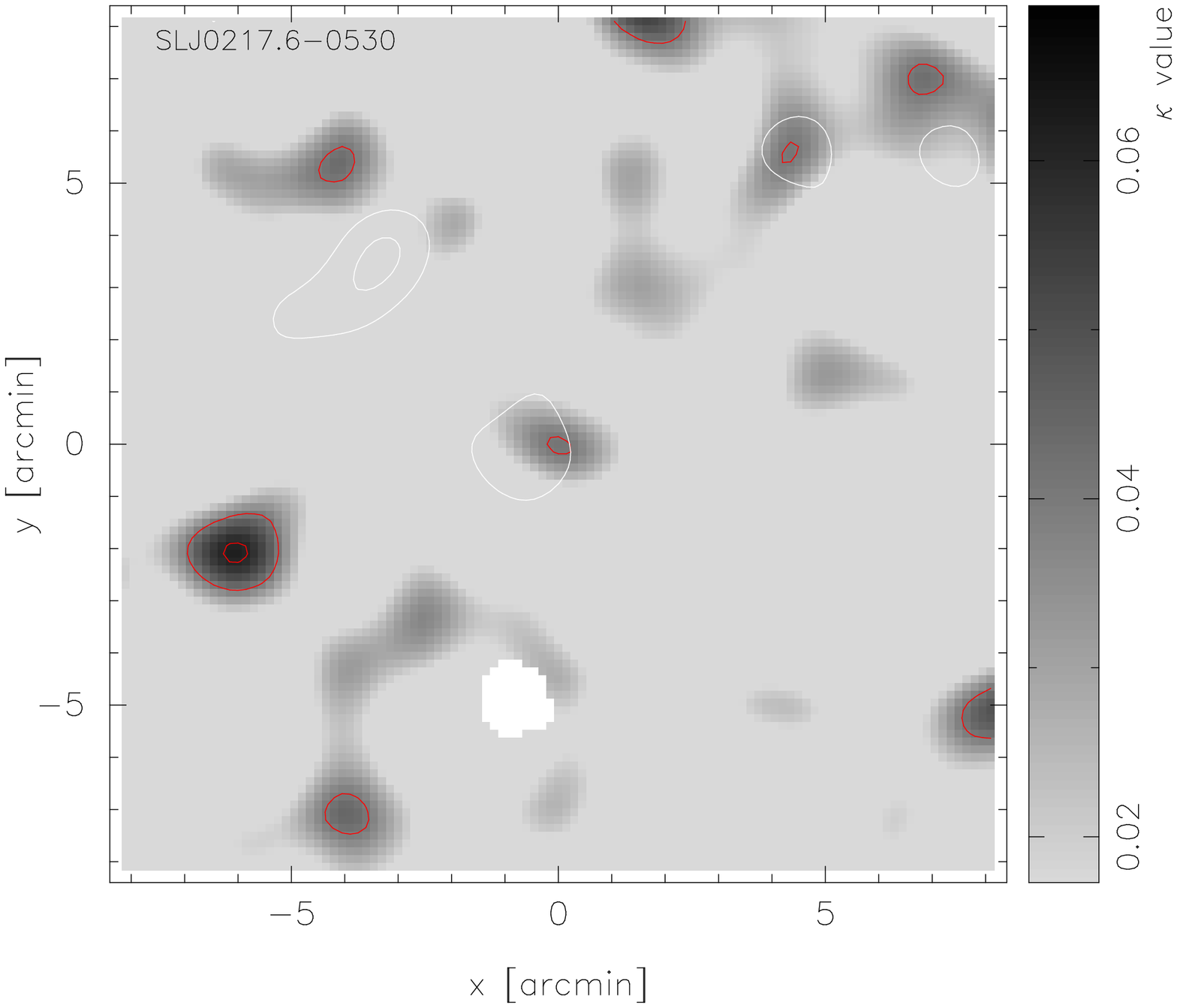}
\caption{Same as Figure \ref{fig:sxds_6} but
for SL~J0217.6$-$0530.} 
\end{figure*}

\clearpage
\begin{figure*}
\includegraphics[height=160mm,clip,angle=-90]{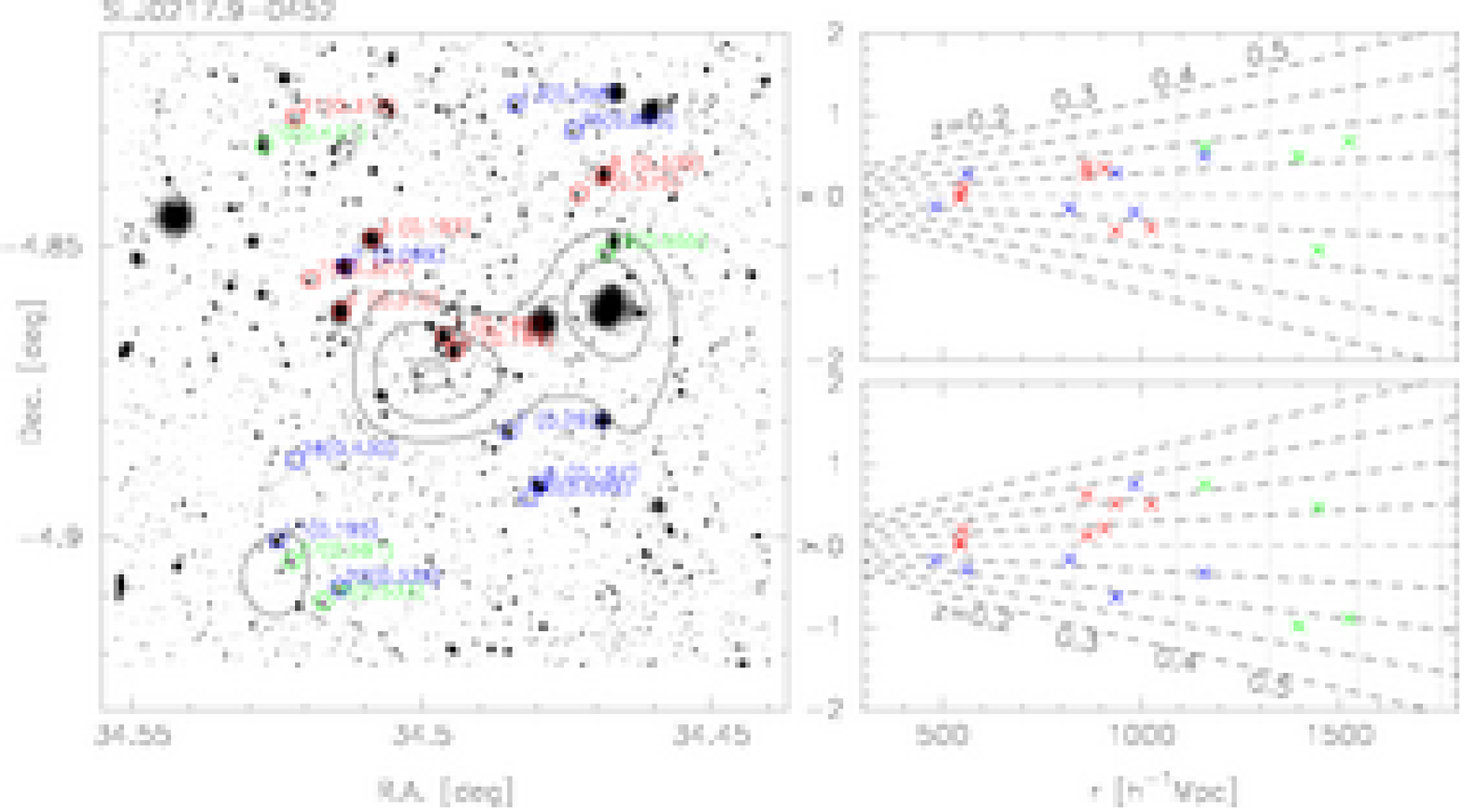}
\vspace{2mm}\\
\includegraphics[width=75mm,clip,angle=-90]{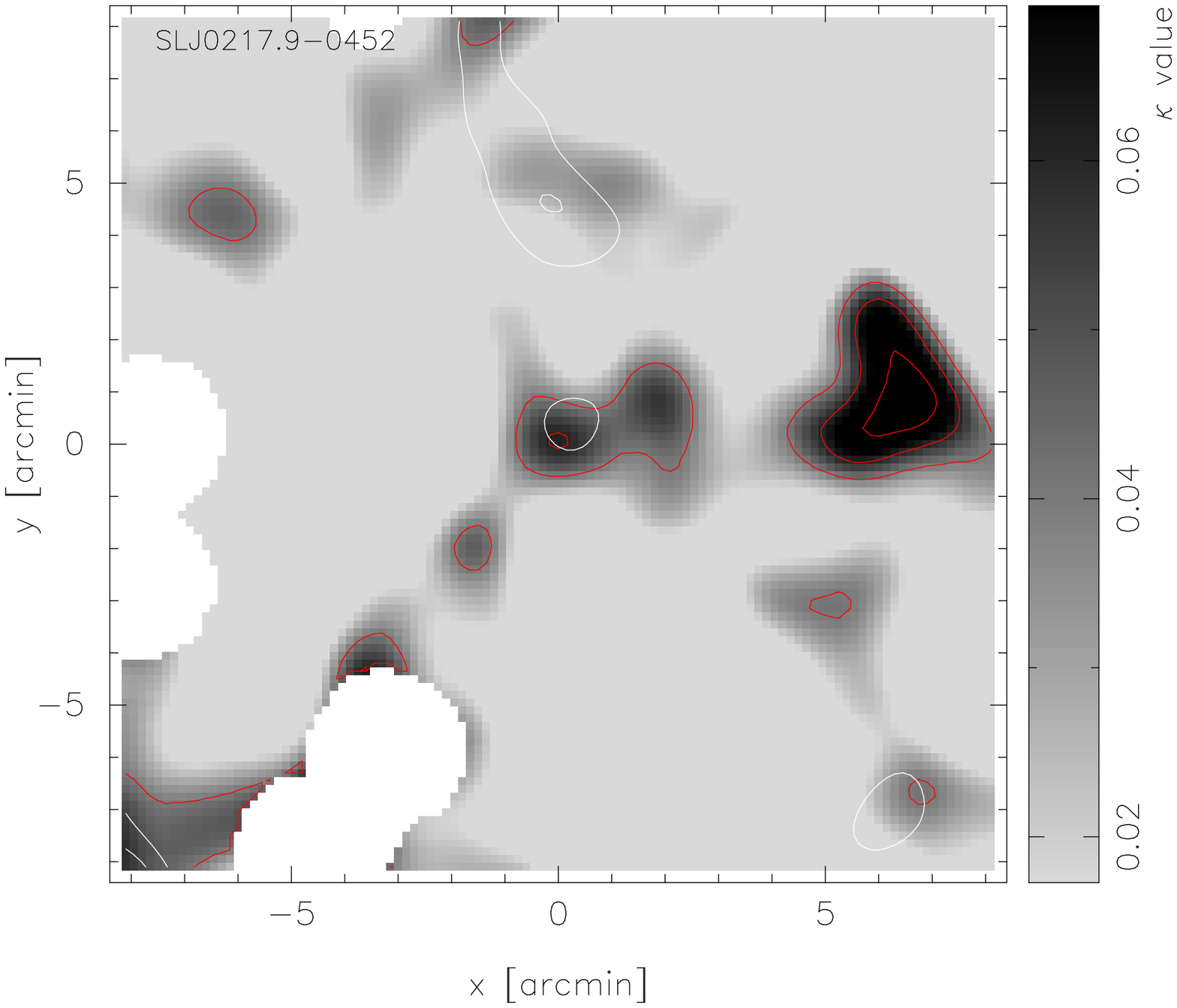}
\caption{Same as Figure \ref{fig:sxds_6} but for SL~J0217.9$-$0452.}
\end{figure*}

\clearpage
\begin{figure*}
\includegraphics[height=160mm,clip,angle=-90]{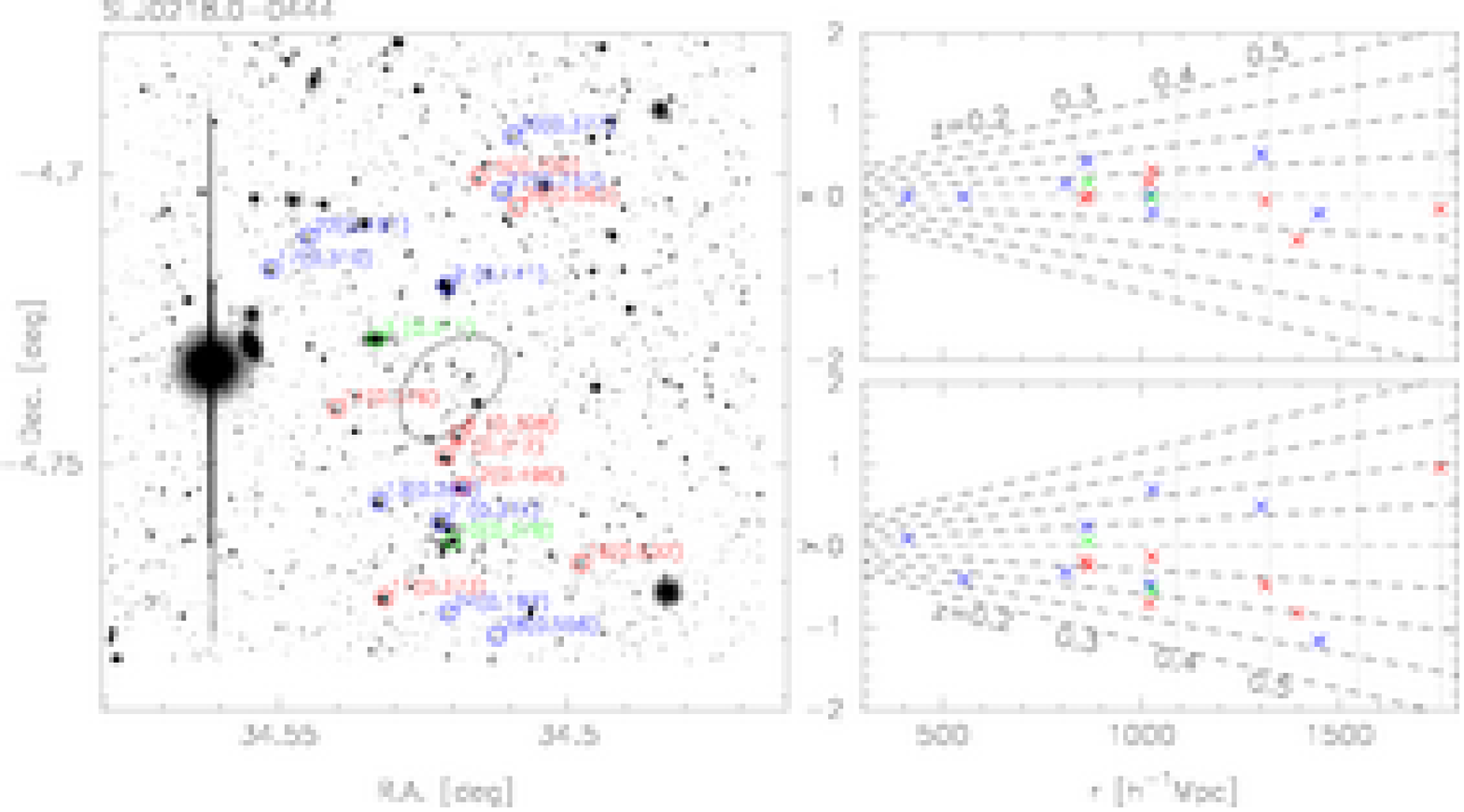}
\vspace{2mm}\\
\includegraphics[width=75mm,clip,angle=-90]{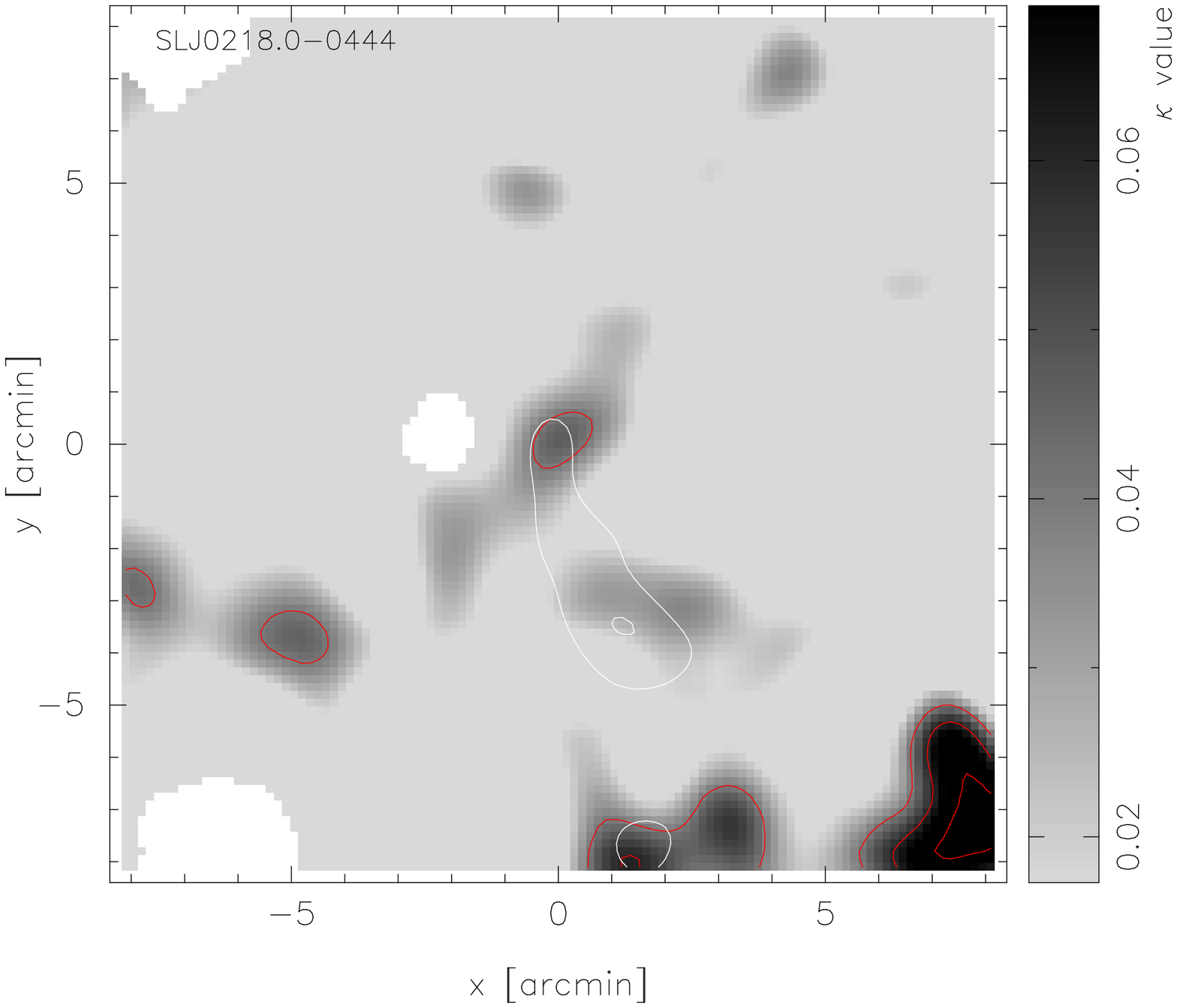}
\caption{Same as Figure \ref{fig:sxds_6} but for SL~J0218.0$-$0444.}
\end{figure*}

\clearpage
\begin{figure*}
\includegraphics[height=160mm,clip,angle=-90]{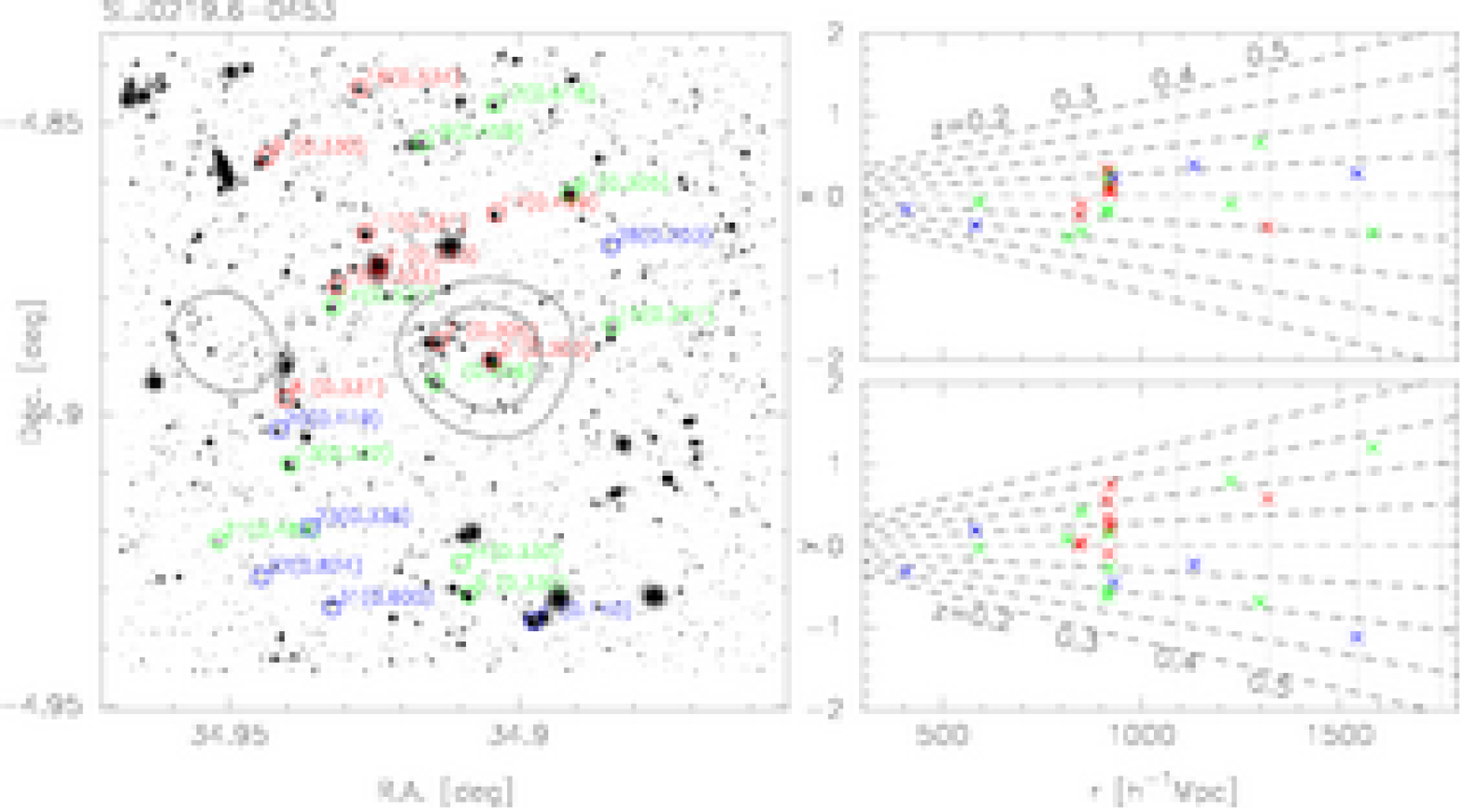}
\vspace{2mm}\\
\includegraphics[width=75mm,clip,angle=-90]{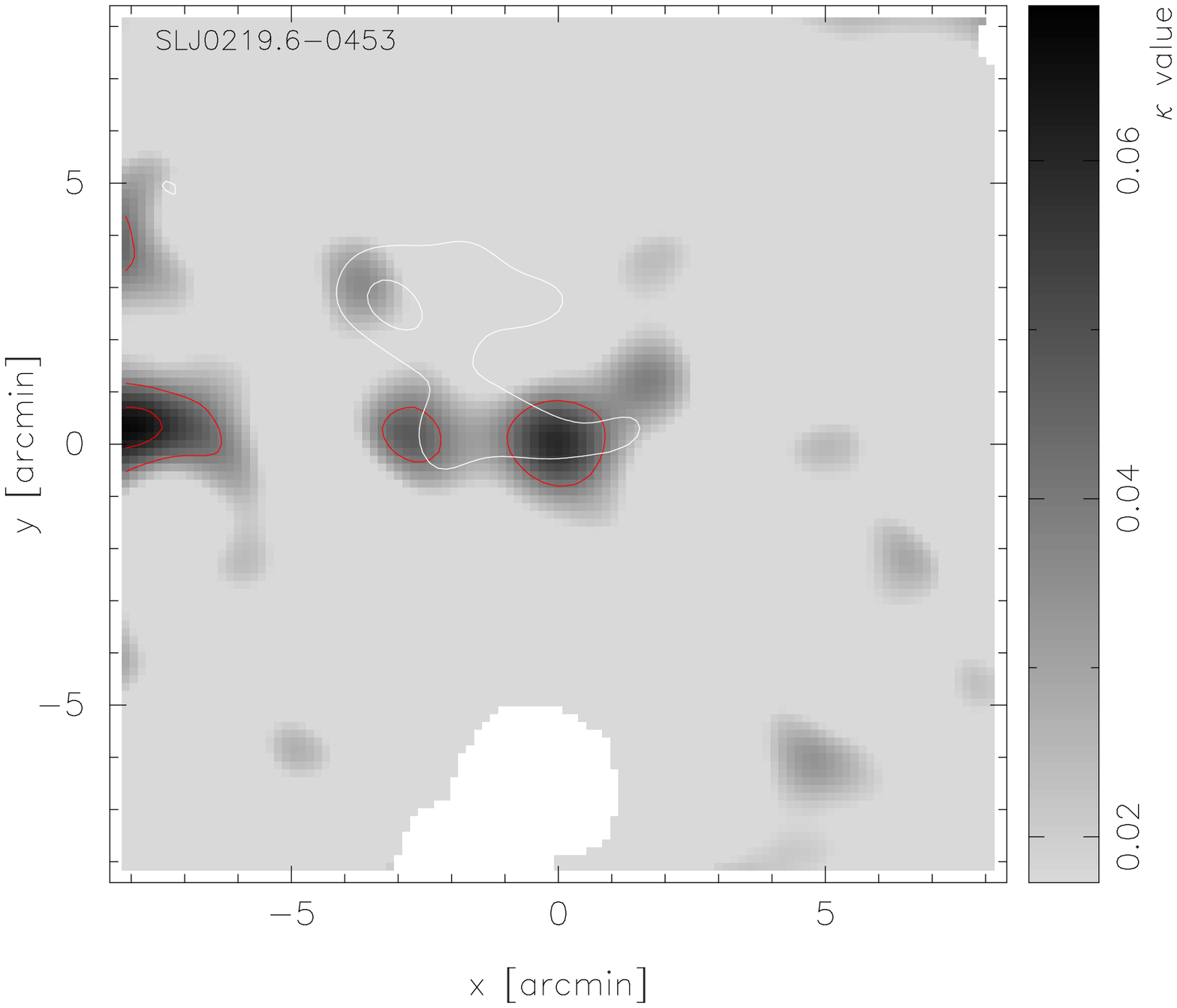}
\hspace{2mm}
\includegraphics[width=75mm,clip,angle=-90]{fig12c.ps}
\caption{Same as Figure \ref{fig:sxds_6} but for
SL~J0219.6$-$0453.
\label{fig:sxds_7}}
\end{figure*}

\clearpage
\begin{figure*}
\includegraphics[height=160mm,clip,angle=-90]{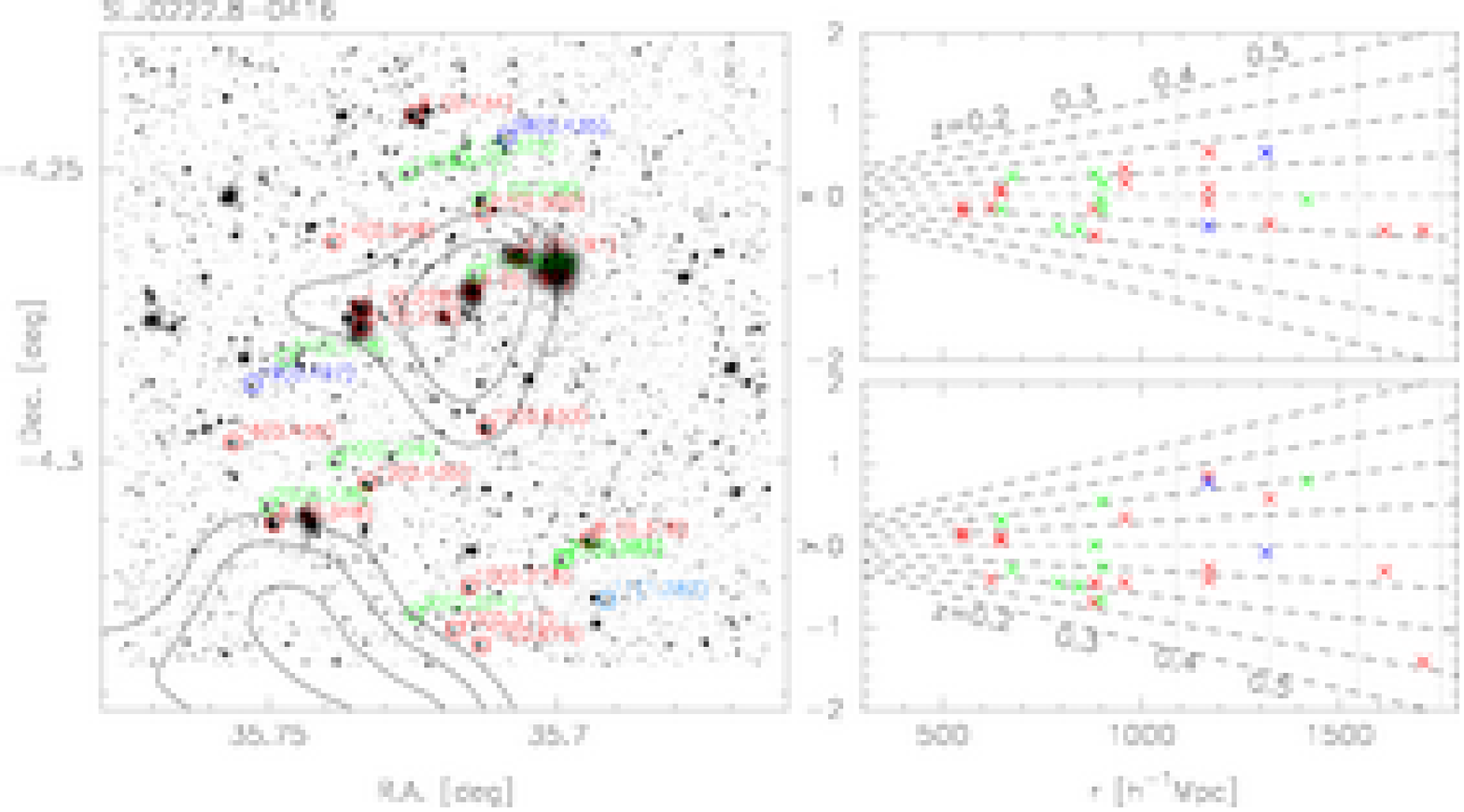}
\vspace{2mm}\\
\includegraphics[width=75mm,clip,angle=-90]{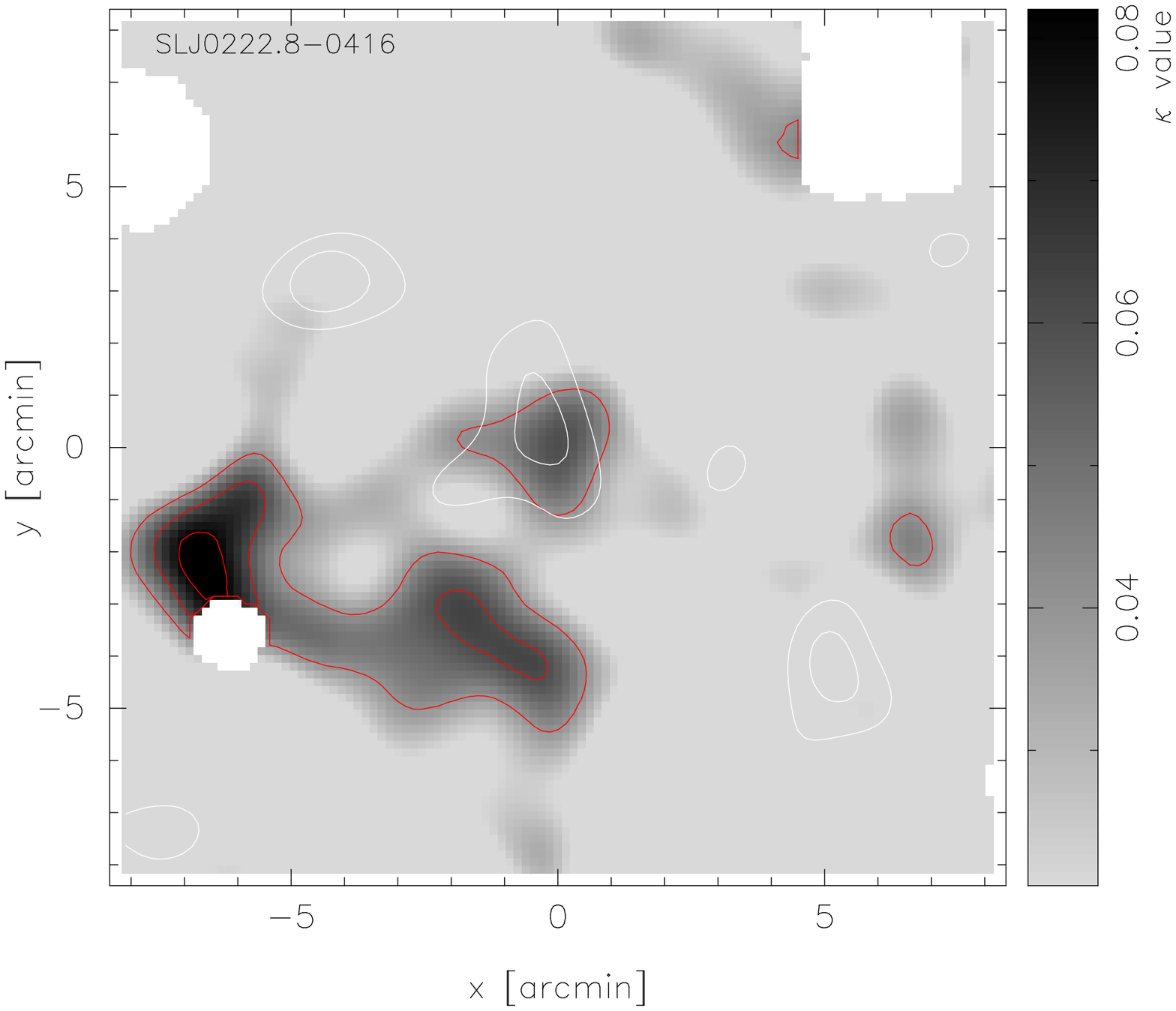}
\hspace{2mm}
\includegraphics[width=75mm,clip,angle=-90]{fig13c.ps}
\caption{Same as Figure \ref{fig:sxds_6} but for SL~J0222.8$-$0416.
\label{fig:saclay_23}}
\end{figure*}

\clearpage
\begin{figure*}
\includegraphics[height=160mm,clip,angle=-90]{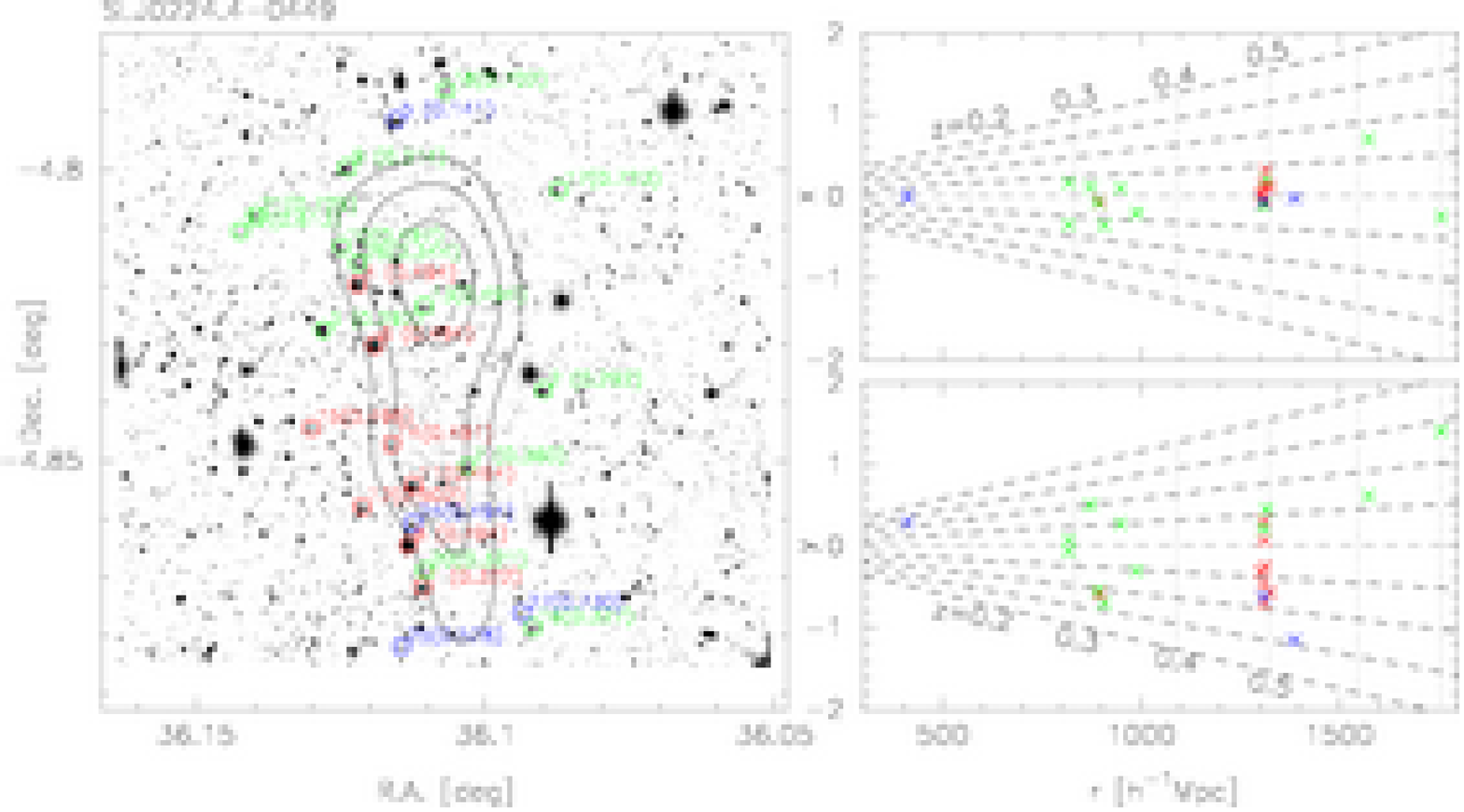}
\vspace{2mm}\\
\includegraphics[width=75mm,clip,angle=-90]{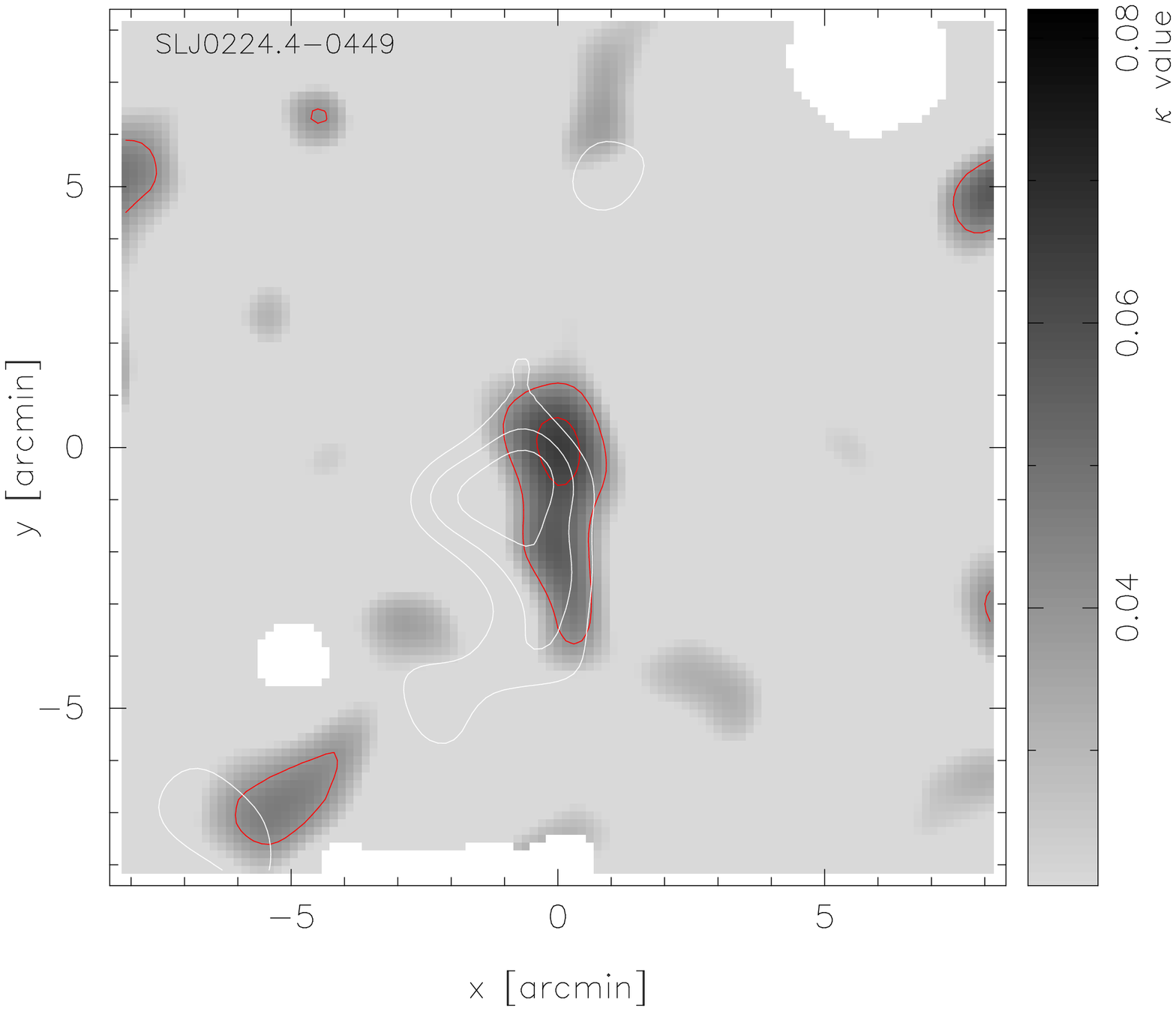}
\hspace{2mm}
\includegraphics[width=75mm,clip,angle=-90]{fig14c.ps}
\caption{Same as Figure \ref{fig:sxds_6} but for SL~J0224.4$-$0449.}
\end{figure*}

\clearpage
\begin{figure*}
\includegraphics[height=160mm,clip,angle=-90]{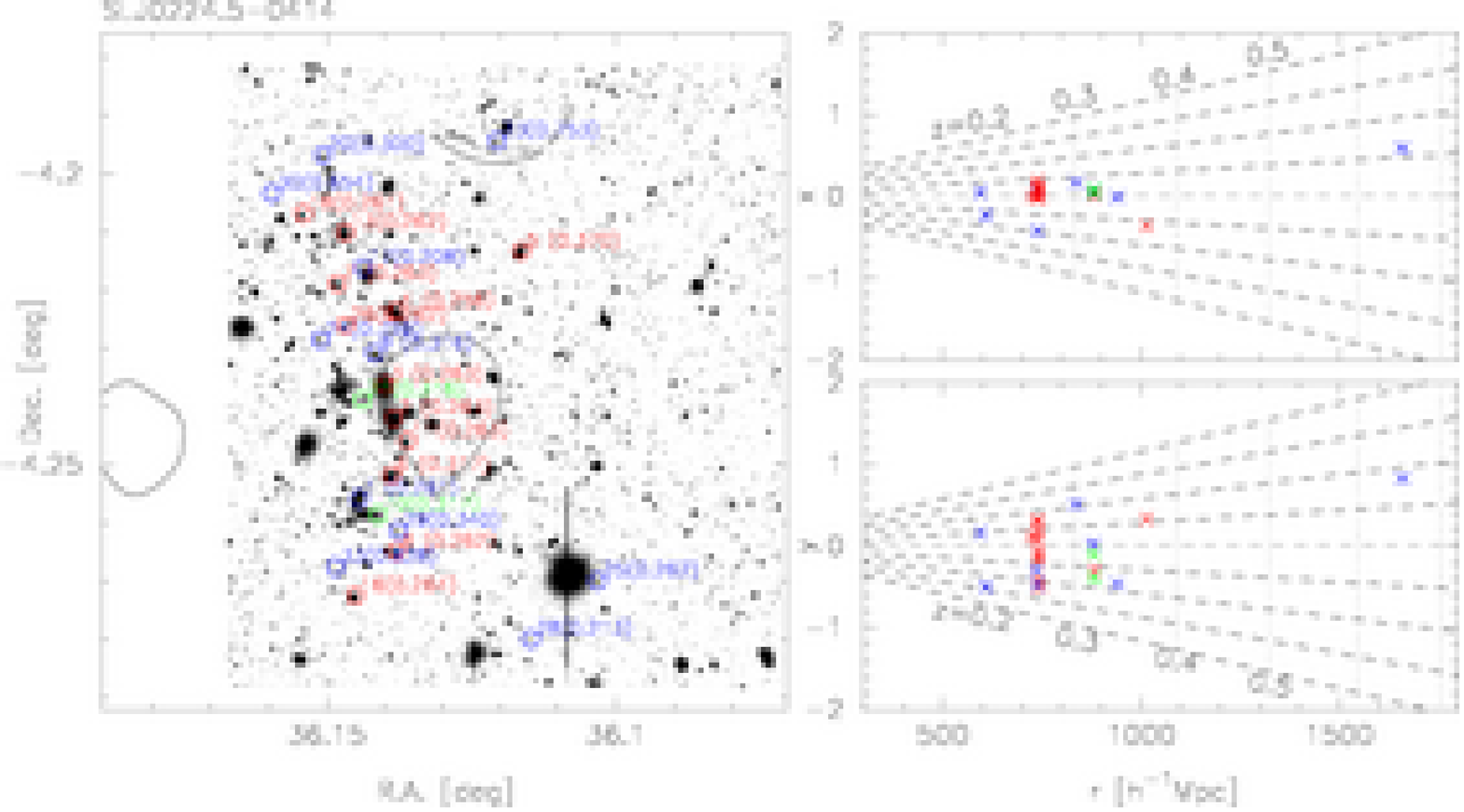}
\vspace{2mm}\\
\includegraphics[width=75mm,clip,angle=-90]{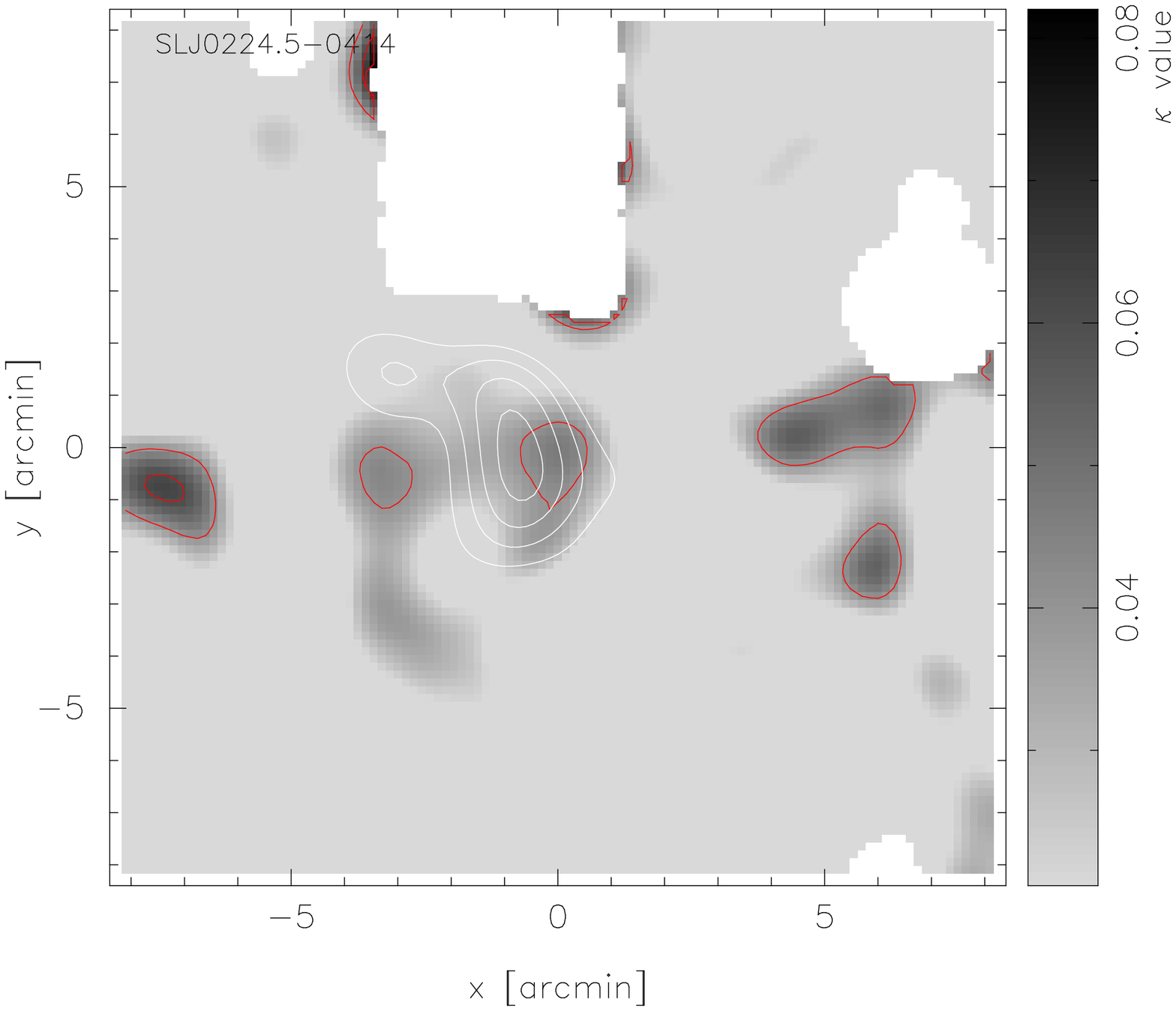}
\hspace{2mm}
\includegraphics[width=75mm,clip,angle=-90]{fig15c.ps}
\caption{Same as Figure \ref{fig:sxds_6} but for SL~J0224.5$-$0414.}
\end{figure*}

\clearpage
\begin{figure*}
\includegraphics[height=160mm,clip,angle=-90]{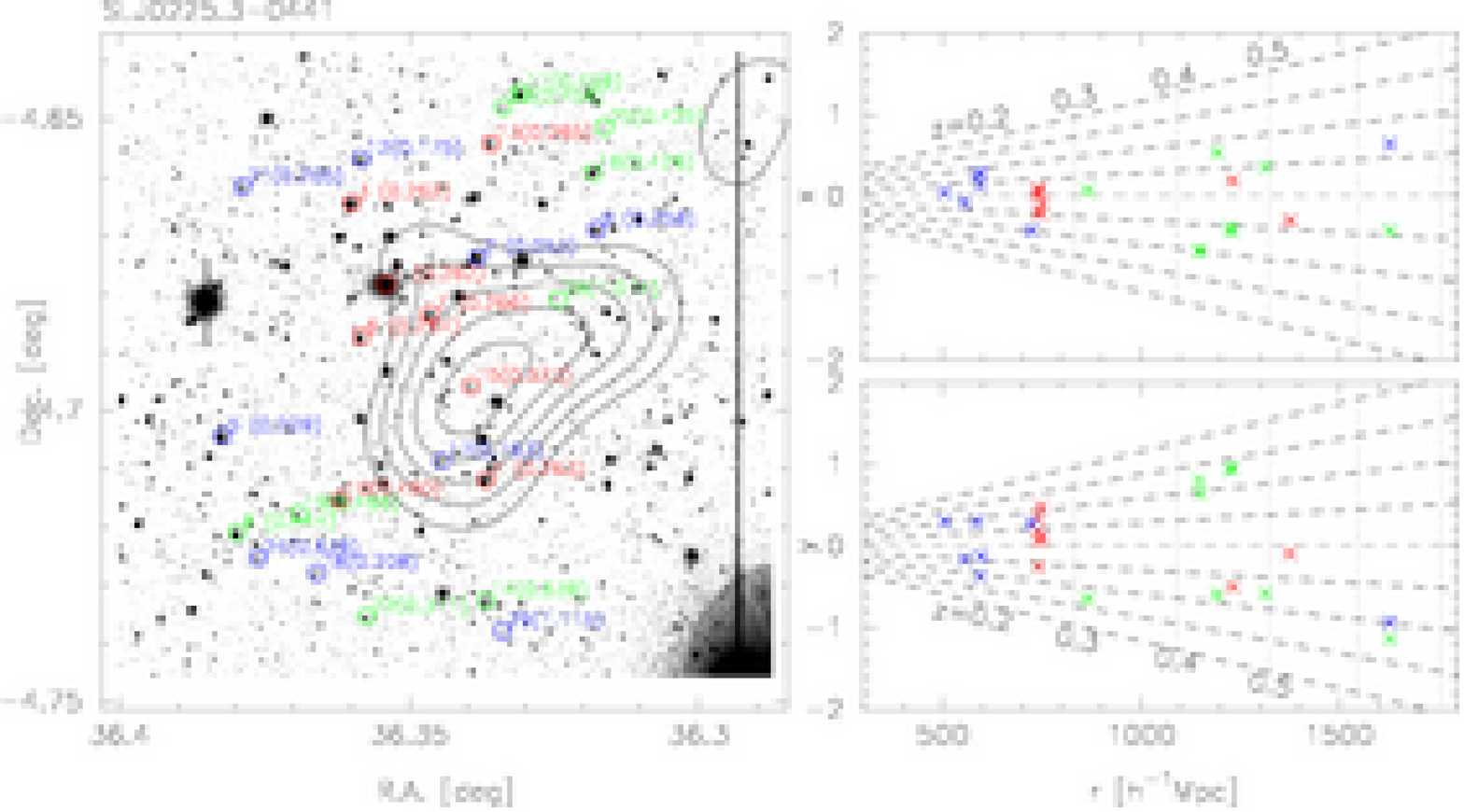}
\vspace{2mm}\\
\includegraphics[width=75mm,clip,angle=-90]{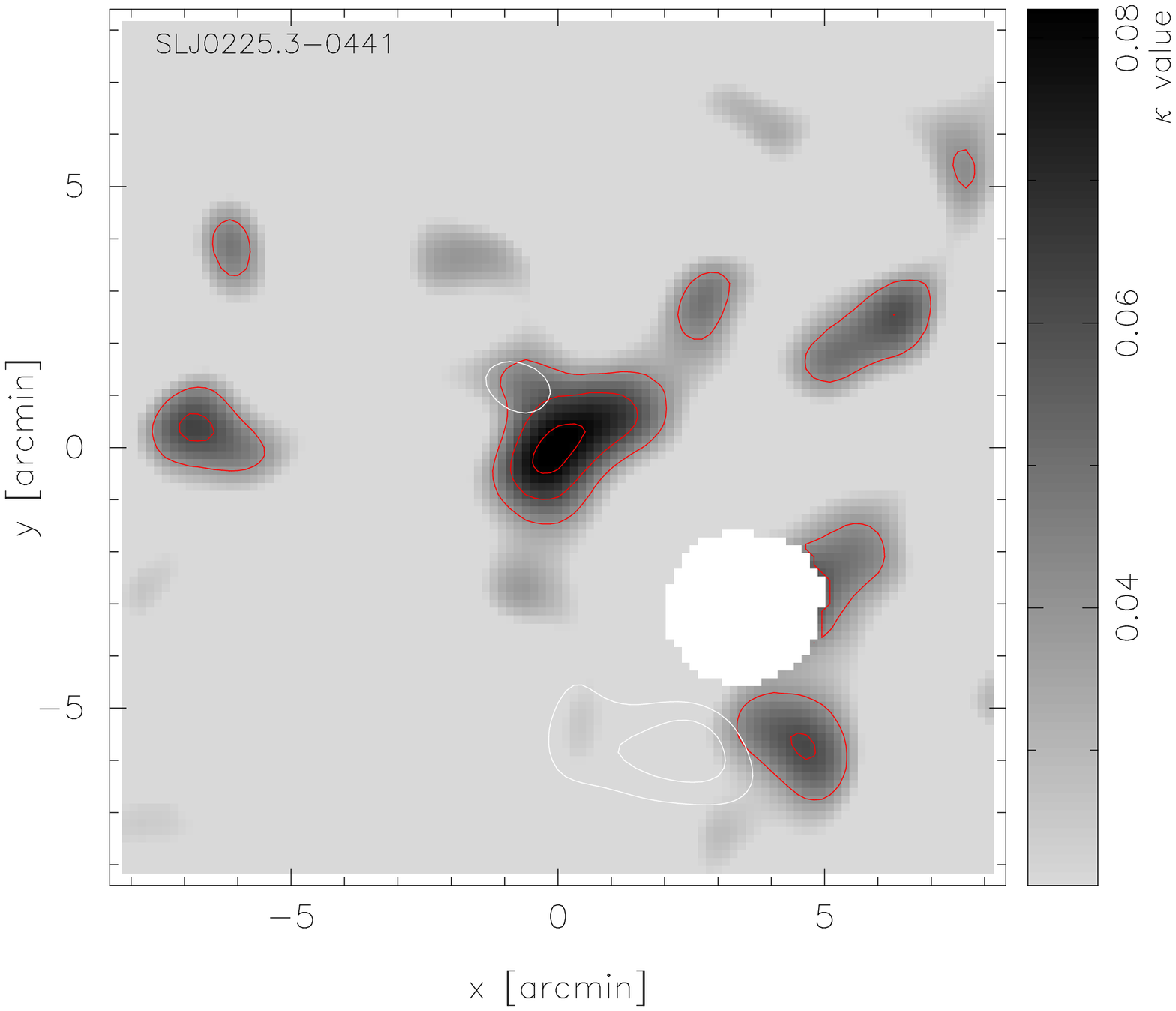}
\hspace{2mm}
\includegraphics[width=75mm,clip,angle=-90]{fig16c.ps}
\caption{Same as Figure \ref{fig:sxds_6} but for SL~J0225.3$-$0441.}
\end{figure*}

\clearpage
\begin{figure*}
\includegraphics[height=160mm,clip,angle=-90]{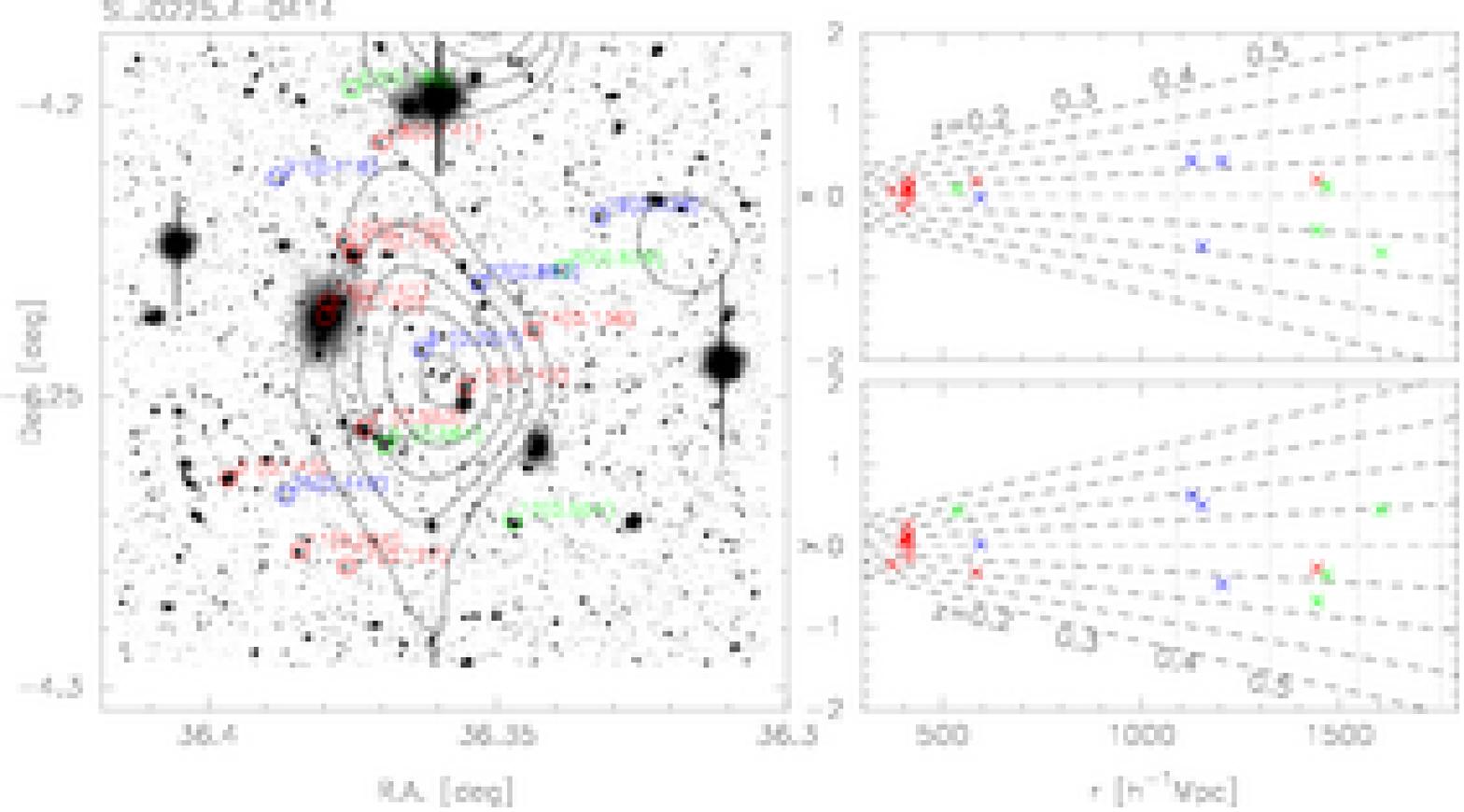}
\vspace{2mm}\\
\includegraphics[width=75mm,clip,angle=-90]{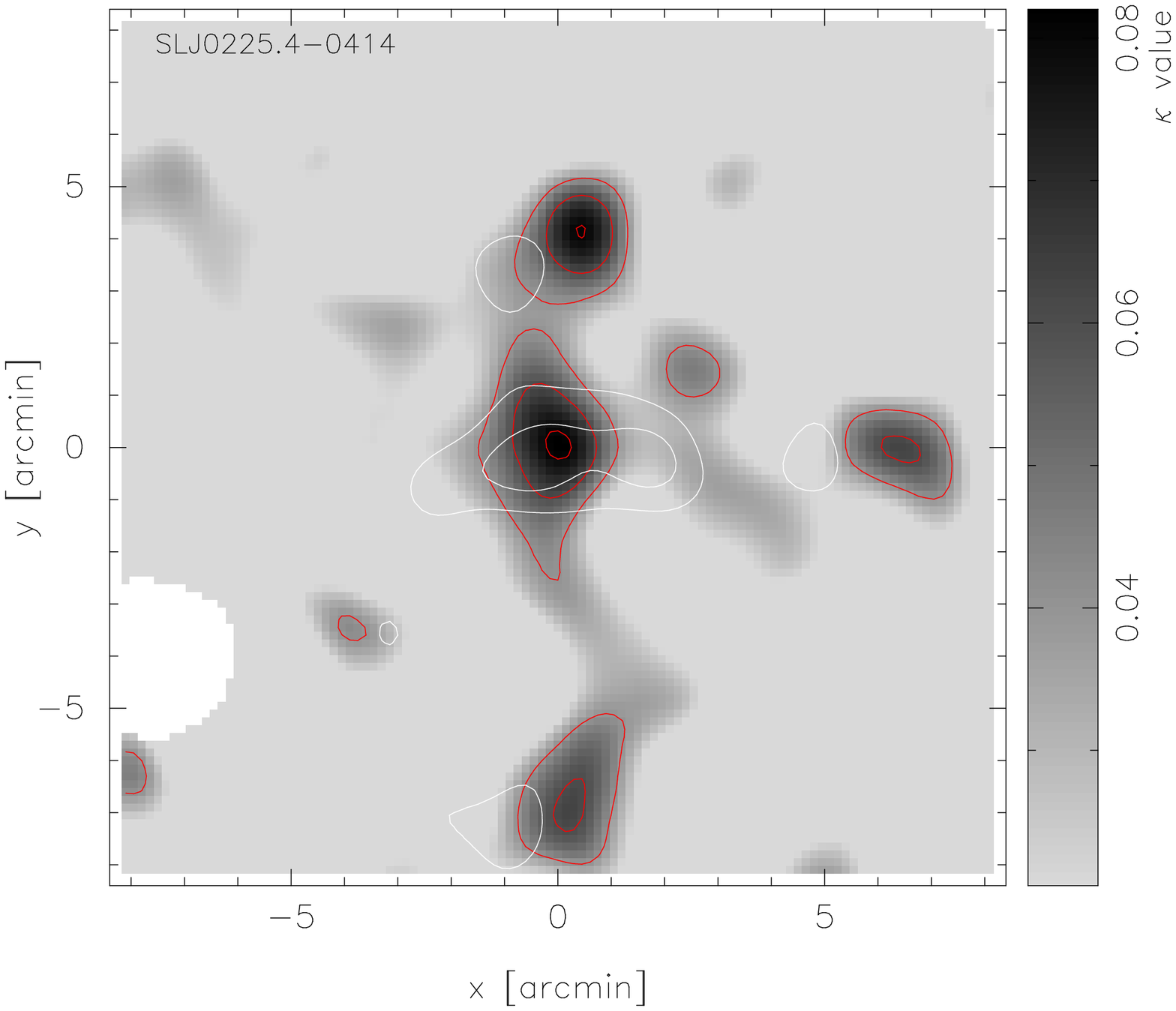}
\hspace{2mm}
\includegraphics[width=75mm,clip,angle=-90]{fig17c.ps}
\caption{Same as Figure \ref{fig:sxds_6} but for SL~J0225.4$-$0414.}
\end{figure*}

\clearpage
\begin{figure*}
\includegraphics[height=160mm,clip,angle=-90]{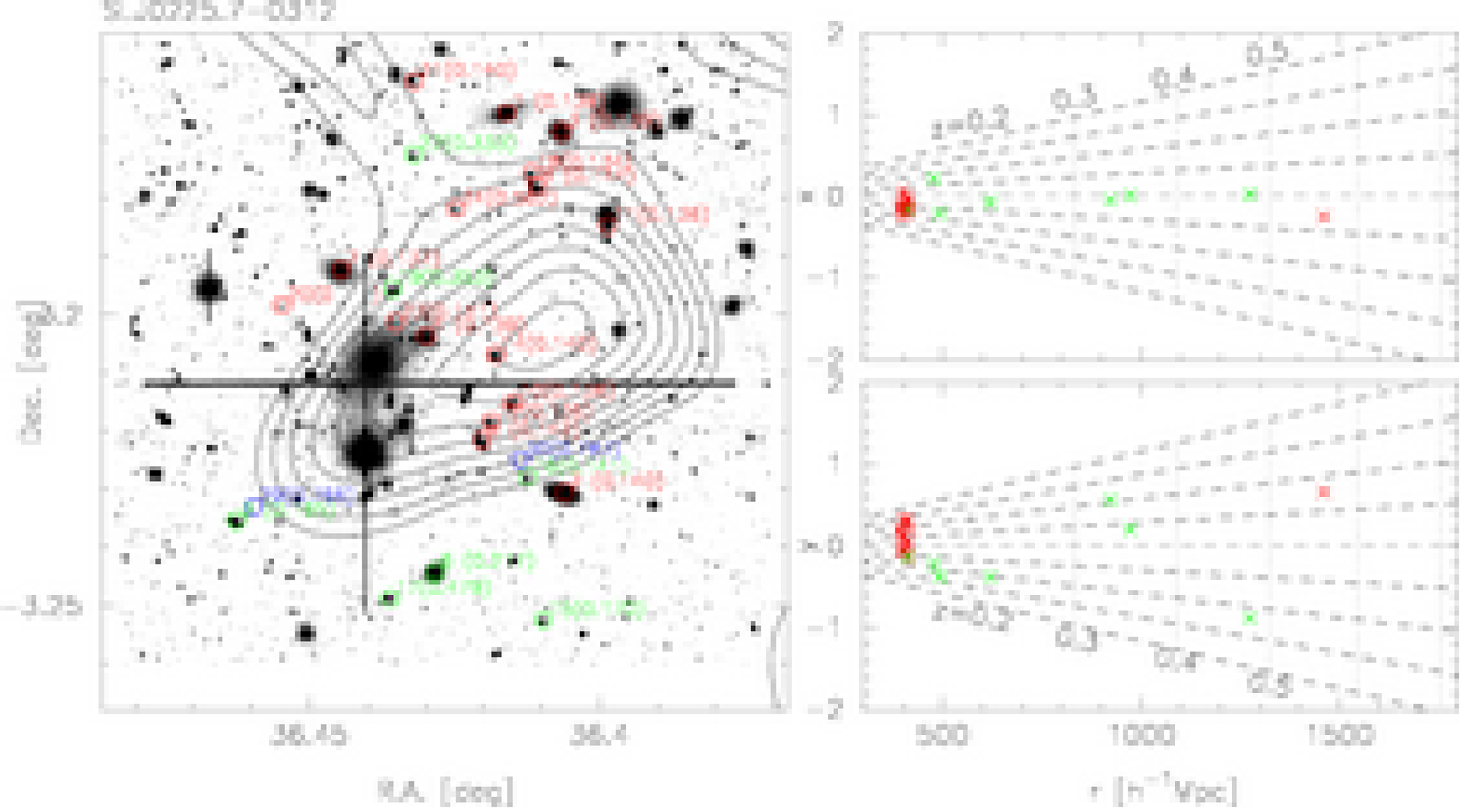}
\vspace{2mm}\\
\includegraphics[width=75mm,clip,angle=-90]{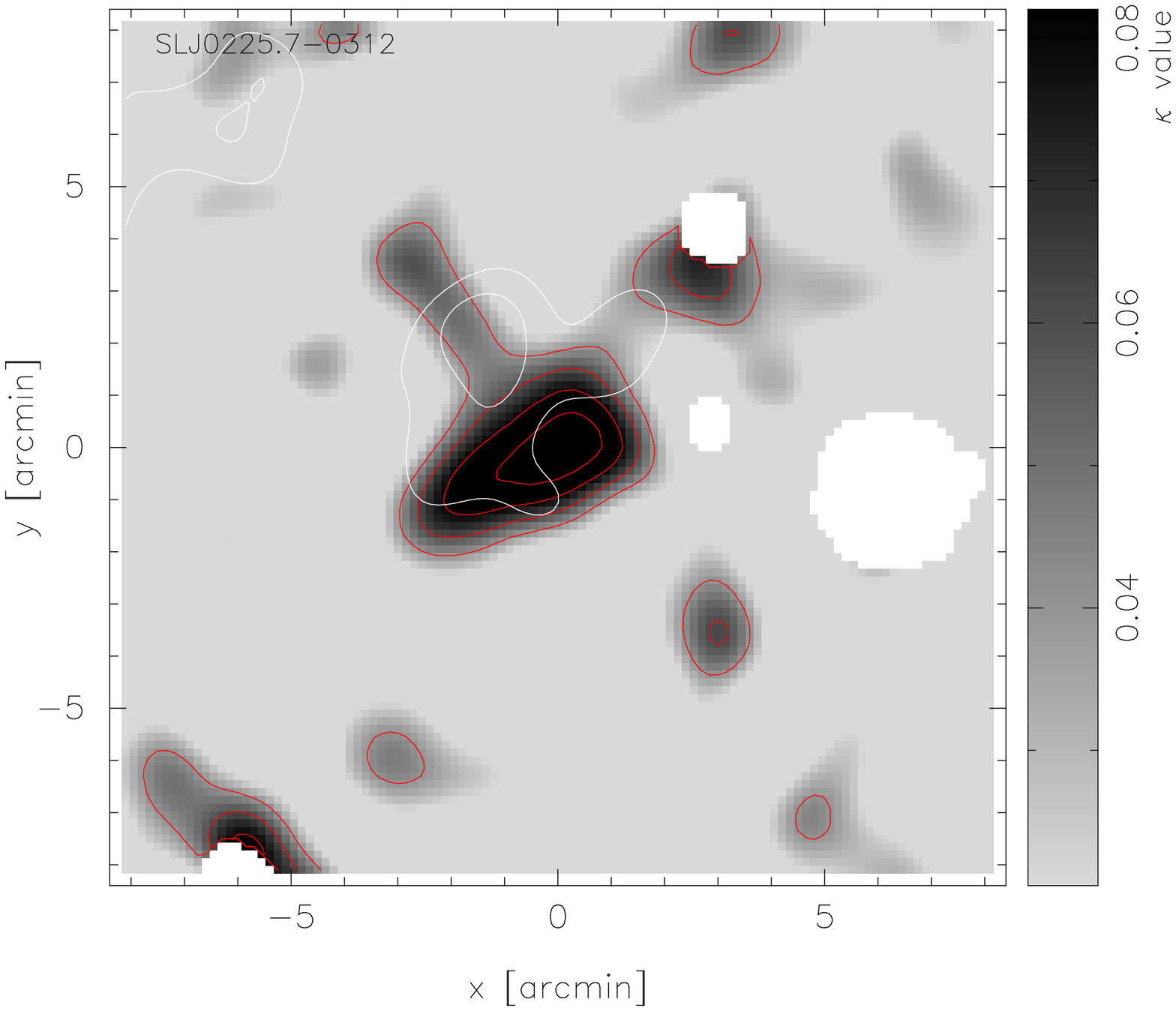}
\hspace{2mm}
\includegraphics[width=75mm,clip,angle=-90]{fig18c.ps}
\caption{Same as Figure \ref{fig:sxds_6} but for SL~J0225.7$-$0312.}
\end{figure*}

\clearpage
\begin{figure*}
\includegraphics[height=160mm,clip,angle=-90]{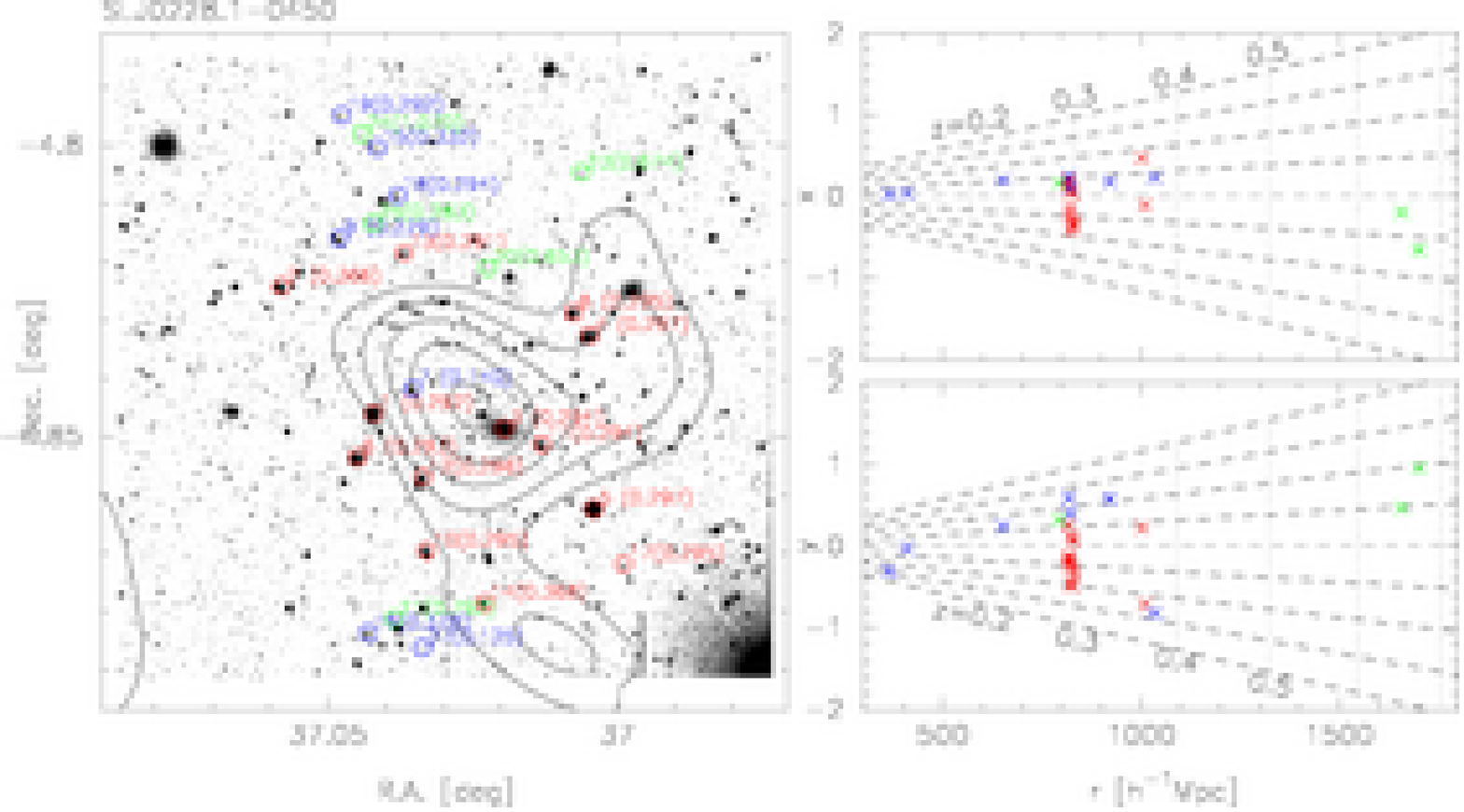}
\vspace{2mm}\\
\includegraphics[width=75mm,clip,angle=-90]{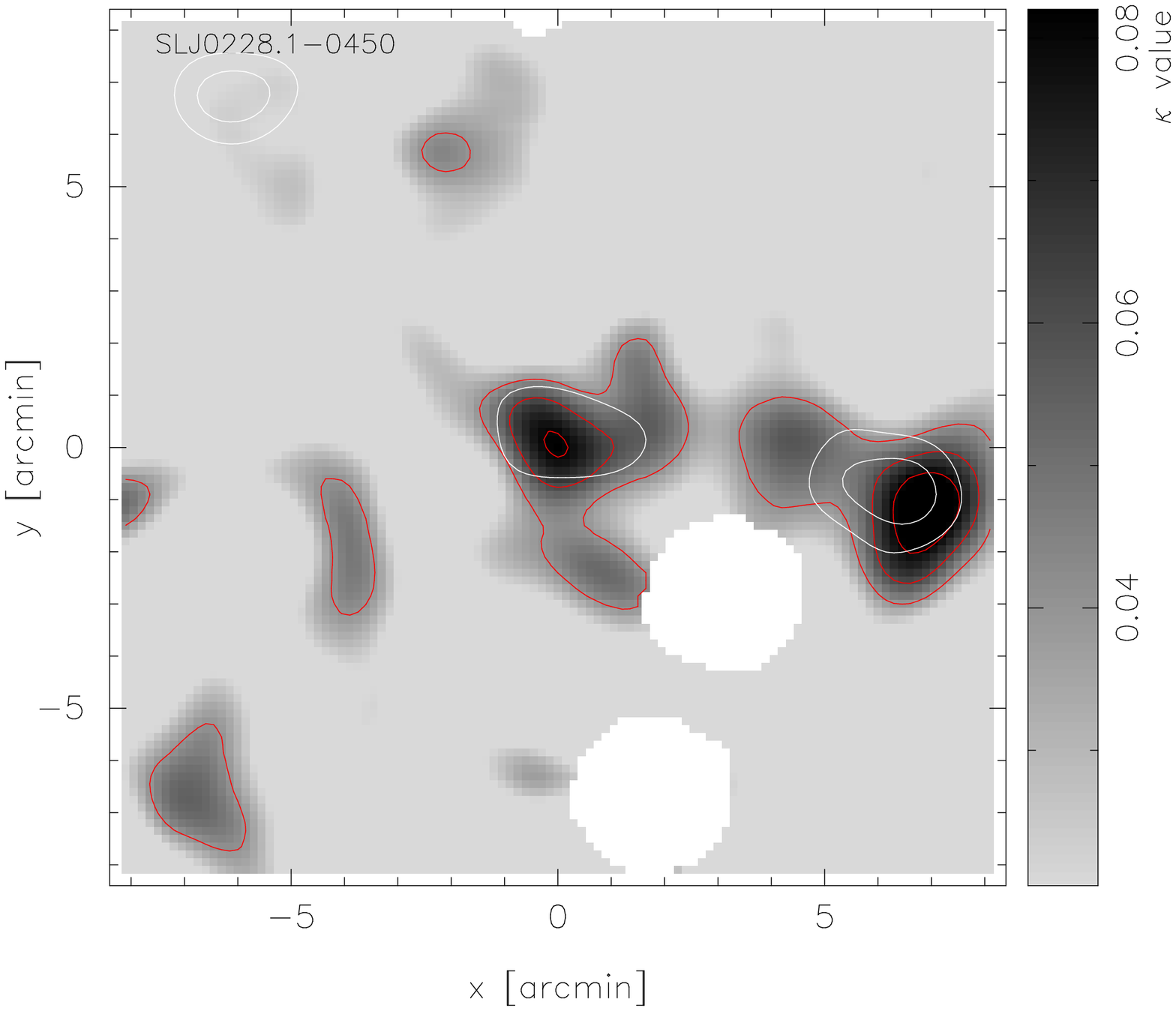}
\hspace{2mm}
\includegraphics[width=75mm,clip,angle=-90]{fig19c.ps}
\caption{Same as Figure \ref{fig:sxds_6} but for SL~J0228.1$-$0450.}
\end{figure*}

\clearpage
\begin{figure*}
\includegraphics[height=160mm,clip,angle=-90]{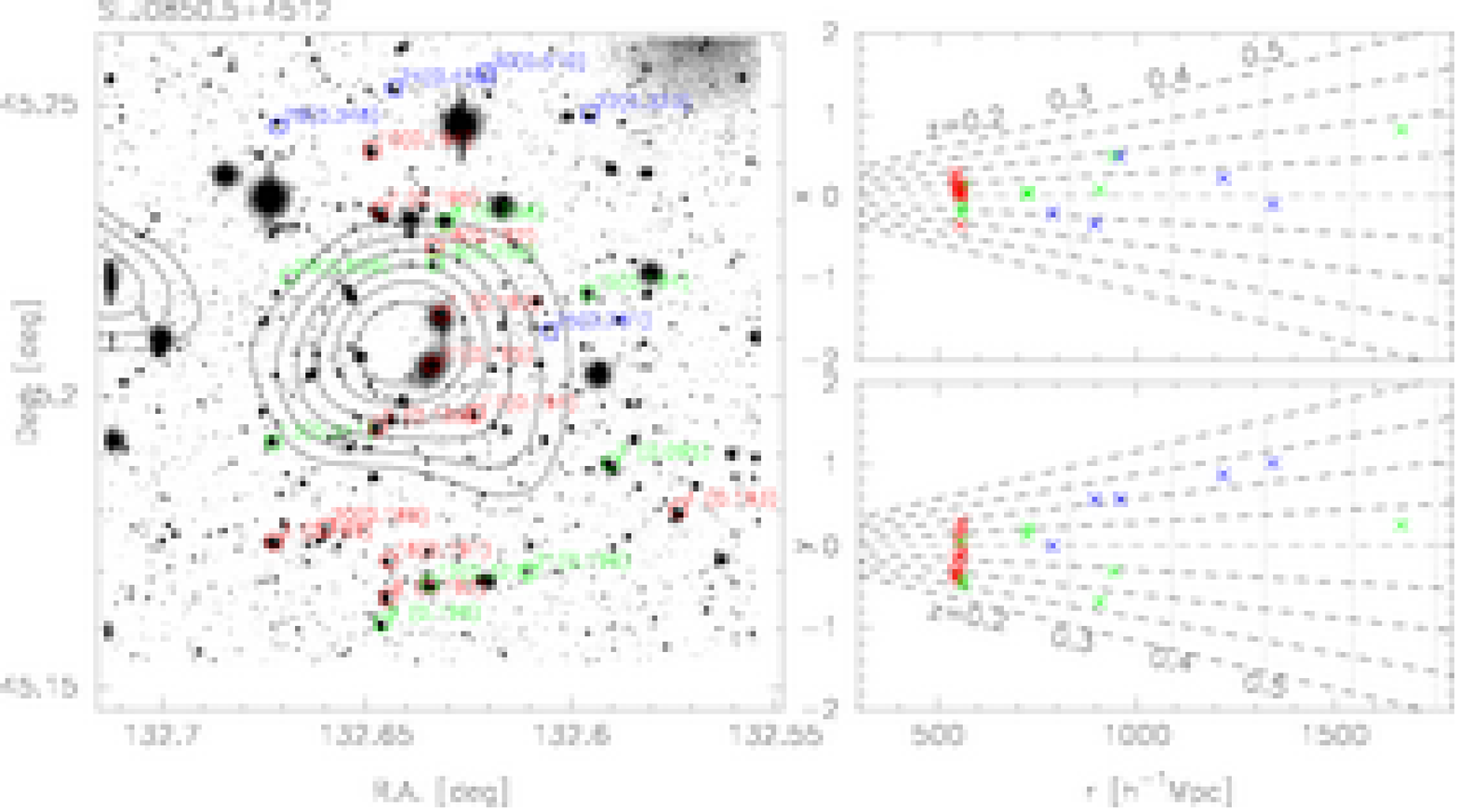}
\vspace{2mm}\\
\includegraphics[width=75mm,clip,angle=-90]{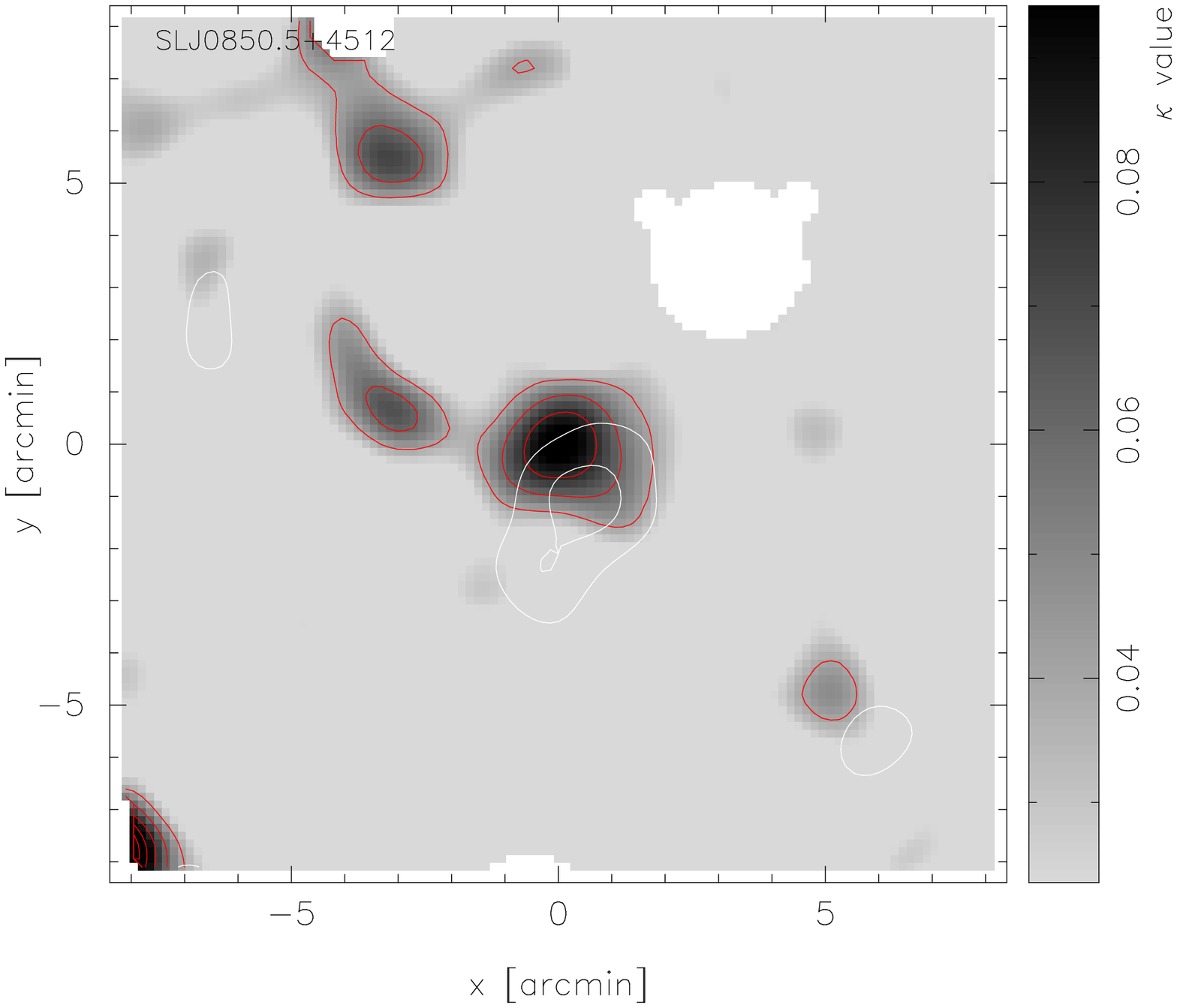}
\hspace{2mm}
\includegraphics[width=75mm,clip,angle=-90]{fig20c.ps}
\caption{Same as Figure \ref{fig:sxds_6} but for SL~J0850.5$+$4512.}
\end{figure*}

\clearpage
\begin{figure*}
\includegraphics[height=160mm,clip,angle=-90]{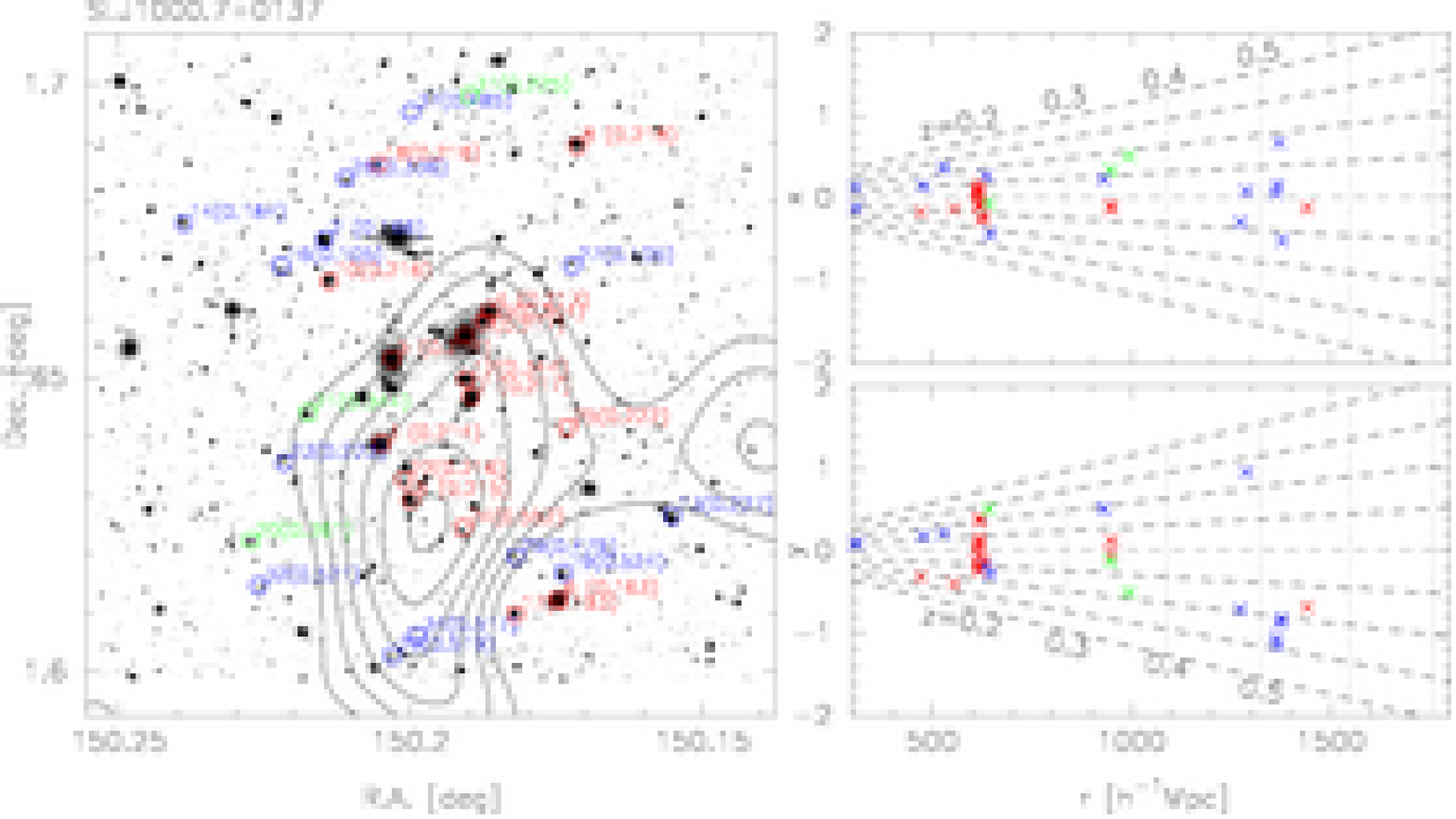}
\vspace{2mm}\\
\includegraphics[width=75mm,clip,angle=-90]{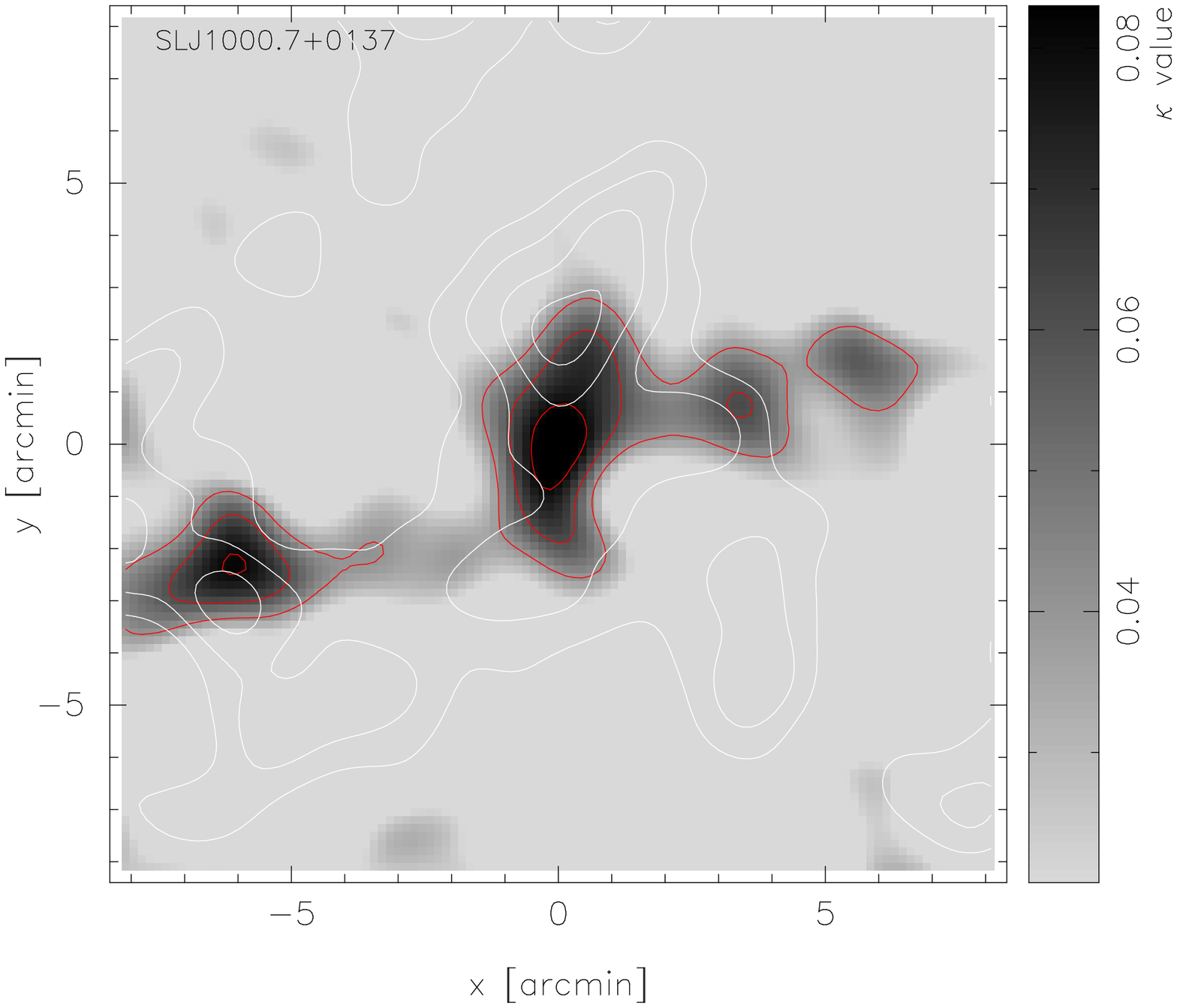}
\hspace{2mm}
\includegraphics[width=75mm,clip,angle=-90]{fig21c.ps}
\caption{Same as Figure \ref{fig:sxds_6} but for SL~J1000.7$+$0137.}
\end{figure*}

\clearpage
\begin{figure*}
\includegraphics[height=160mm,clip,angle=-90]{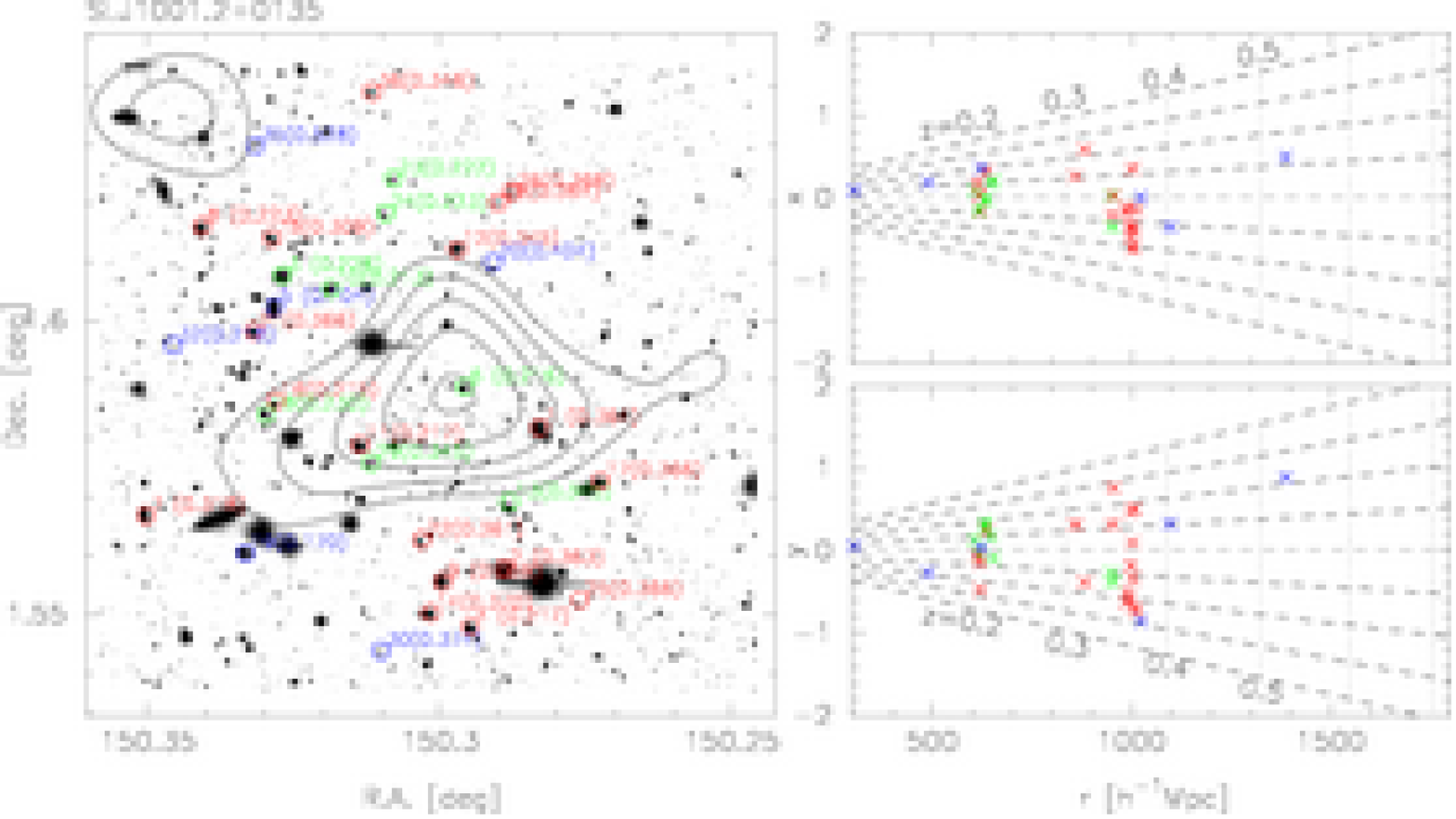}
\vspace{2mm}\\
\includegraphics[width=75mm,clip,angle=-90]{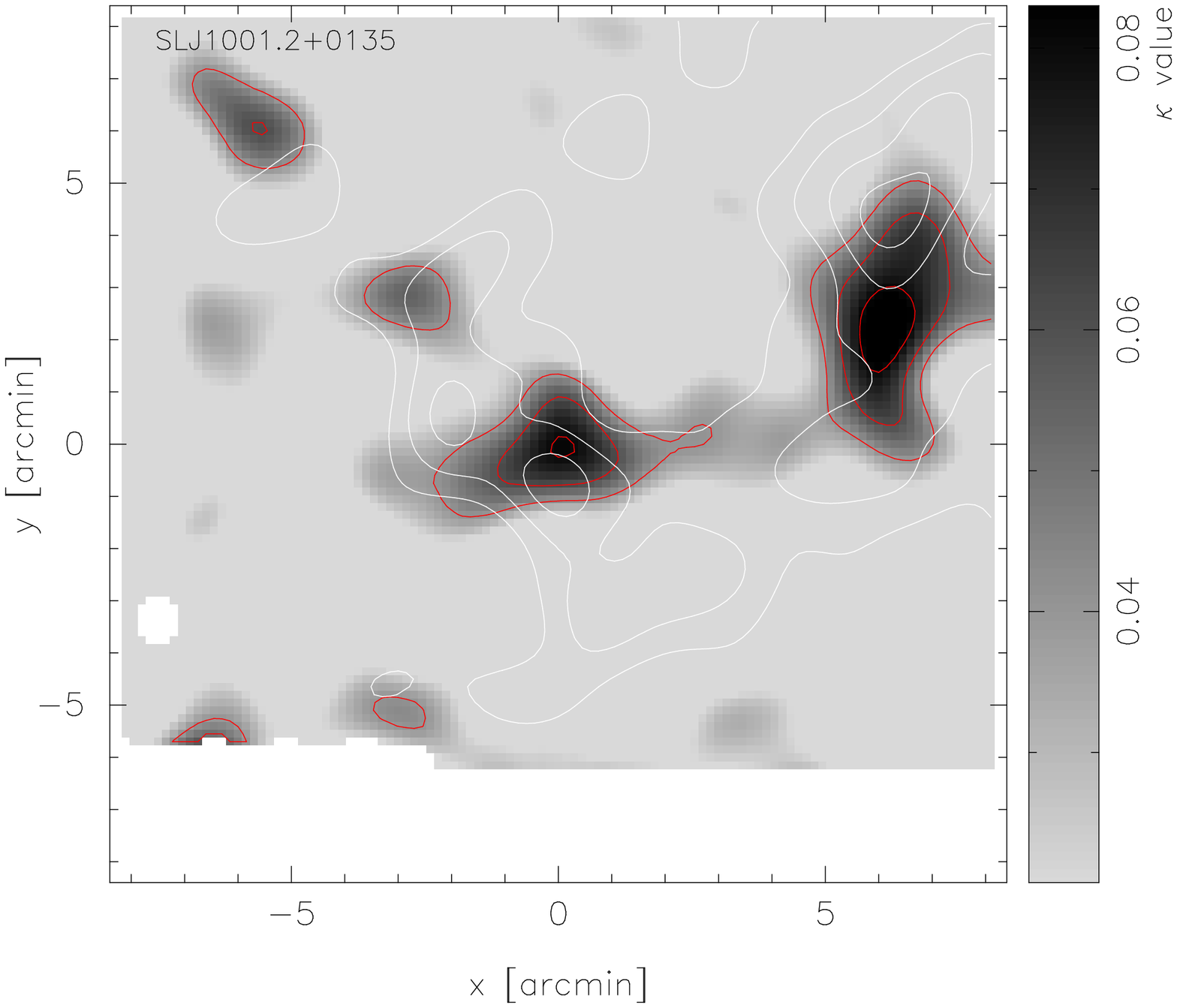}
\caption{Same as Figure \ref{fig:sxds_6} but for SL~J1001.2$+$0135.}
\end{figure*}

\clearpage
\begin{figure*}
\includegraphics[height=160mm,clip,angle=-90]{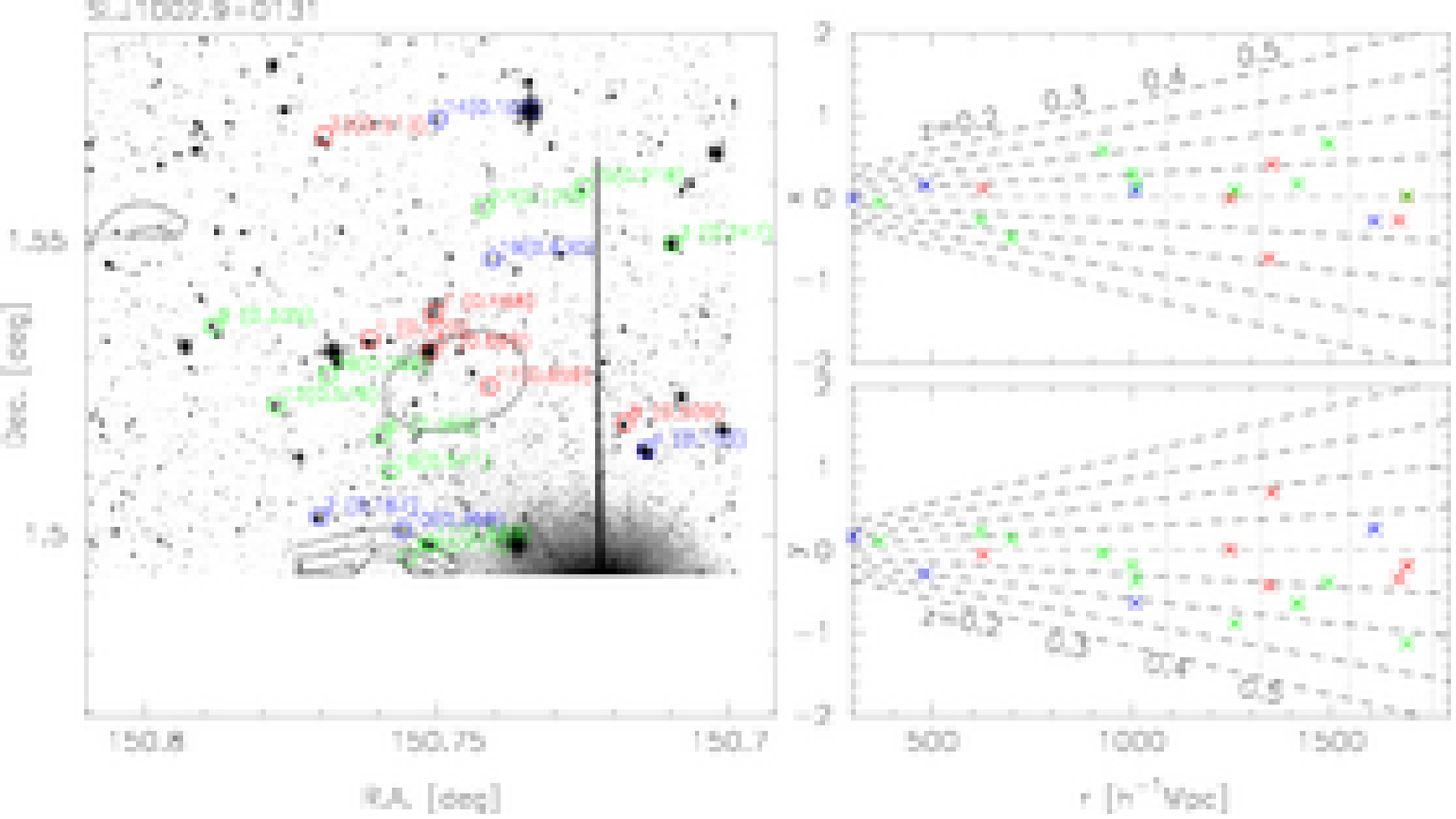}
\vspace{2mm}\\
\includegraphics[width=75mm,clip,angle=-90]{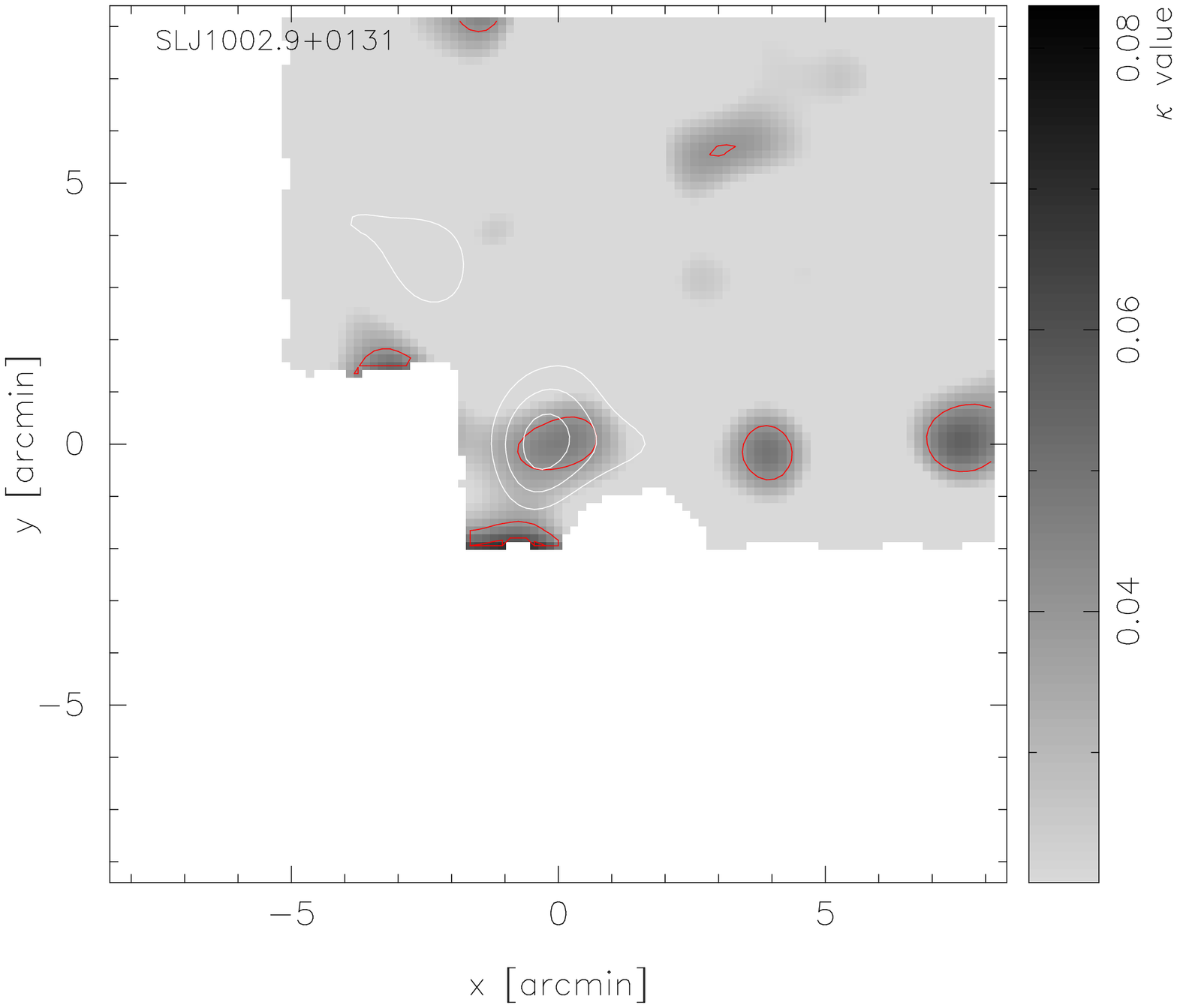}
\caption{Same as Figure \ref{fig:sxds_6} but for SL~J1002.9$+$0131.}
\end{figure*}

\clearpage
\begin{figure*}
\includegraphics[height=160mm,clip,angle=-90]{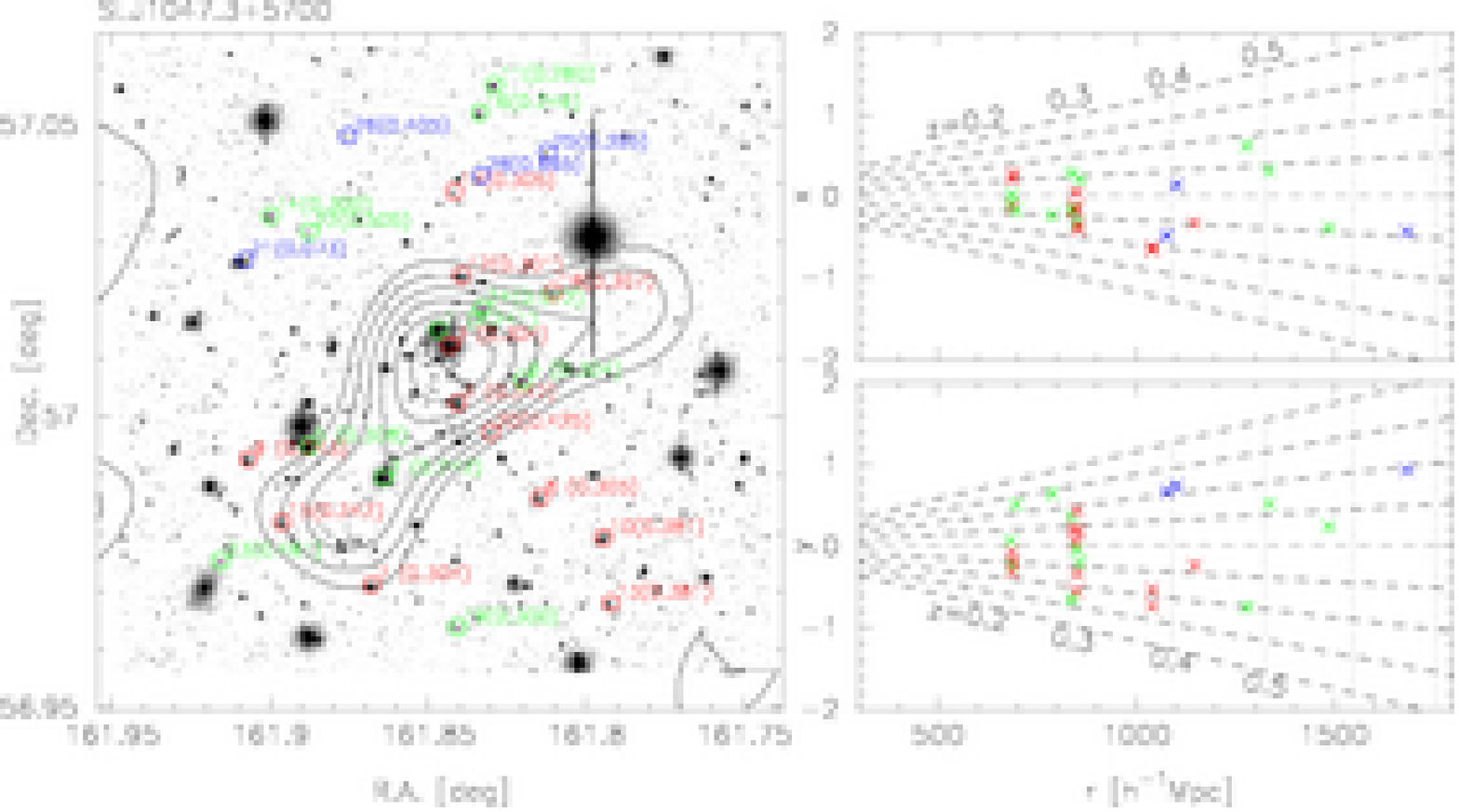}
\vspace{2mm}\\
\includegraphics[width=75mm,clip,angle=-90]{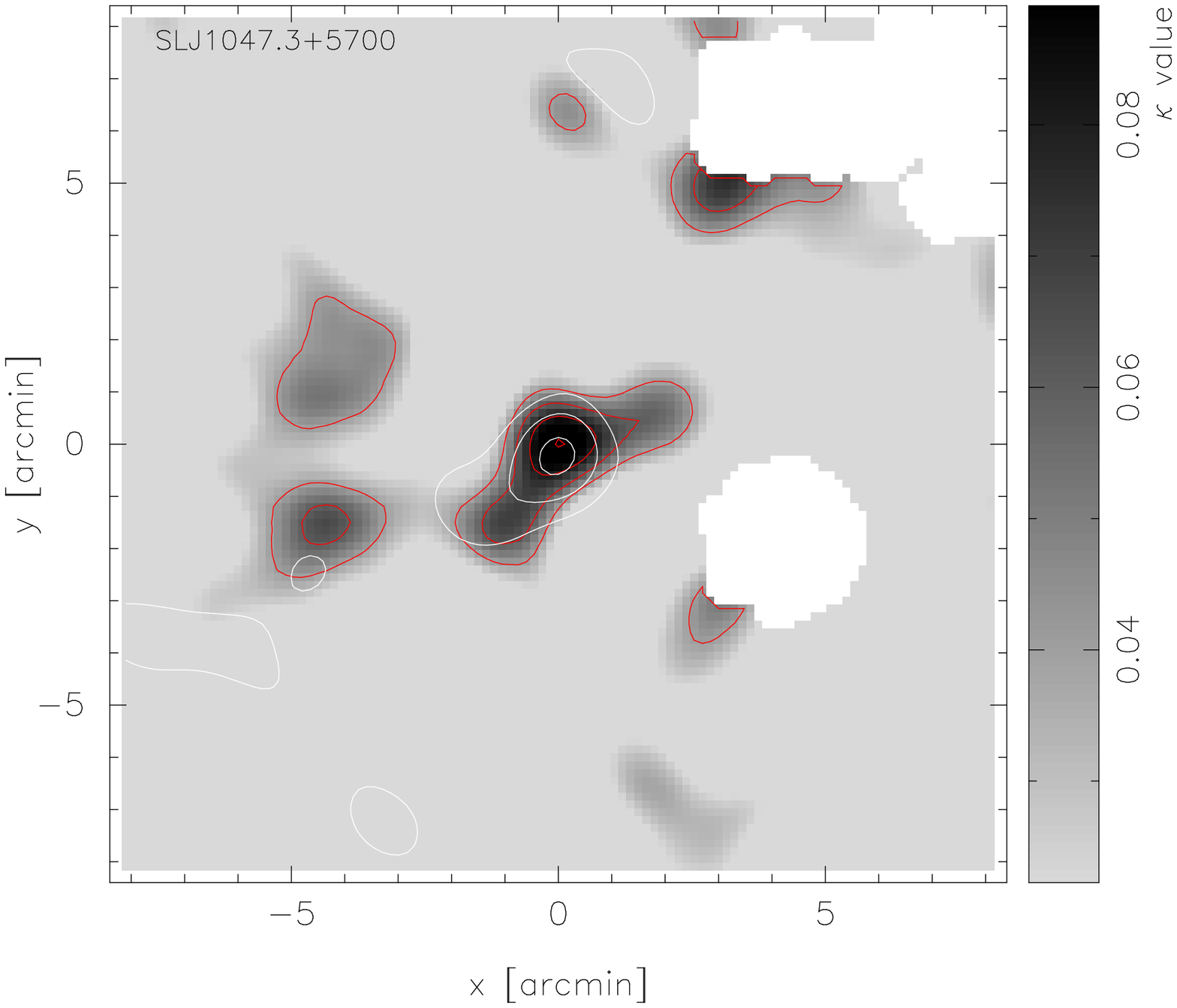}
\caption{Same as Figure \ref{fig:sxds_6} but for SL~J1047.3$+$5700.}
\end{figure*}

\clearpage
\begin{figure*}
\includegraphics[height=160mm,clip,angle=-90]{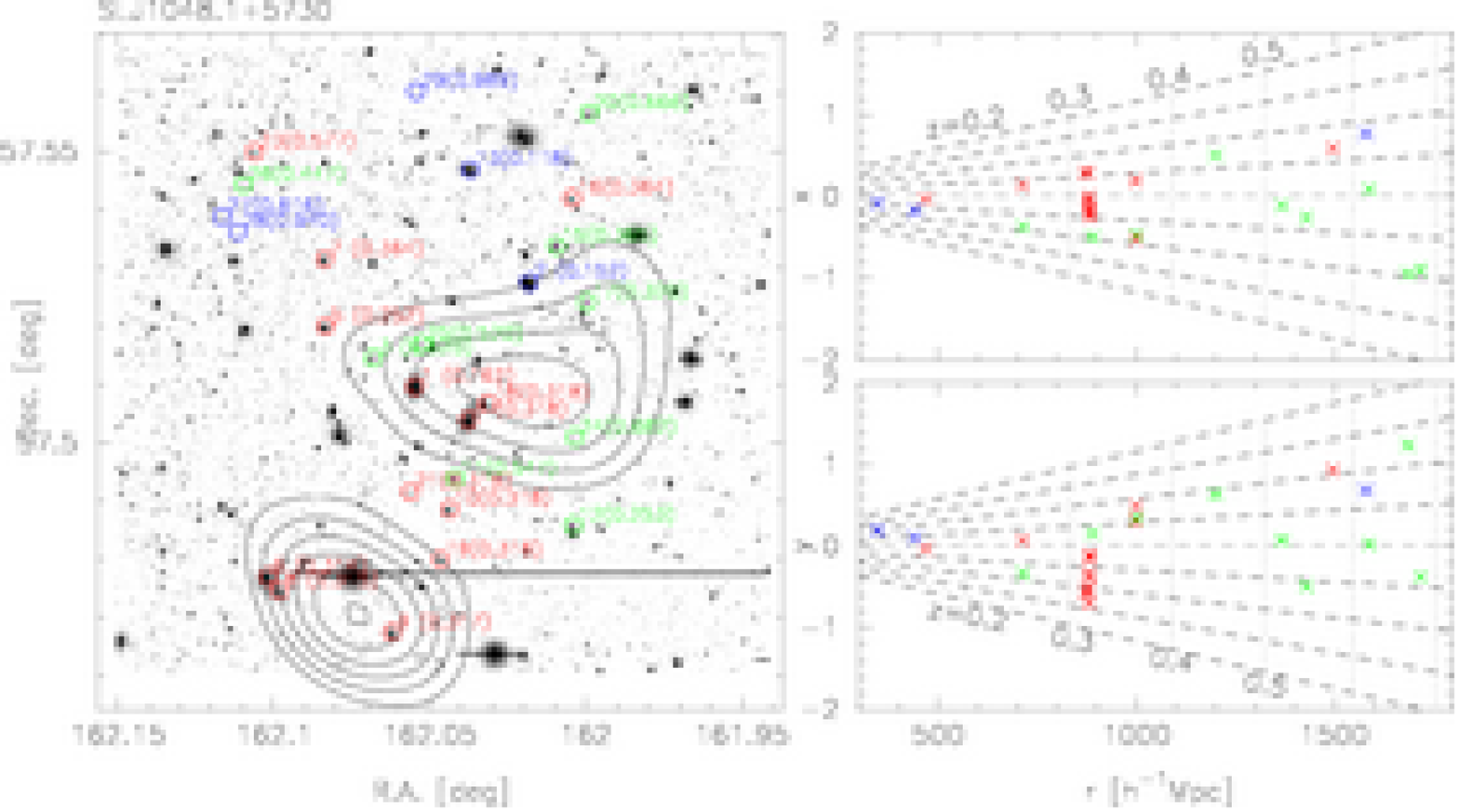}
\vspace{2mm}\\
\includegraphics[width=75mm,clip,angle=-90]{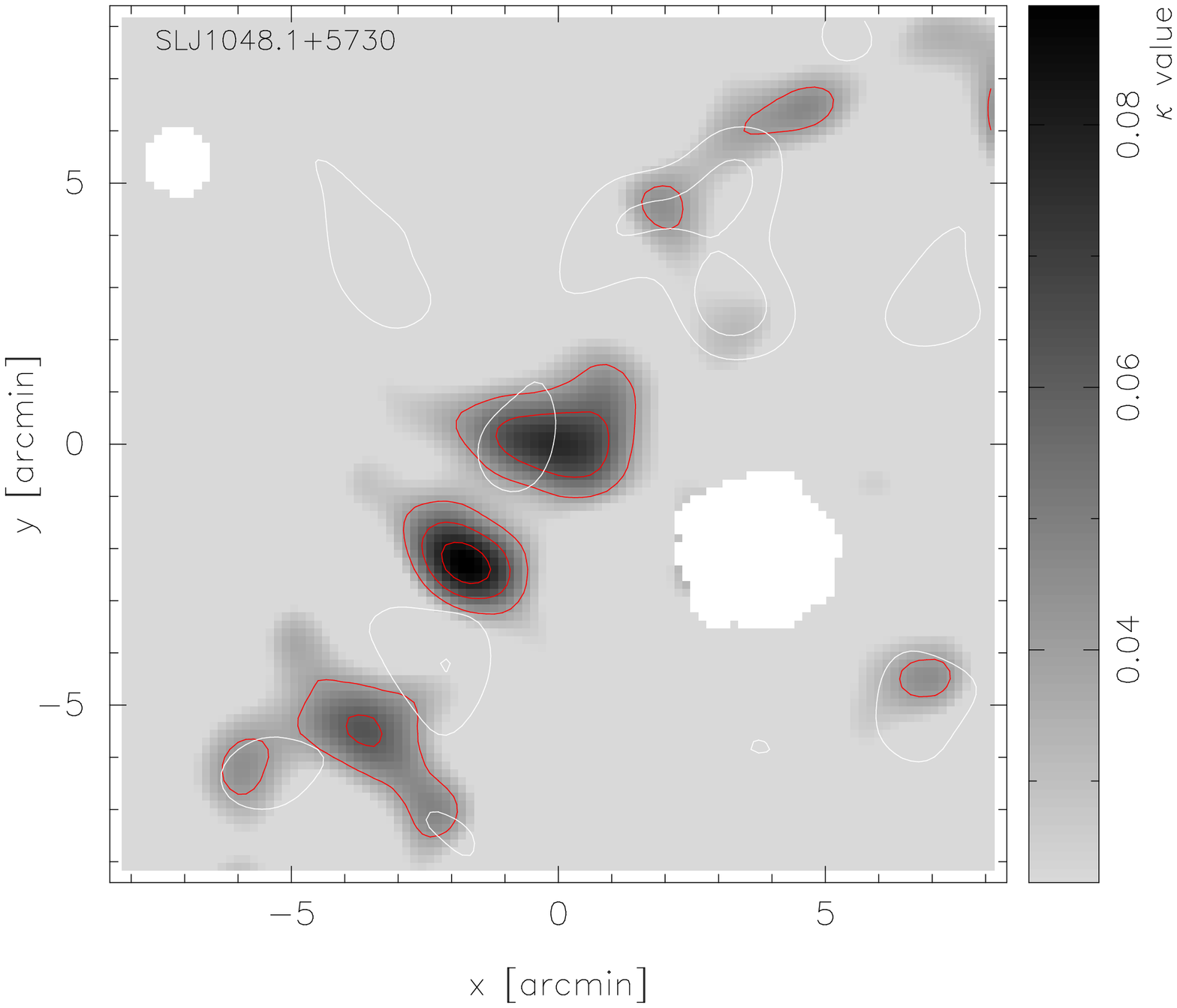}
\hspace{2mm}
\includegraphics[width=75mm,clip,angle=-90]{fig25c.ps}
\caption{Same as Figure \ref{fig:sxds_6} but for SL~J1048.1$+$5730.}
\end{figure*}

\clearpage
\begin{figure*}
\includegraphics[height=160mm,clip,angle=-90]{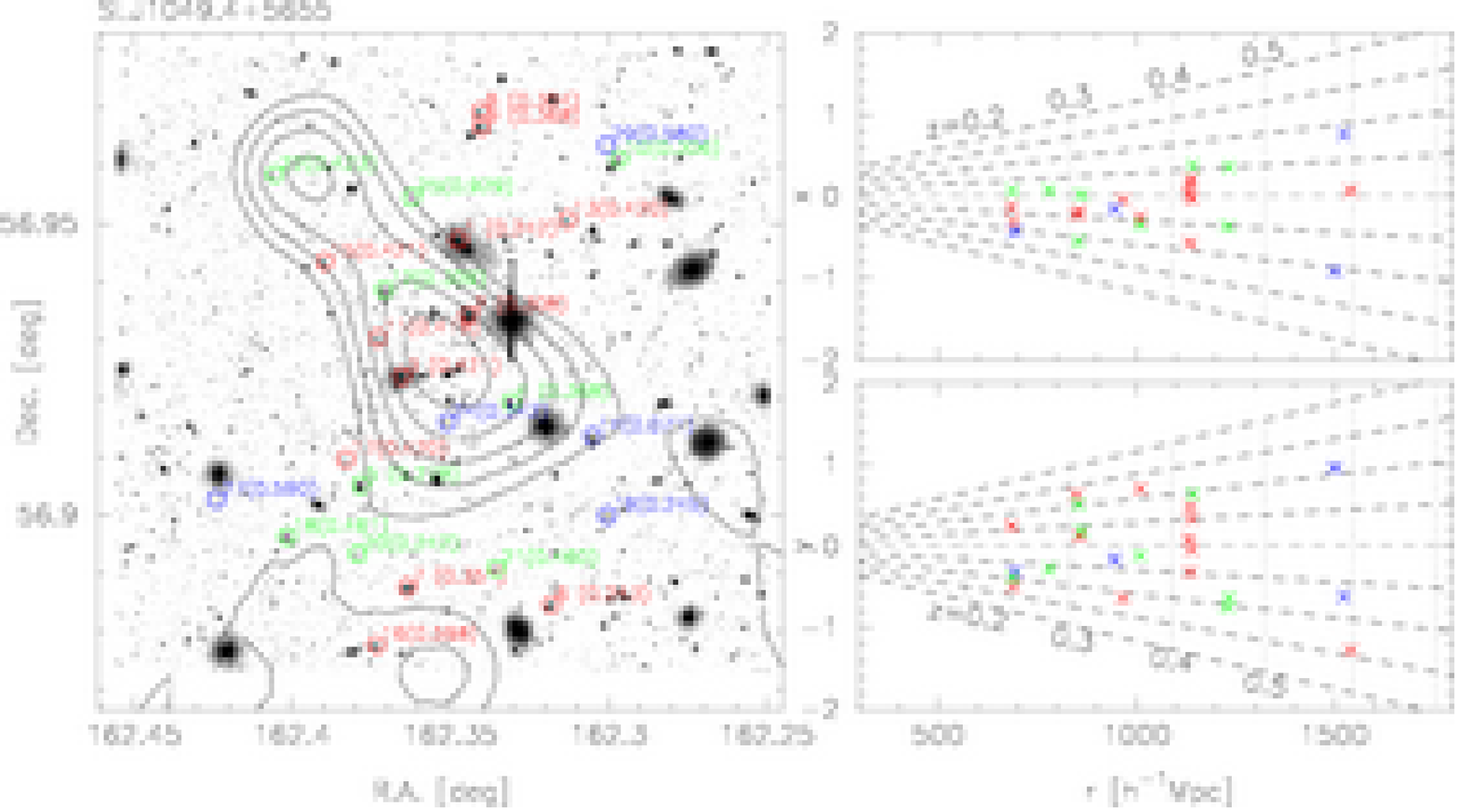}
\vspace{2mm}\\
\includegraphics[width=75mm,clip,angle=-90]{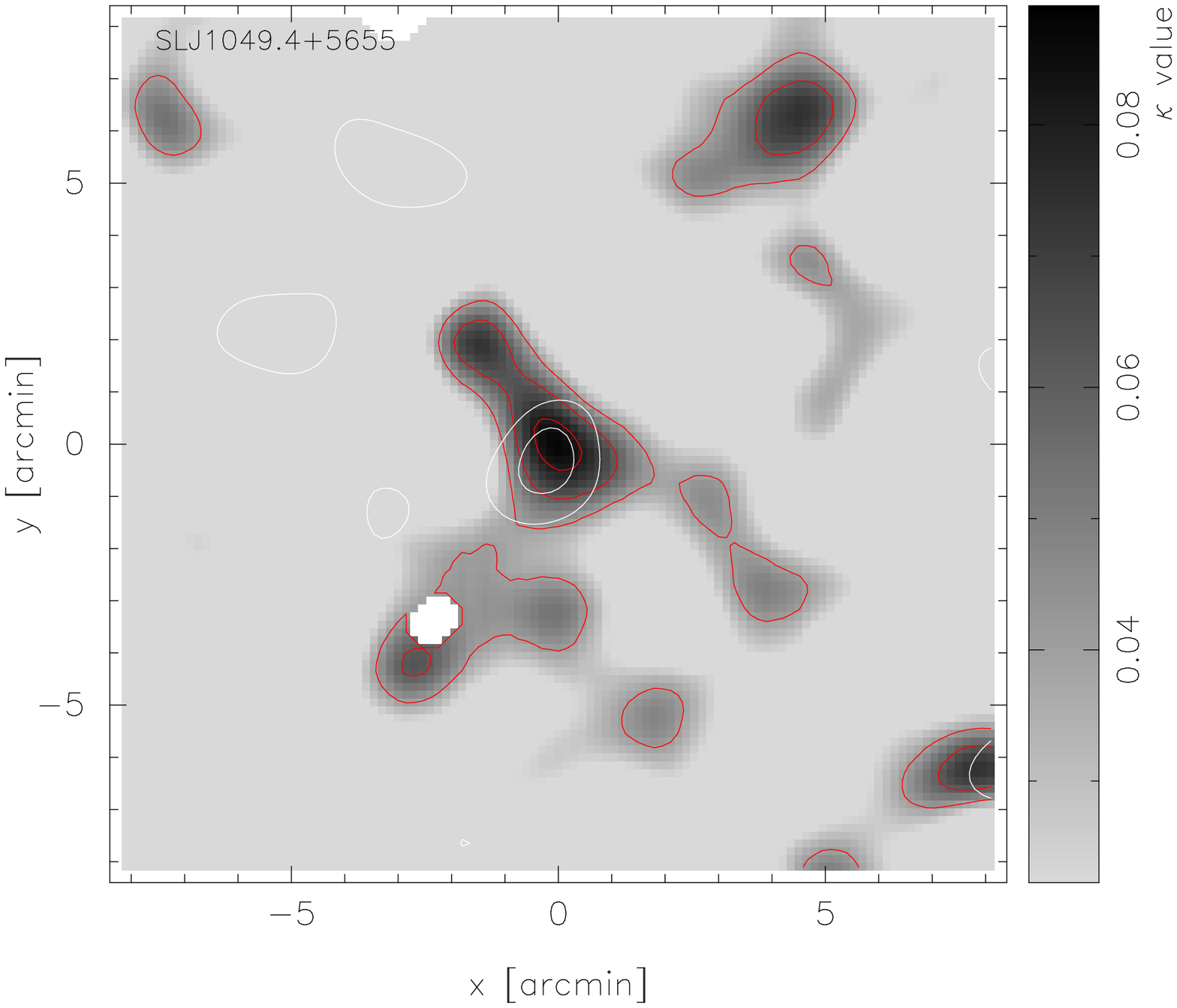}
\hspace{2mm}
\includegraphics[width=75mm,clip,angle=-90]{fig26c.ps}
\caption{Same as Figure \ref{fig:sxds_6} but for SL~J1049.4$+$5655.}
\end{figure*}

\clearpage
\begin{figure*}
\includegraphics[height=160mm,clip,angle=-90]{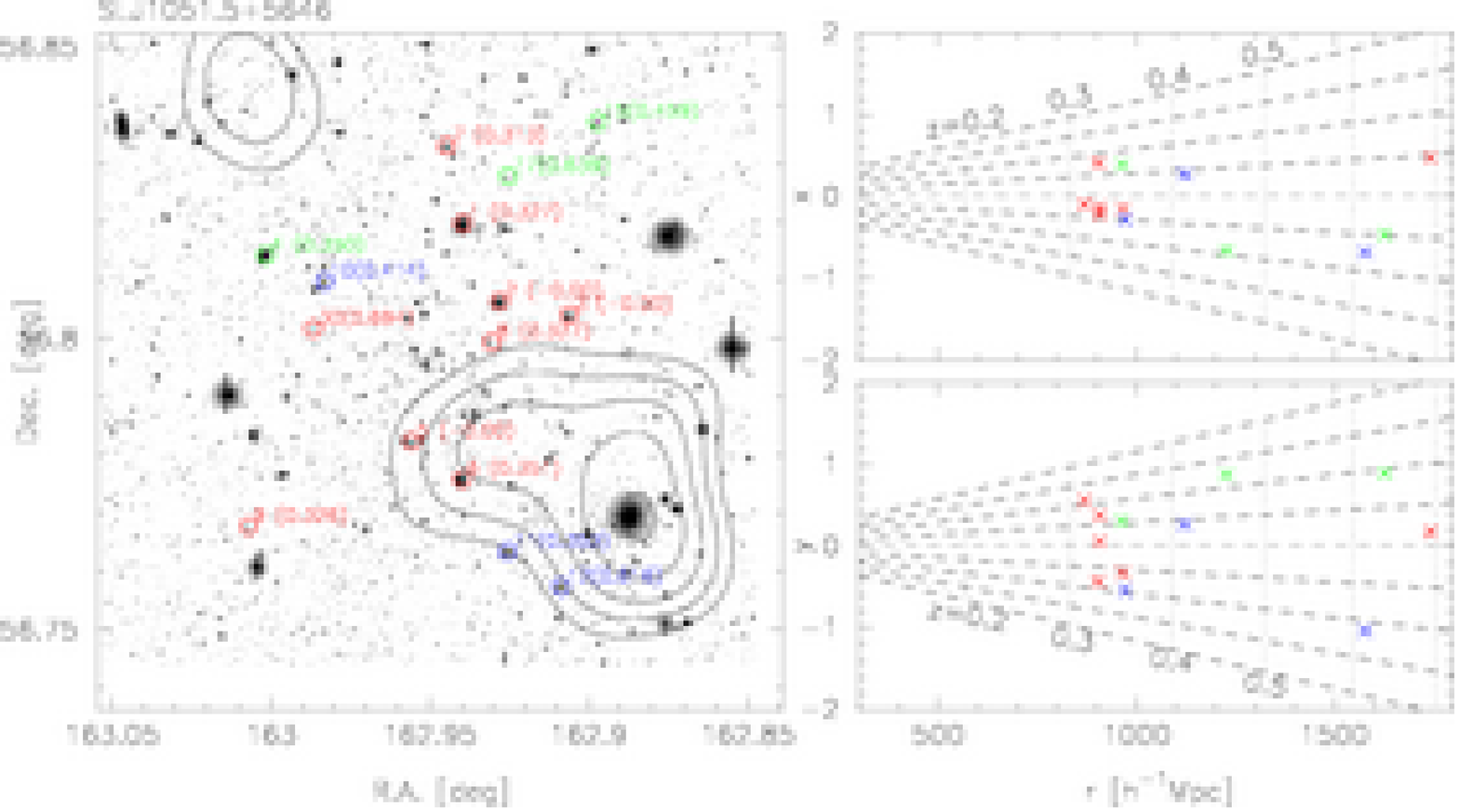}
\vspace{2mm}\\
\includegraphics[width=75mm,clip,angle=-90]{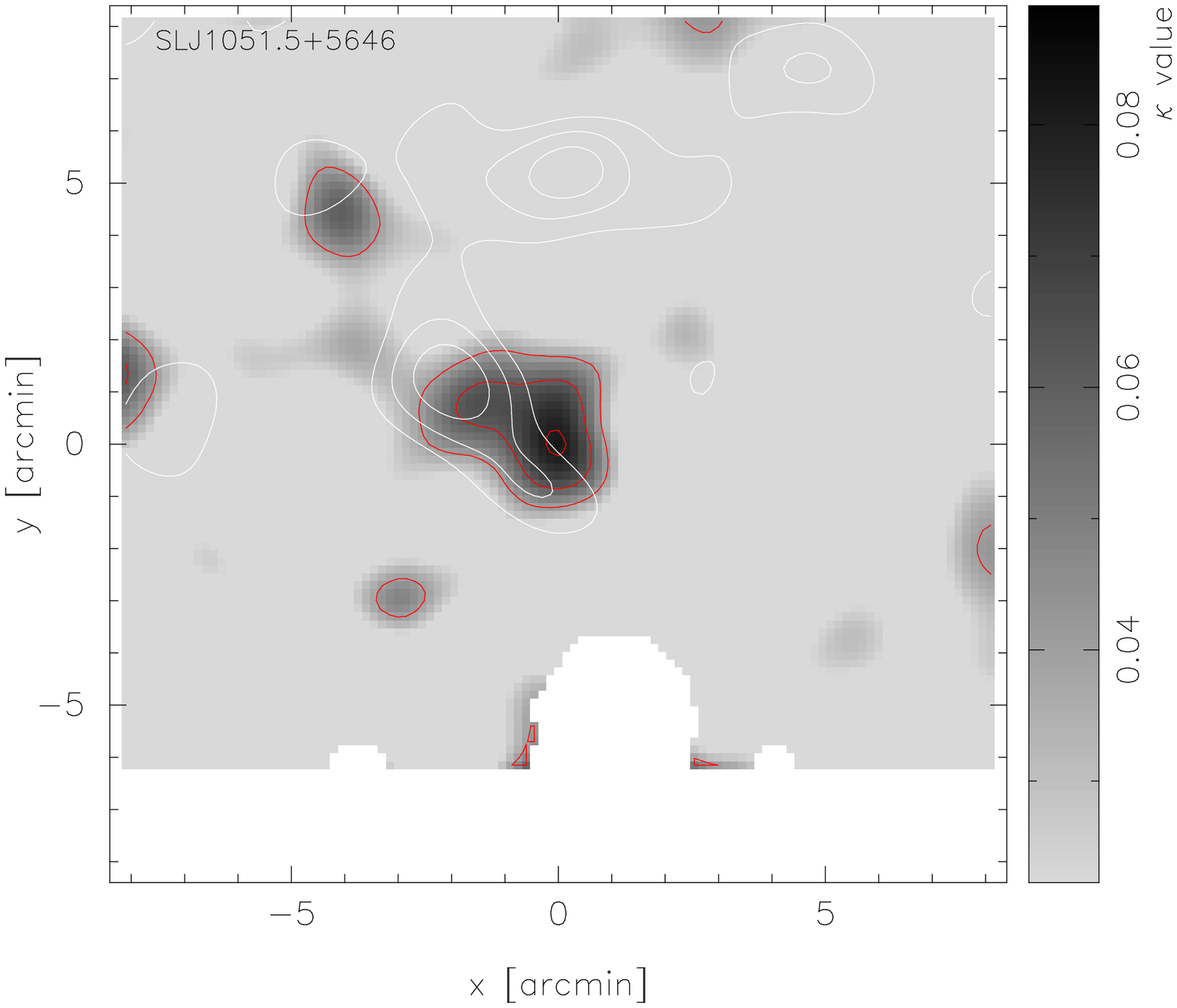}
\caption{Same as Figure \ref{fig:sxds_6} but for SL~J1051.5$+$5646.}
\end{figure*}

\clearpage
\begin{figure*}
\includegraphics[height=160mm,clip,angle=-90]{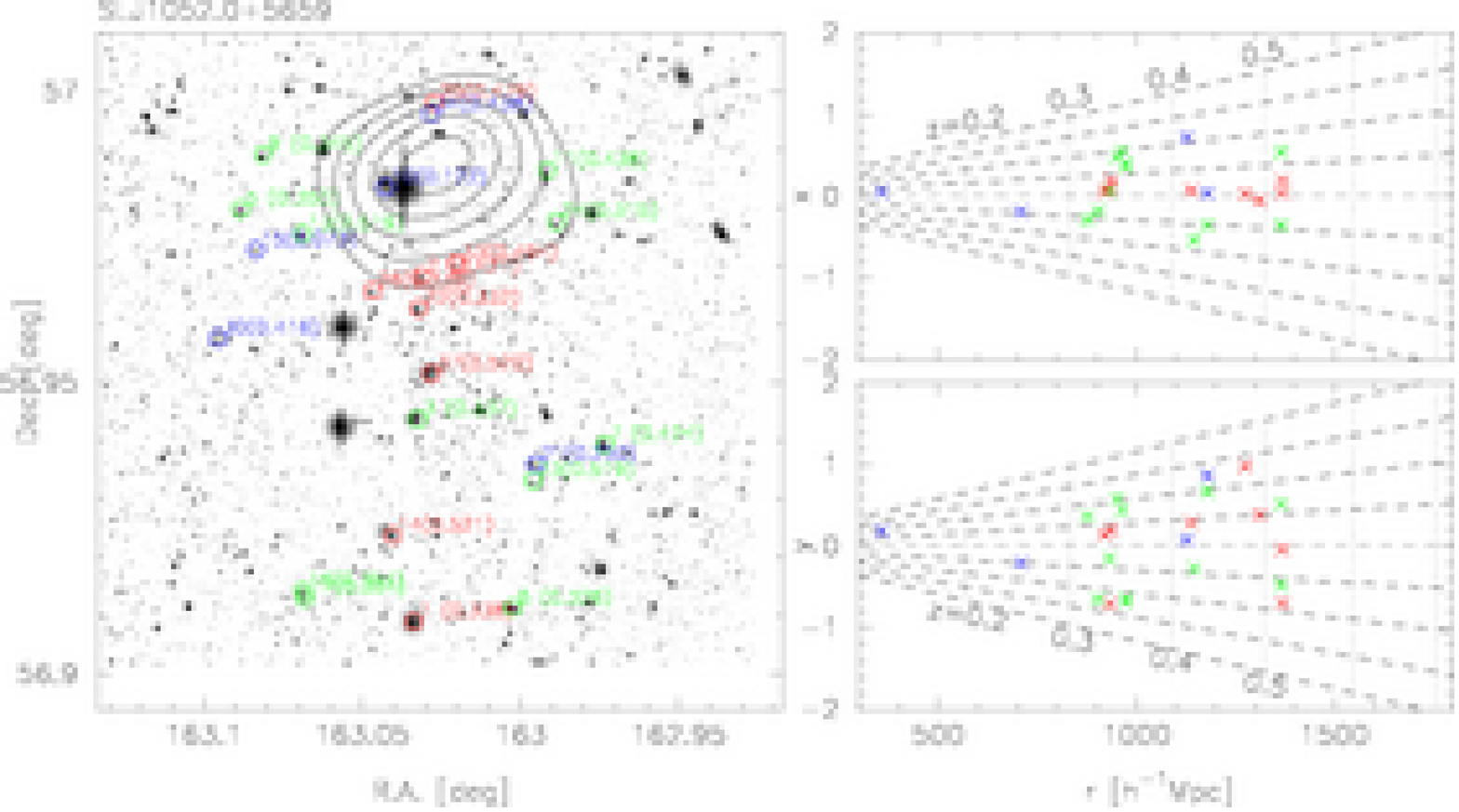}
\vspace{2mm}\\
\includegraphics[width=75mm,clip,angle=-90]{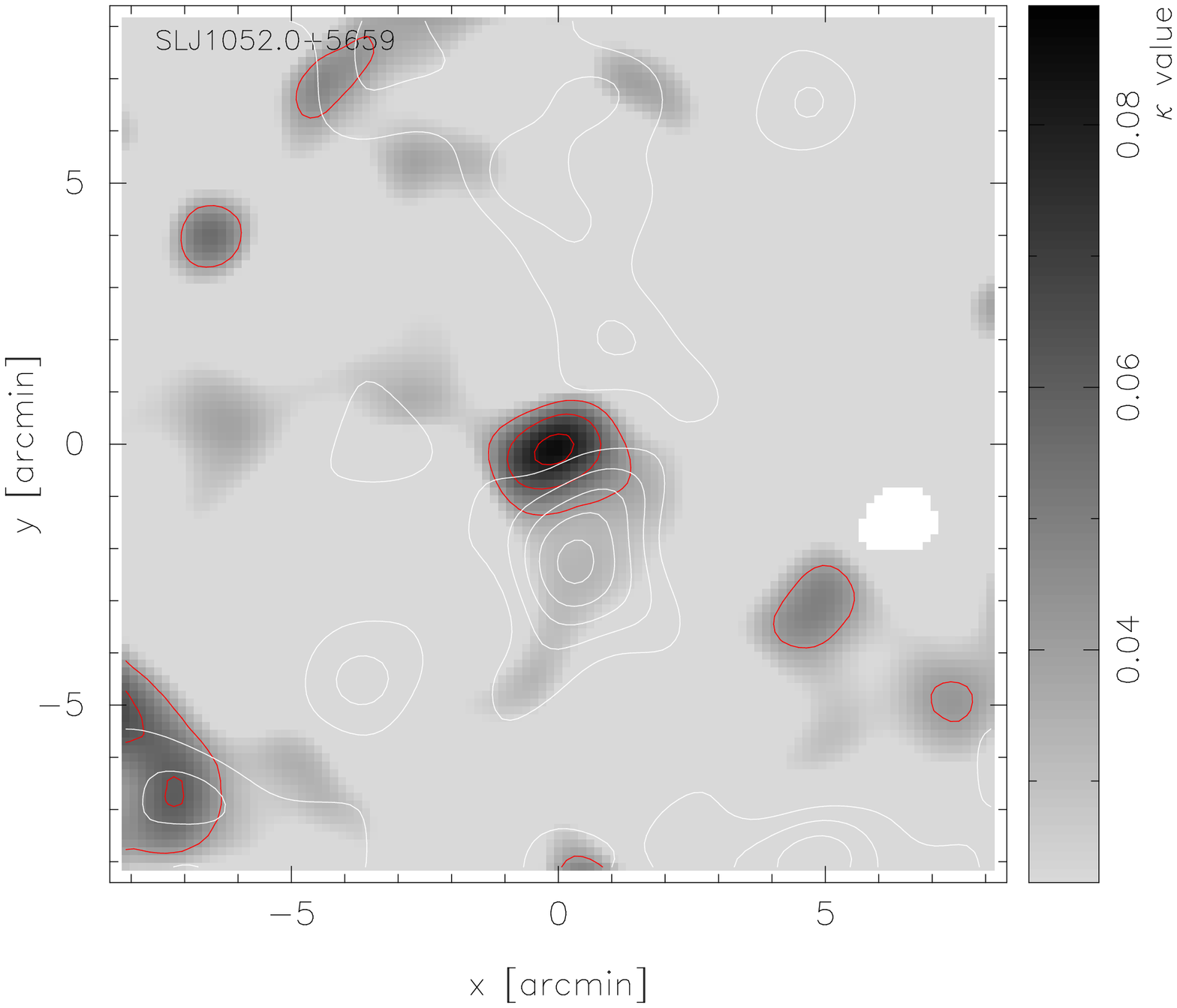}
\caption{Same as Figure \ref{fig:sxds_6} but for SL~J1052.0$+$5659.}
\end{figure*}

\clearpage
\begin{figure*}
\includegraphics[height=160mm,clip,angle=-90]{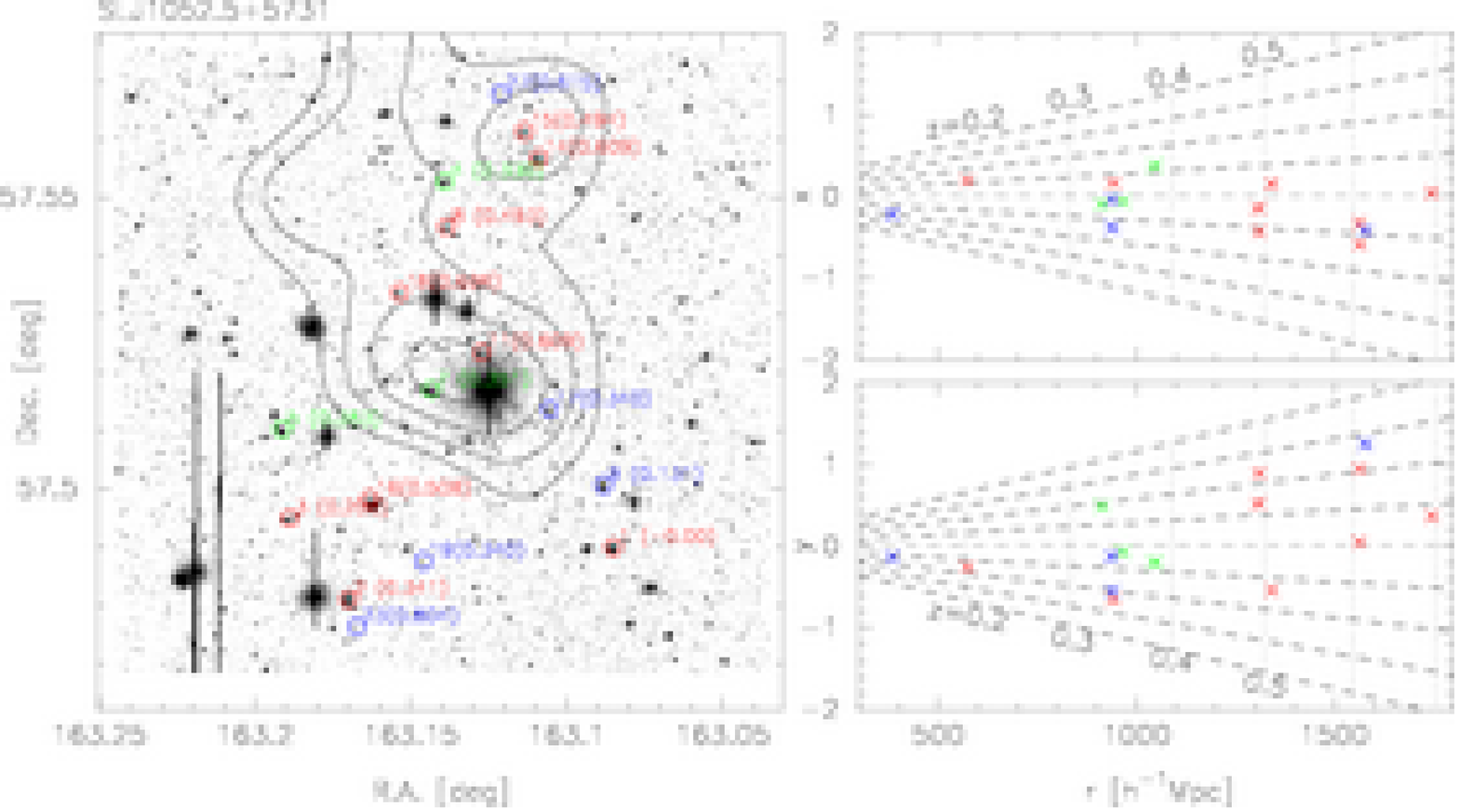}
\vspace{2mm}\\
\includegraphics[width=75mm,clip,angle=-90]{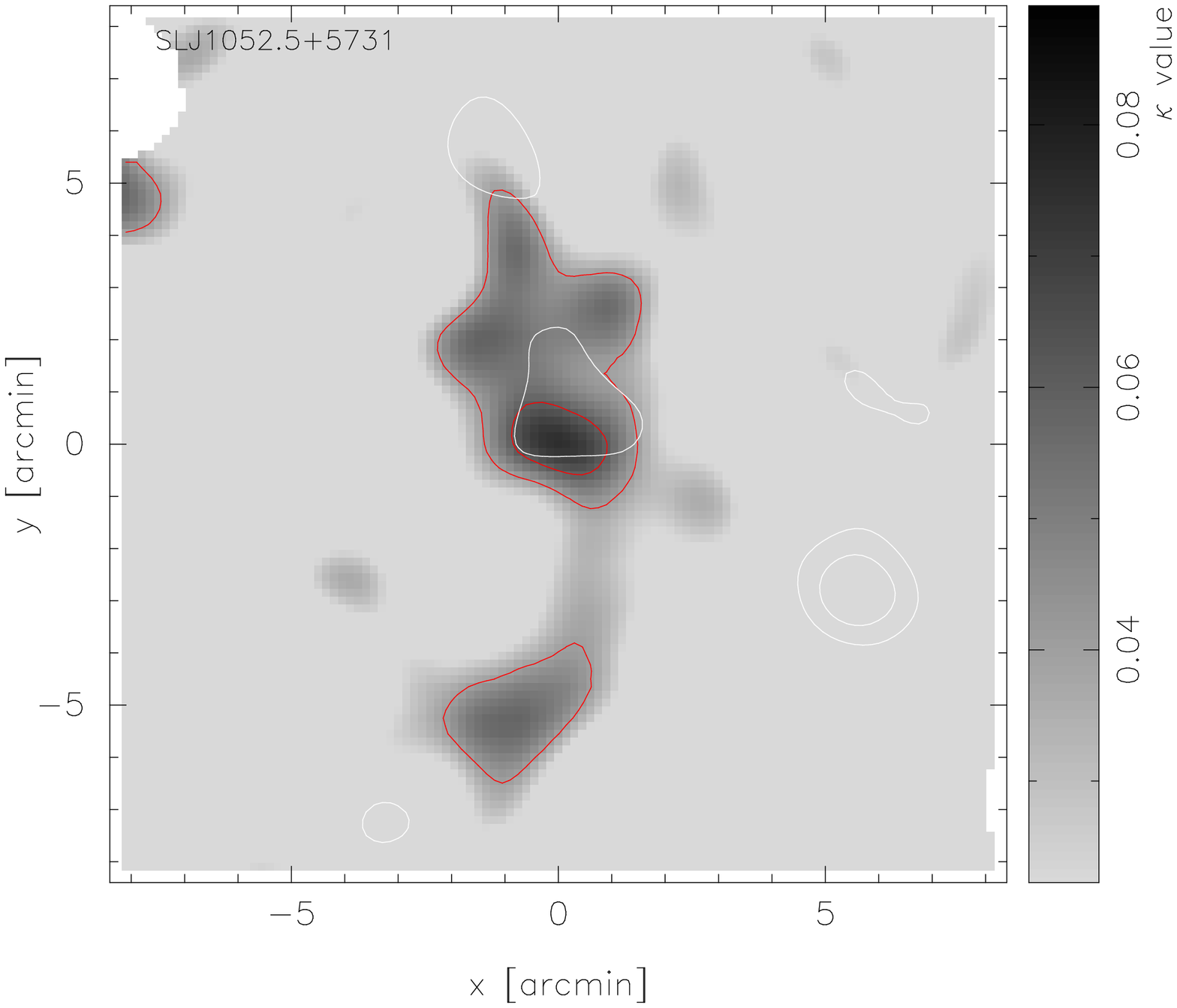}
\caption{Same as Figure \ref{fig:sxds_6} but for SL~J1052.5$+$5731.}
\end{figure*}

\clearpage
\begin{figure*}
\includegraphics[height=160mm,clip,angle=-90]{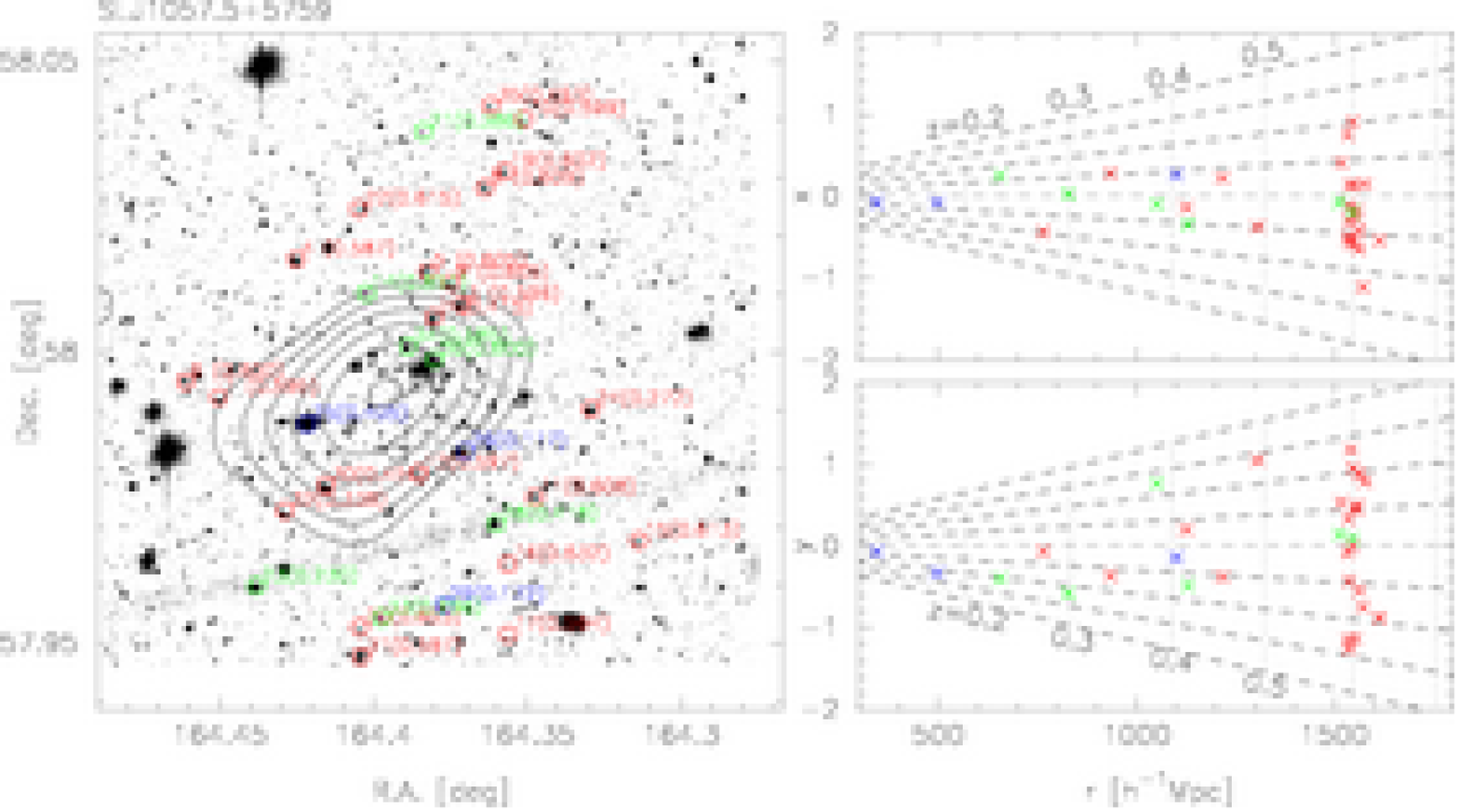}
\vspace{2mm}\\
\includegraphics[width=75mm,clip,angle=-90]{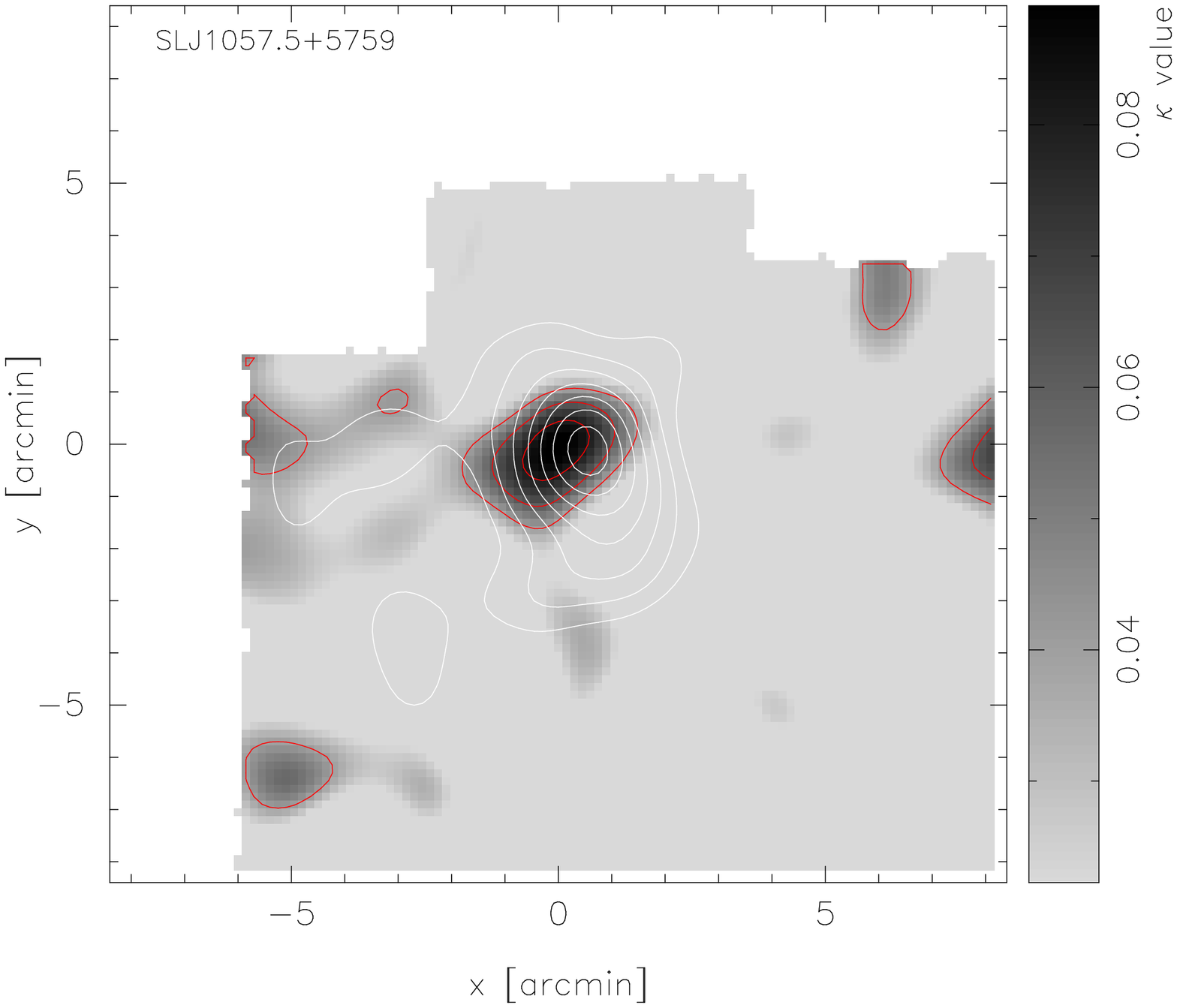}
\hspace{2mm}
\includegraphics[width=75mm,clip,angle=-90]{fig30c.ps}
\caption{Same as Figure \ref{fig:sxds_6} but for SL~J1057.5$+$5759.}
\end{figure*}

\clearpage
\begin{figure*}
\includegraphics[height=160mm,clip,angle=-90]{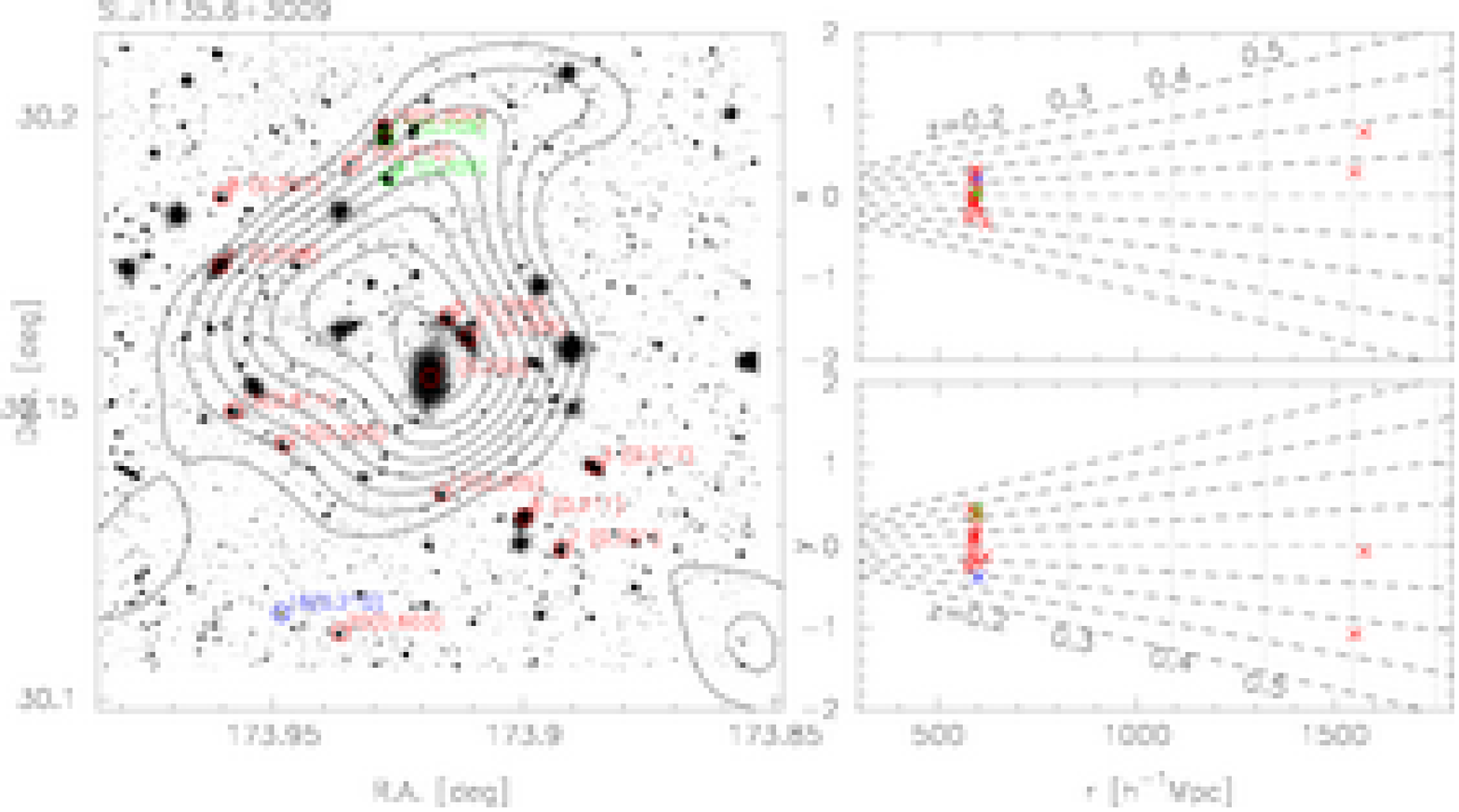}
\vspace{2mm}\\
\includegraphics[width=75mm,clip,angle=-90]{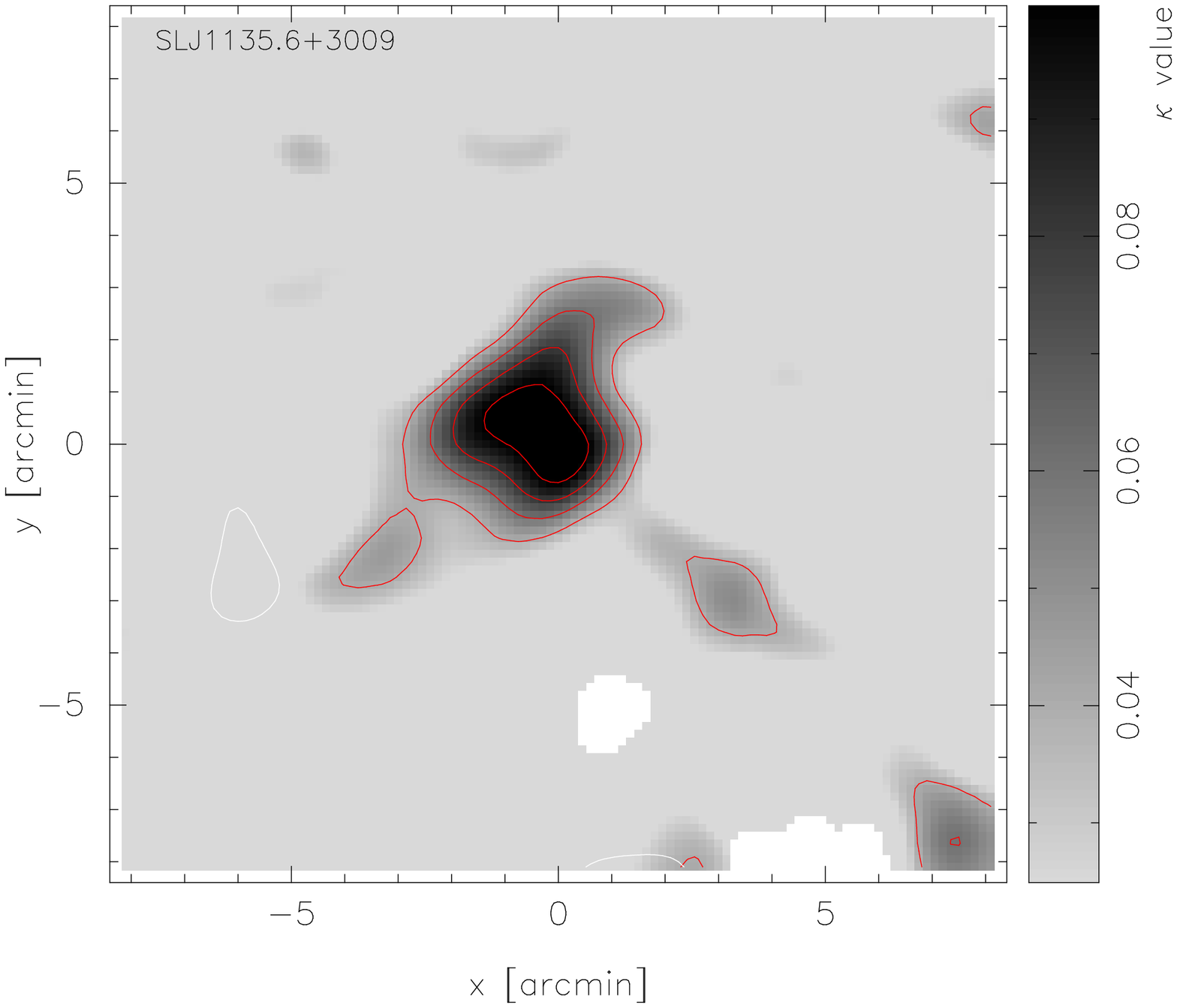}
\hspace{2mm}
\includegraphics[width=75mm,clip,angle=-90]{fig31c.ps}
\caption{Same as Figure \ref{fig:sxds_6} but for SL~J1135.6$+$3009.}
\end{figure*}

\clearpage
\begin{figure*}
\includegraphics[height=160mm,clip,angle=-90]{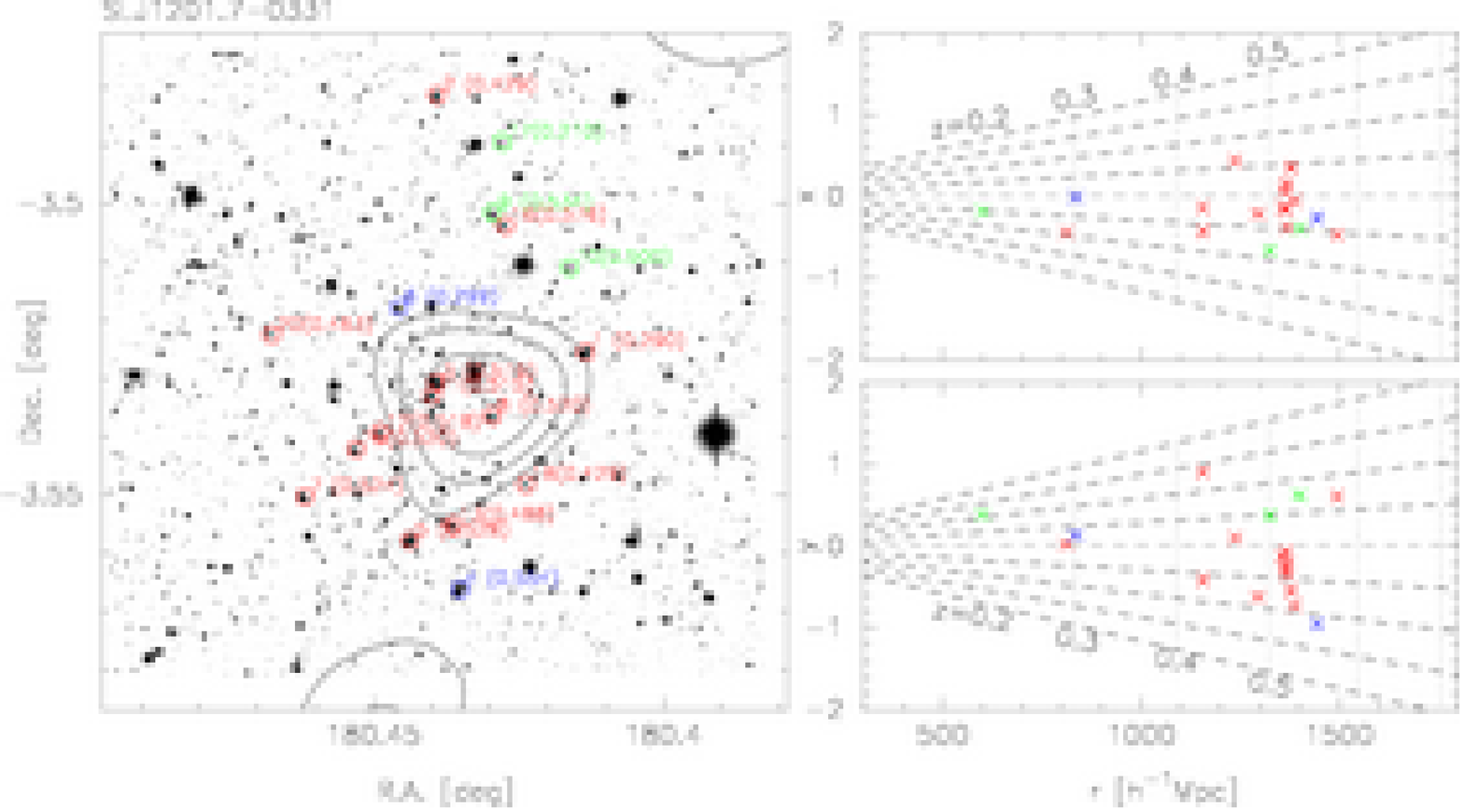}
\vspace{2mm}\\
\includegraphics[width=75mm,clip,angle=-90]{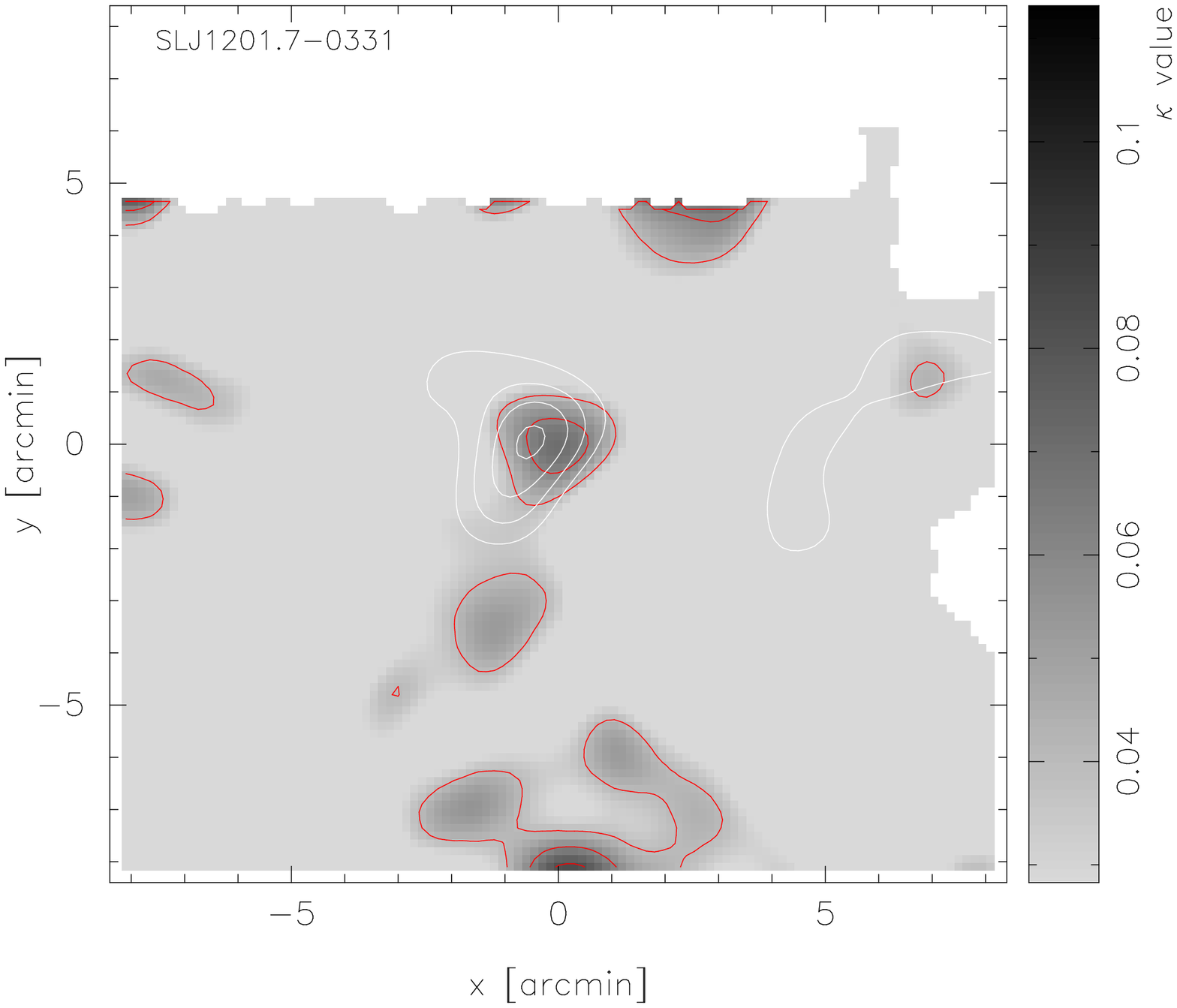}
\hspace{2mm}
\includegraphics[width=75mm,clip,angle=-90]{fig32c.ps}
\caption{Same as Figure \ref{fig:sxds_6} but for SL~J1201.7$-$0331.}
\end{figure*}

\clearpage
\begin{figure*}
\includegraphics[height=160mm,clip,angle=-90]{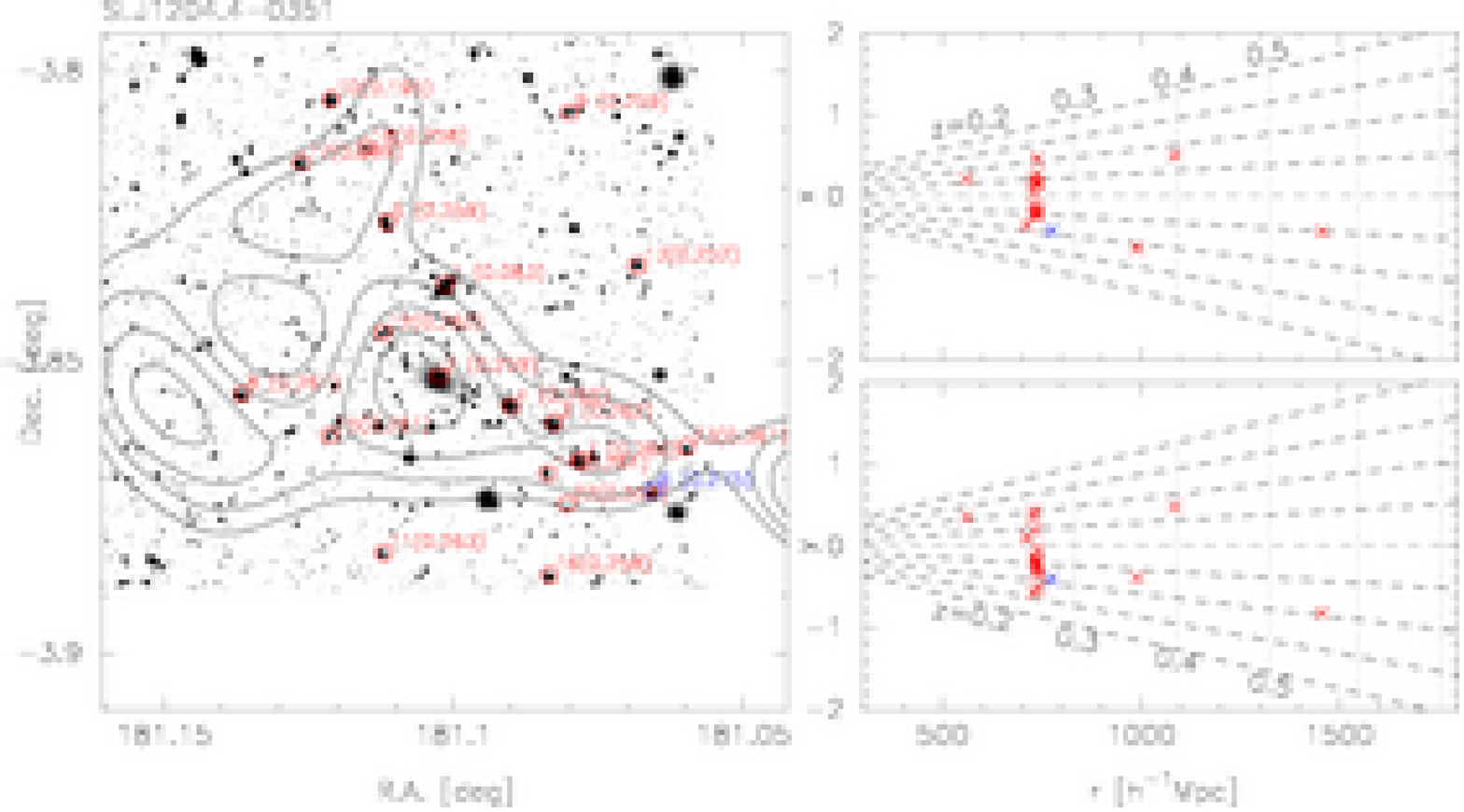}
\vspace{2mm}\\
\includegraphics[width=75mm,clip,angle=-90]{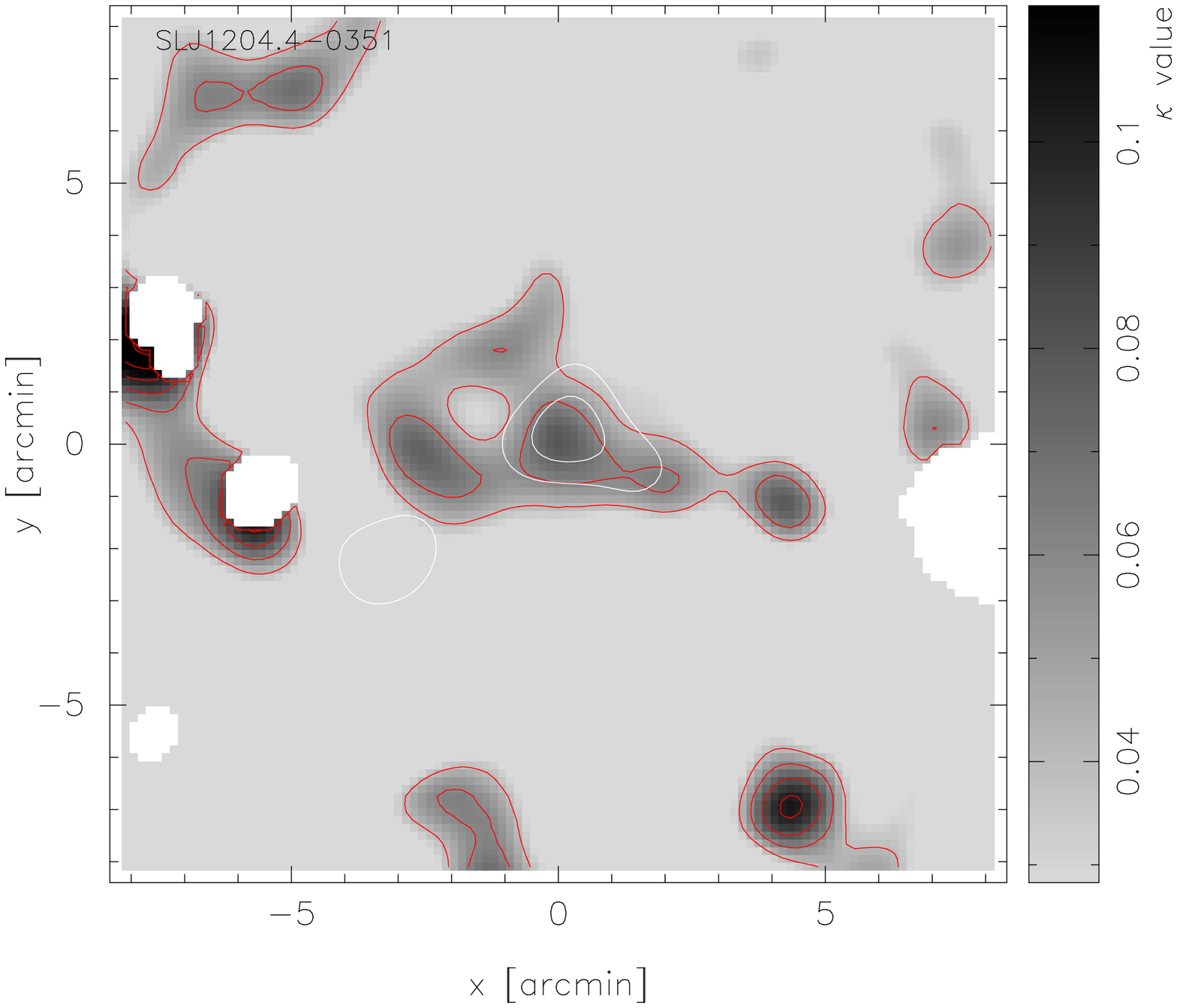}
\hspace{2mm}
\includegraphics[width=75mm,clip,angle=-90]{fig33c.ps}
\caption{Same as Figure \ref{fig:sxds_6} but for
SL~J1204.4$-$0351.
\label{pg1159-h2-170}}
\end{figure*}

\clearpage
\begin{figure*}
\includegraphics[height=160mm,clip,angle=-90]{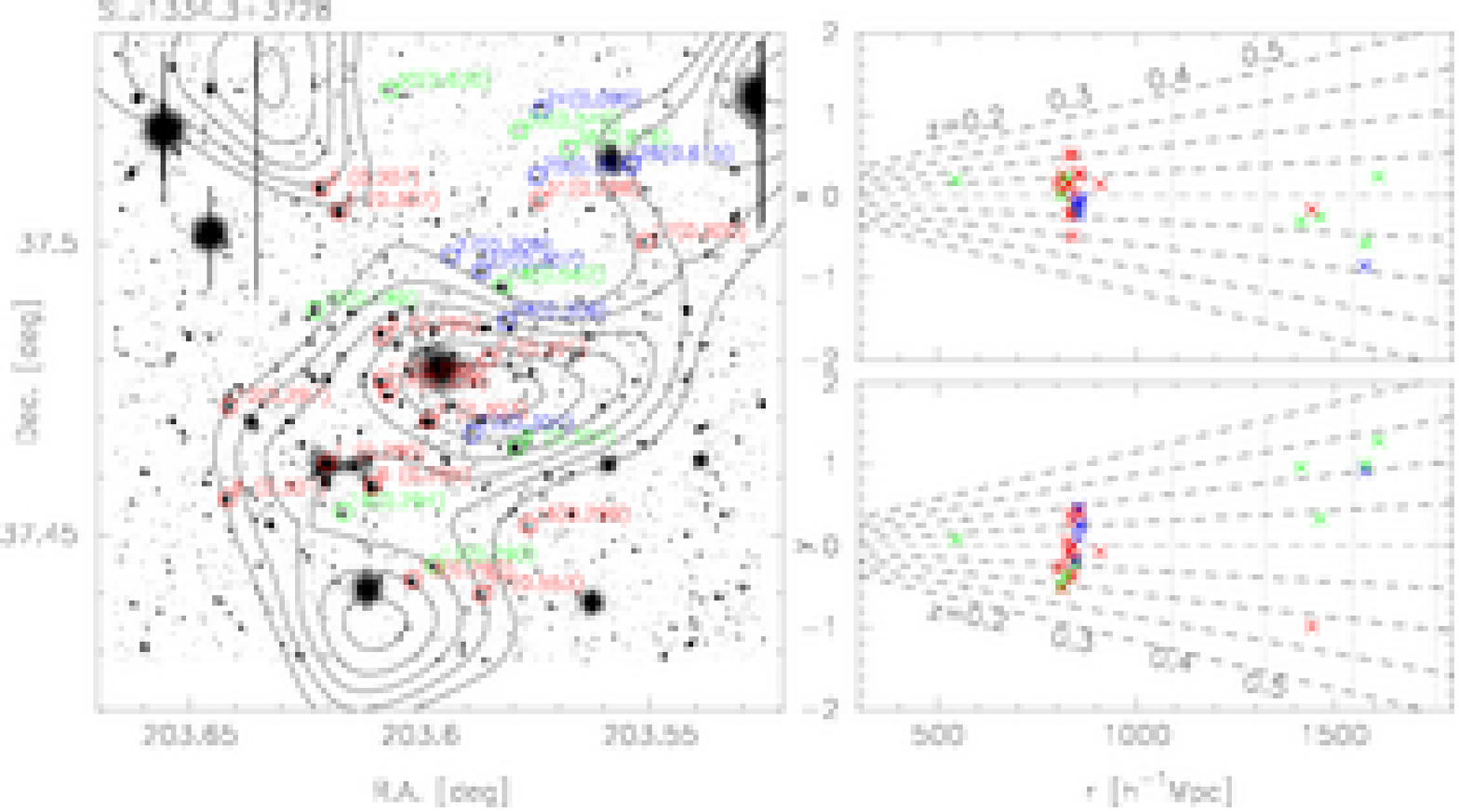}
\vspace{2mm}\\
\includegraphics[width=75mm,clip,angle=-90]{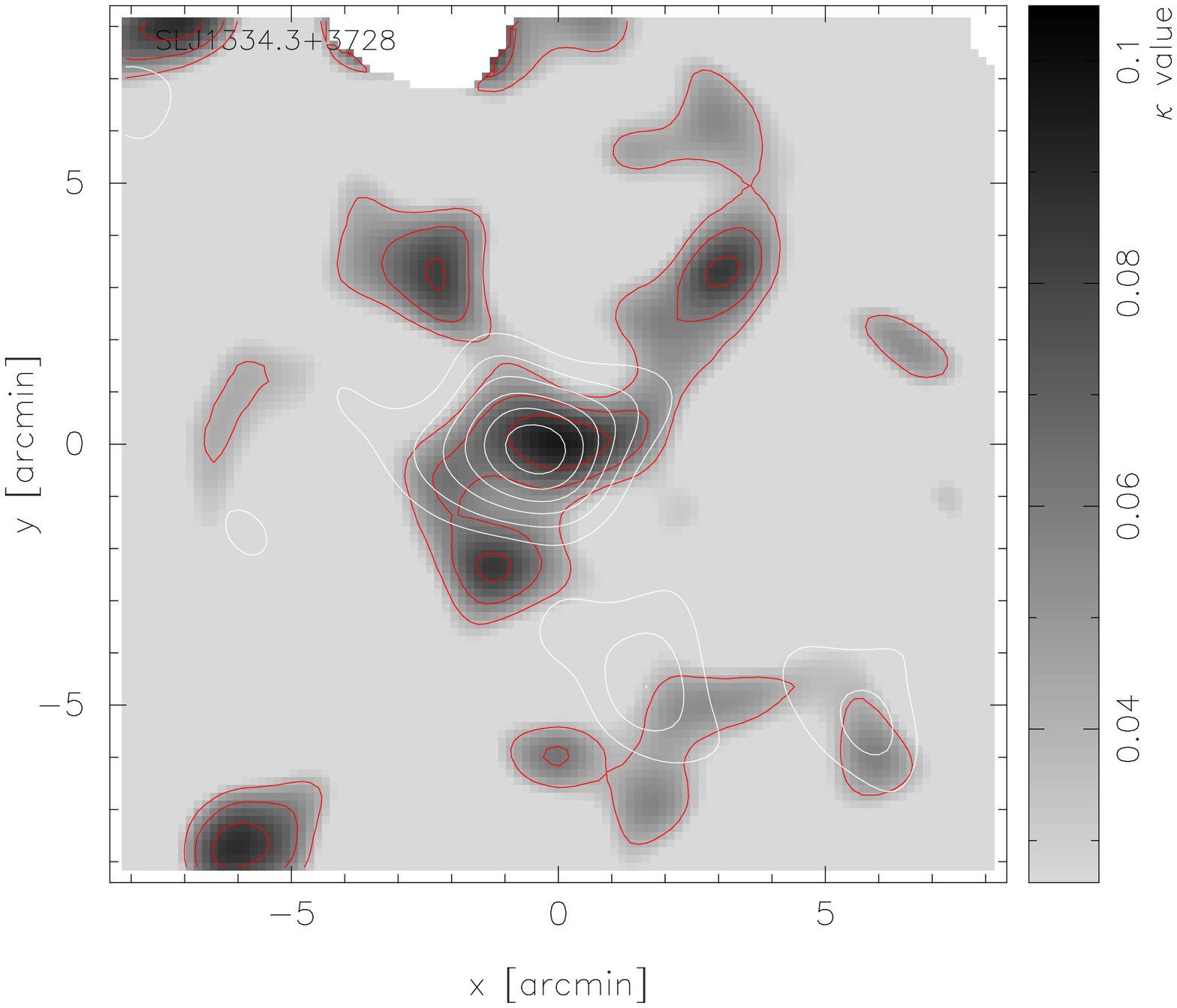}
\hspace{2mm}
\includegraphics[width=75mm,clip,angle=-90]{fig34c.ps}
\caption{Same as Figure \ref{fig:sxds_6} but for SL~J1334.3$+$3728.}
\end{figure*}

\clearpage
\begin{figure*}
\includegraphics[height=160mm,clip,angle=-90]{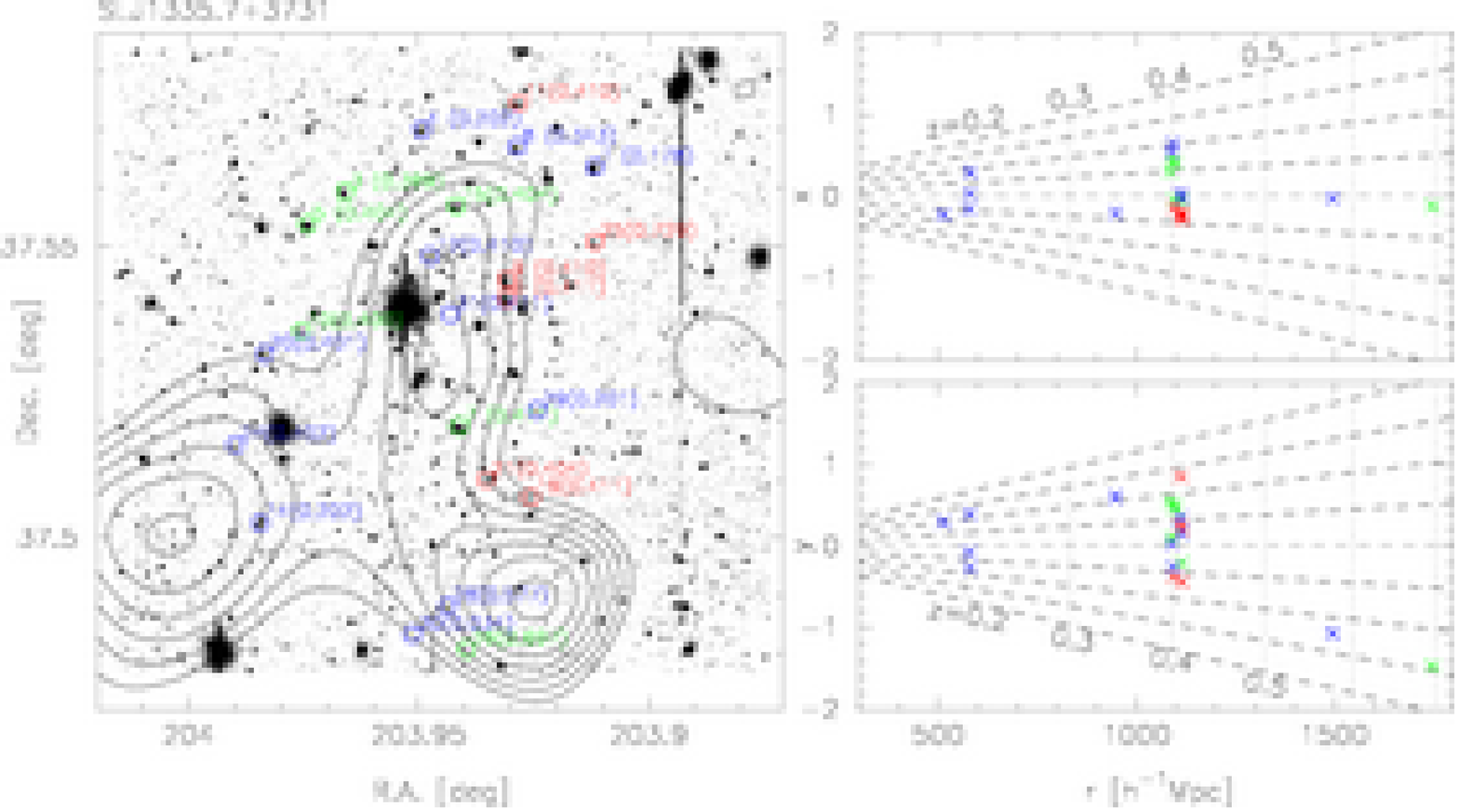}
\vspace{2mm}\\
\includegraphics[width=75mm,clip,angle=-90]{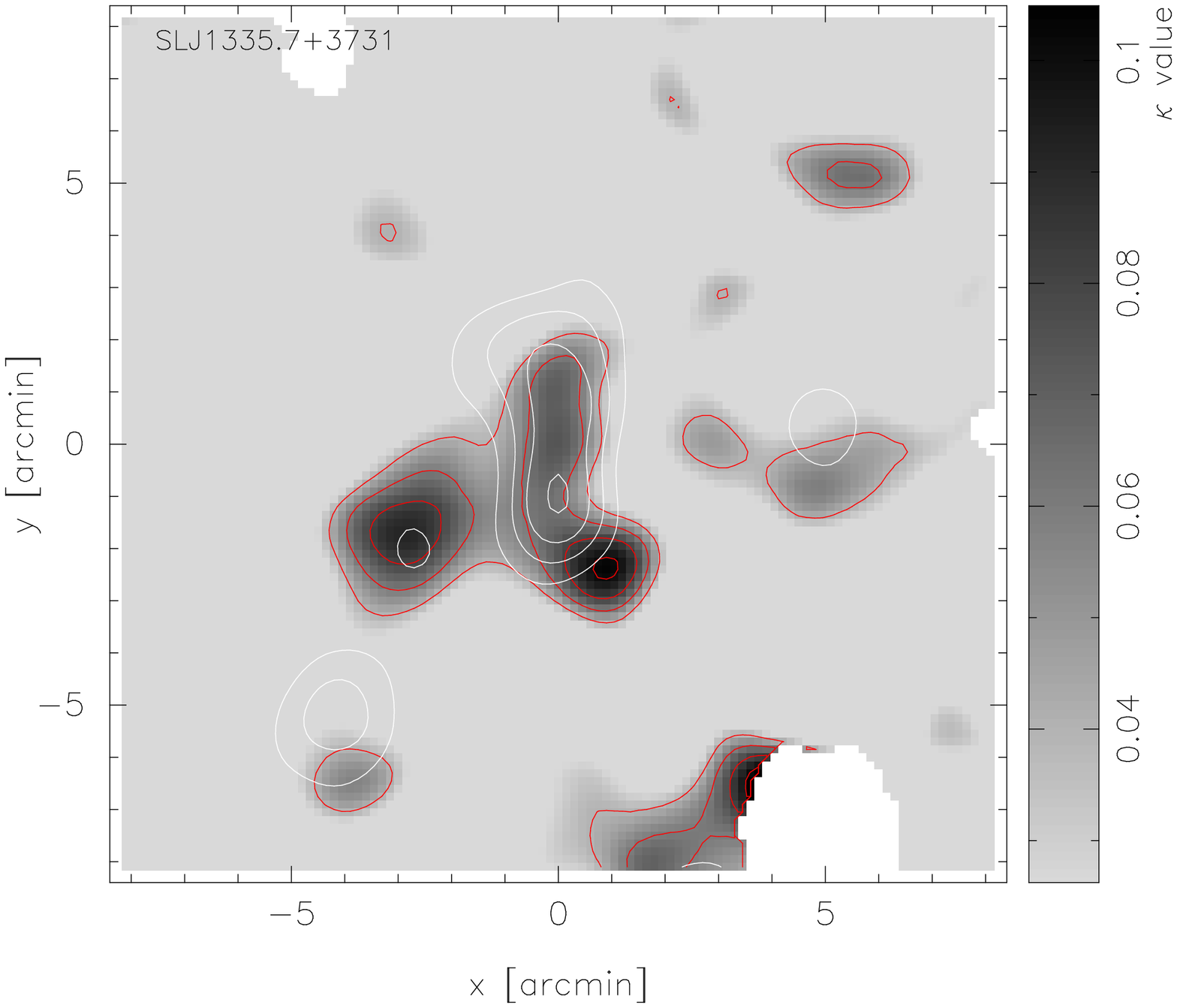}
\hspace{2mm}
\includegraphics[width=75mm,clip,angle=-90]{fig35c.ps}
\caption{Same as Figure \ref{fig:sxds_6} but for SL~J1335.7$+$3731.}
\end{figure*}

\clearpage
\begin{figure*}
\includegraphics[height=160mm,clip,angle=-90]{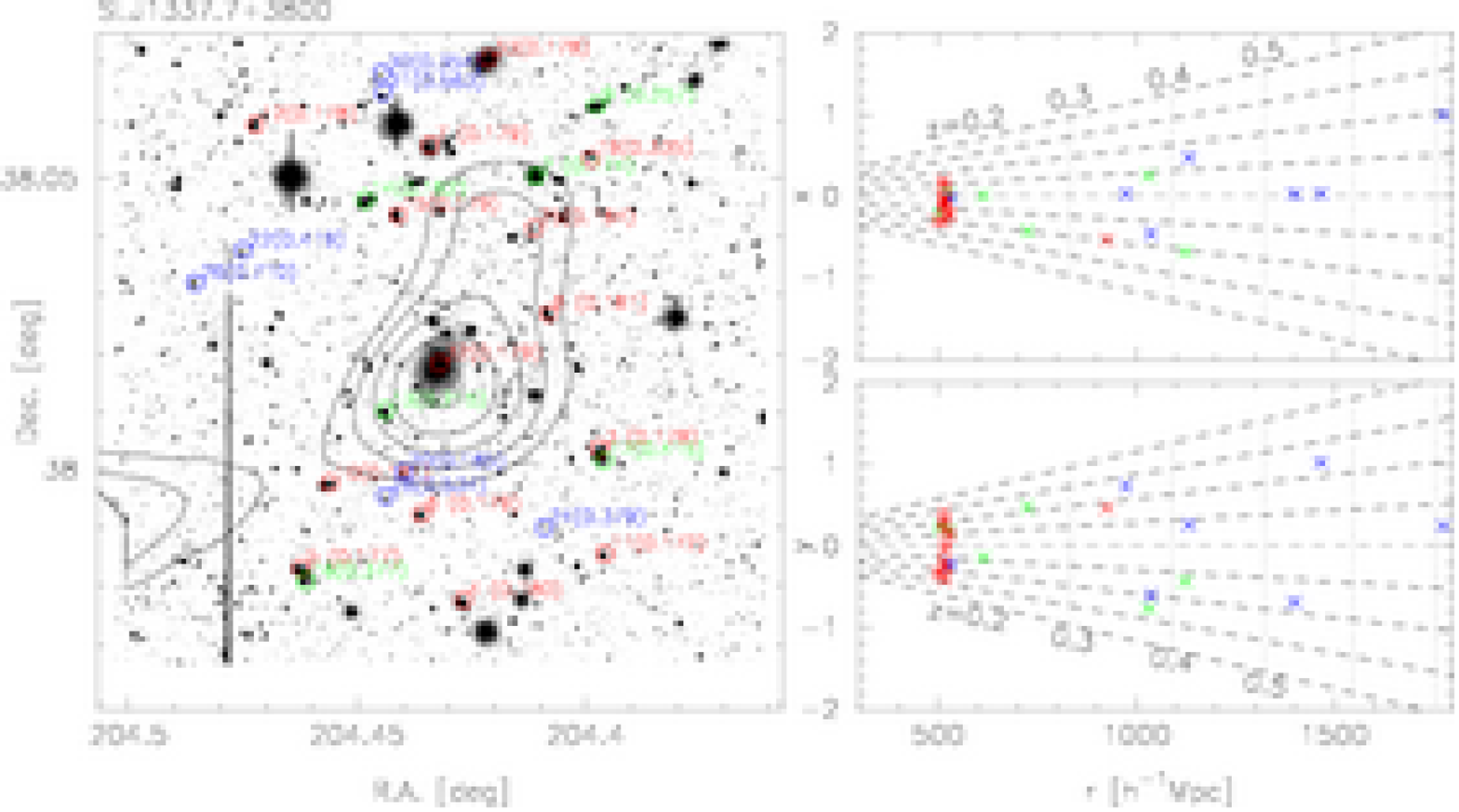}
\vspace{2mm}\\
\includegraphics[width=75mm,clip,angle=-90]{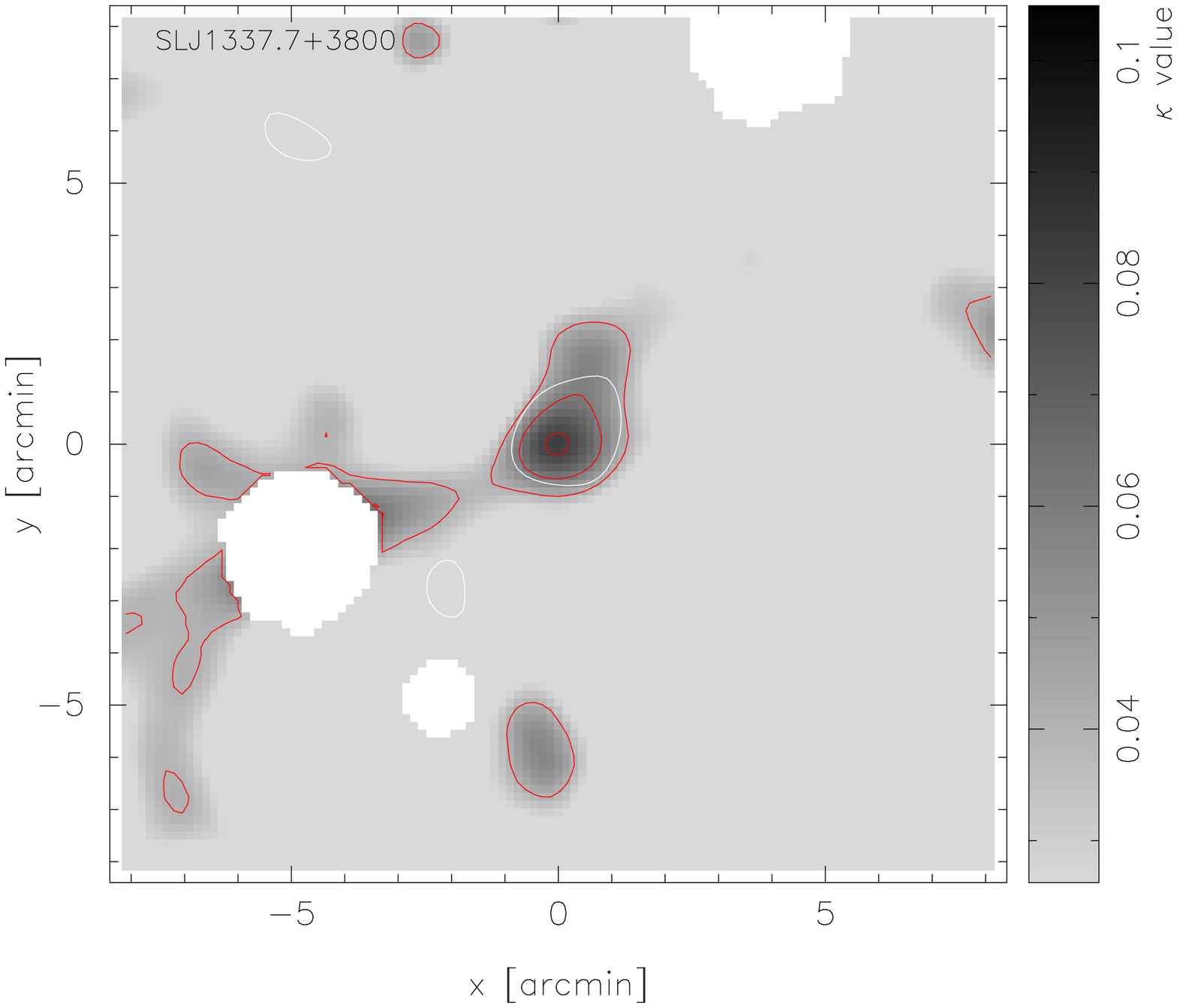}
\hspace{2mm}
\includegraphics[width=75mm,clip,angle=-90]{fig36c.ps}
\caption{Same as Figure \ref{fig:sxds_6} but for SL~J1337.7$+$3800.}
\end{figure*}

\clearpage
\begin{figure*}
\includegraphics[height=160mm,clip,angle=-90]{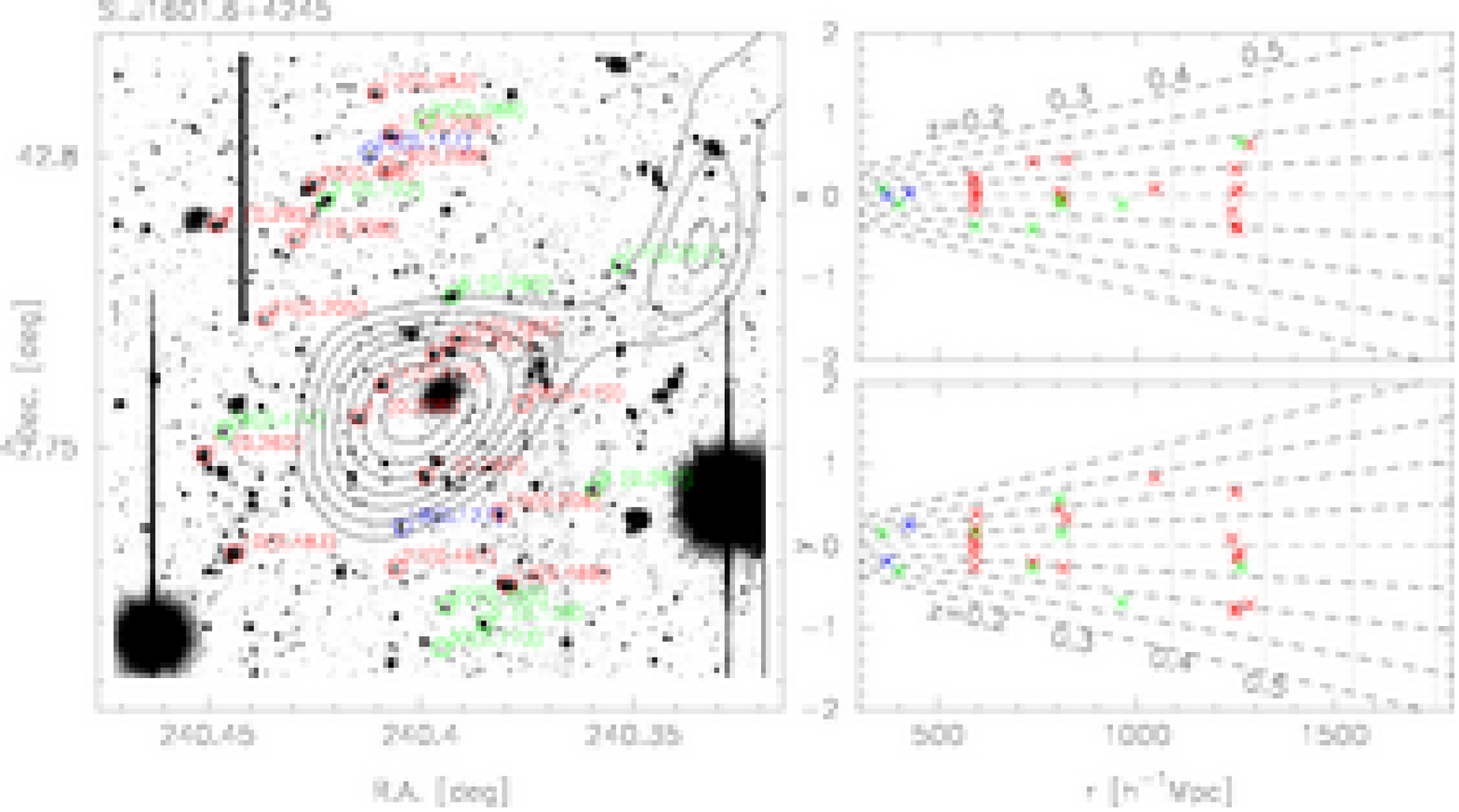}
\vspace{2mm}\\
\includegraphics[width=75mm,clip,angle=-90]{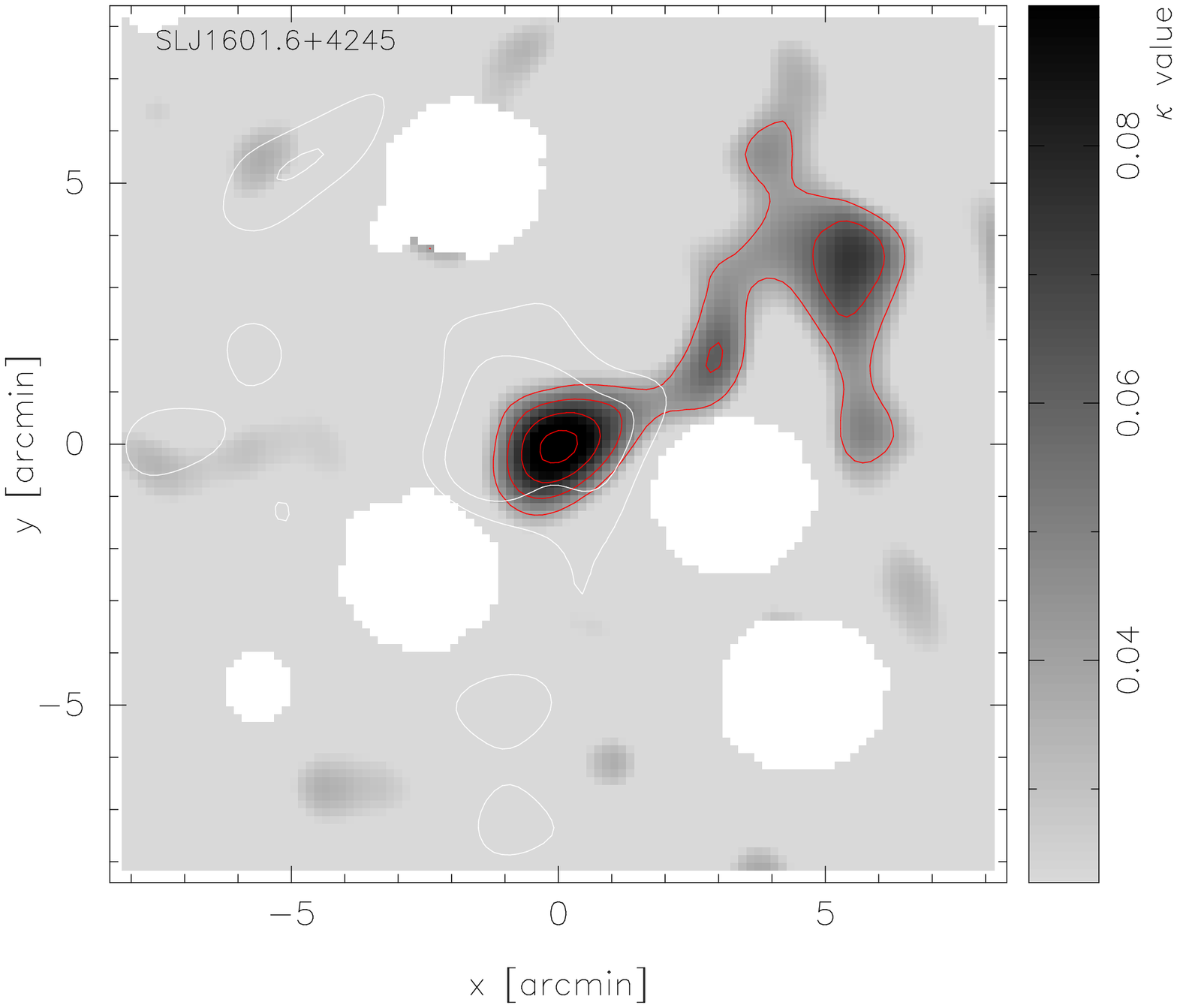}
\caption{Same as Figure \ref{fig:sxds_6} but for SL~J1601.6$+$4245.}
\end{figure*}

\clearpage
\begin{figure*}
\includegraphics[height=160mm,clip,angle=-90]{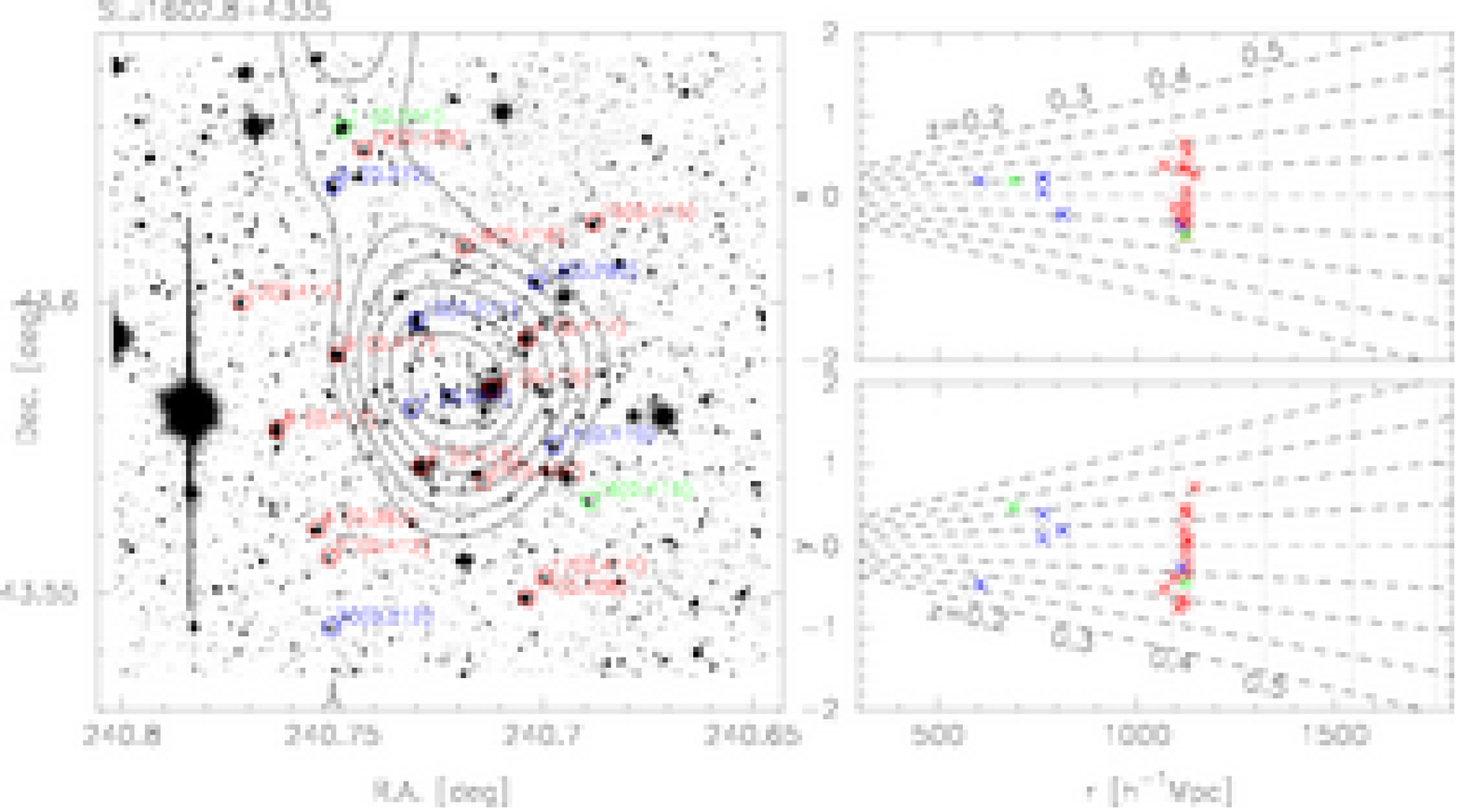}
\vspace{2mm}\\
\includegraphics[width=75mm,clip,angle=-90]{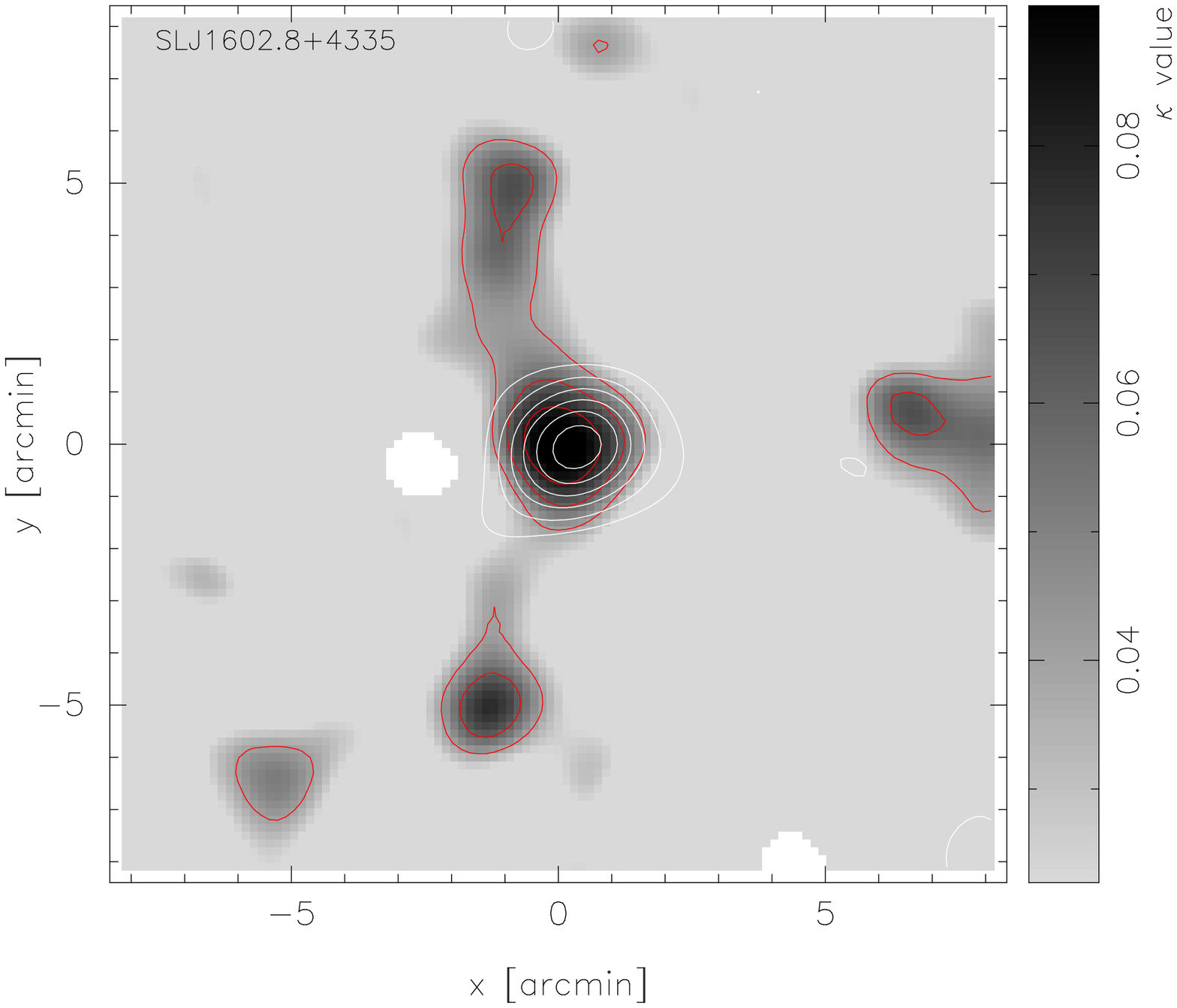}
\hspace{2mm}
\includegraphics[width=75mm,clip,angle=-90]{fig38c.ps}
\caption{Same as Figure \ref{fig:sxds_6} but for SL~J1602.8$+$4335.}
\end{figure*}

\clearpage
\begin{figure*}
\includegraphics[height=160mm,clip,angle=-90]{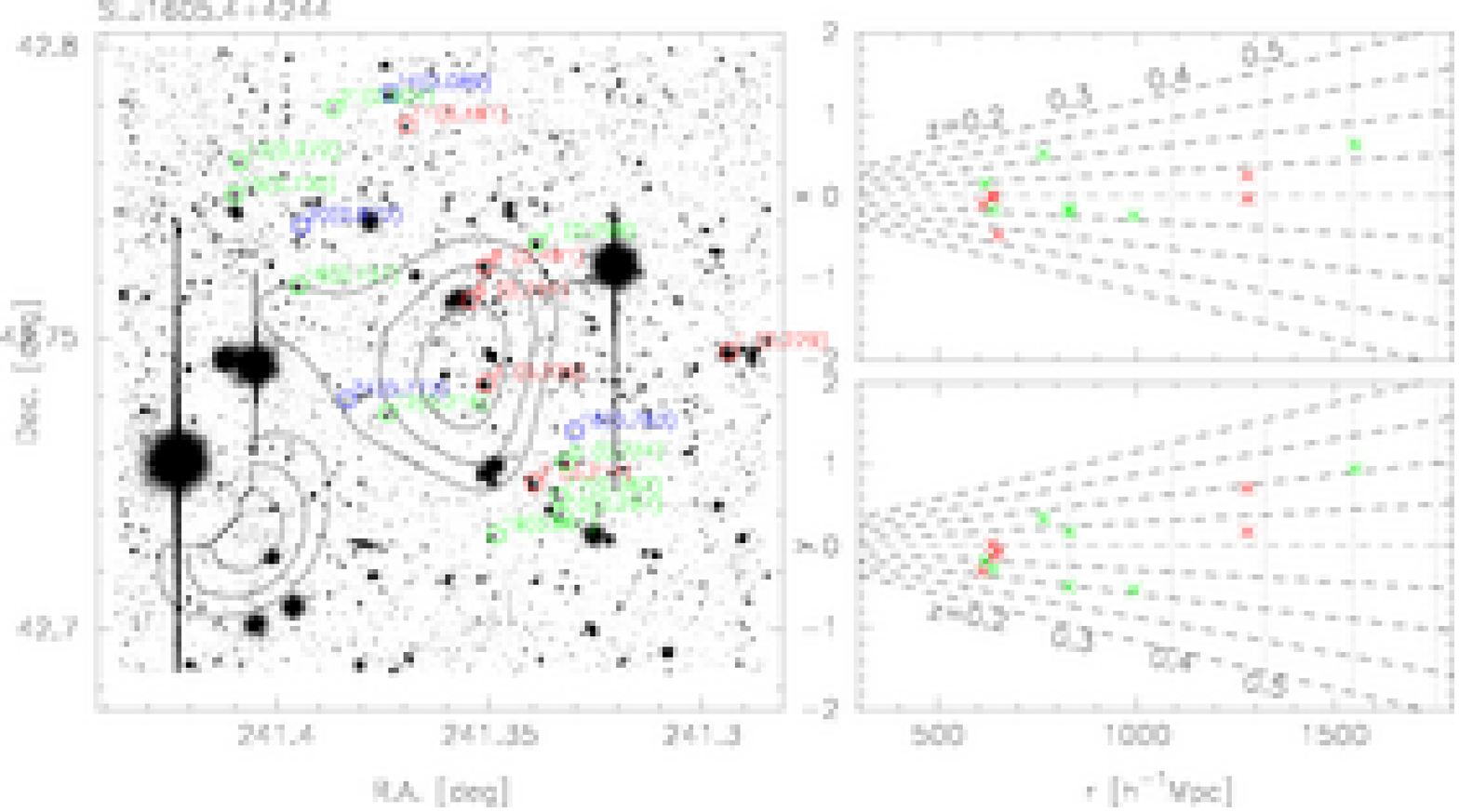}
\vspace{2mm}\\
\includegraphics[width=75mm,clip,angle=-90]{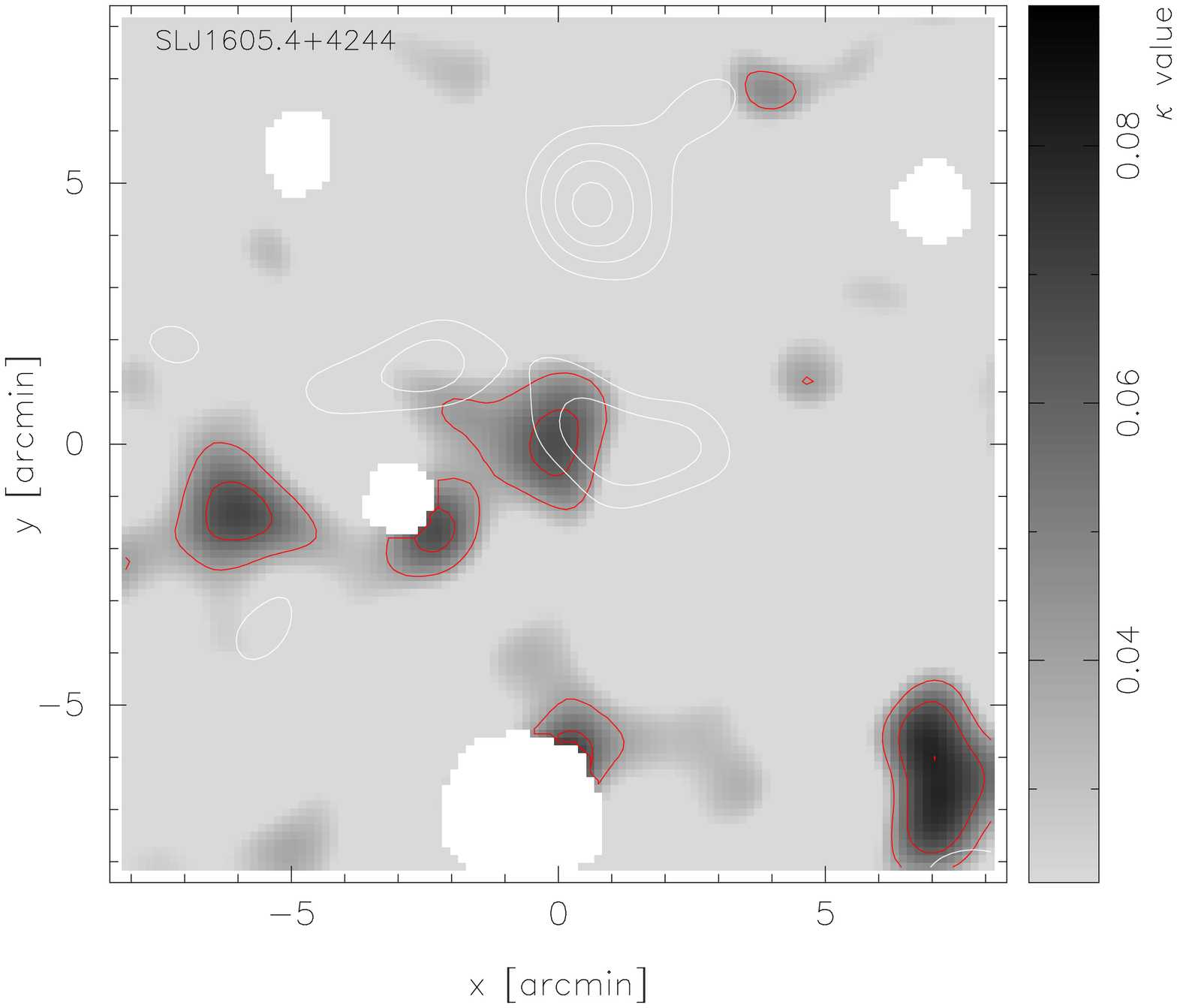}
\hspace{2mm}
\includegraphics[width=75mm,clip,angle=-90]{fig39c.ps}
\caption{Same as Figure \ref{fig:sxds_6} but for SL~J1605.4$+$4244.}
\end{figure*}

\clearpage
\begin{figure*}
\includegraphics[height=160mm,clip,angle=-90]{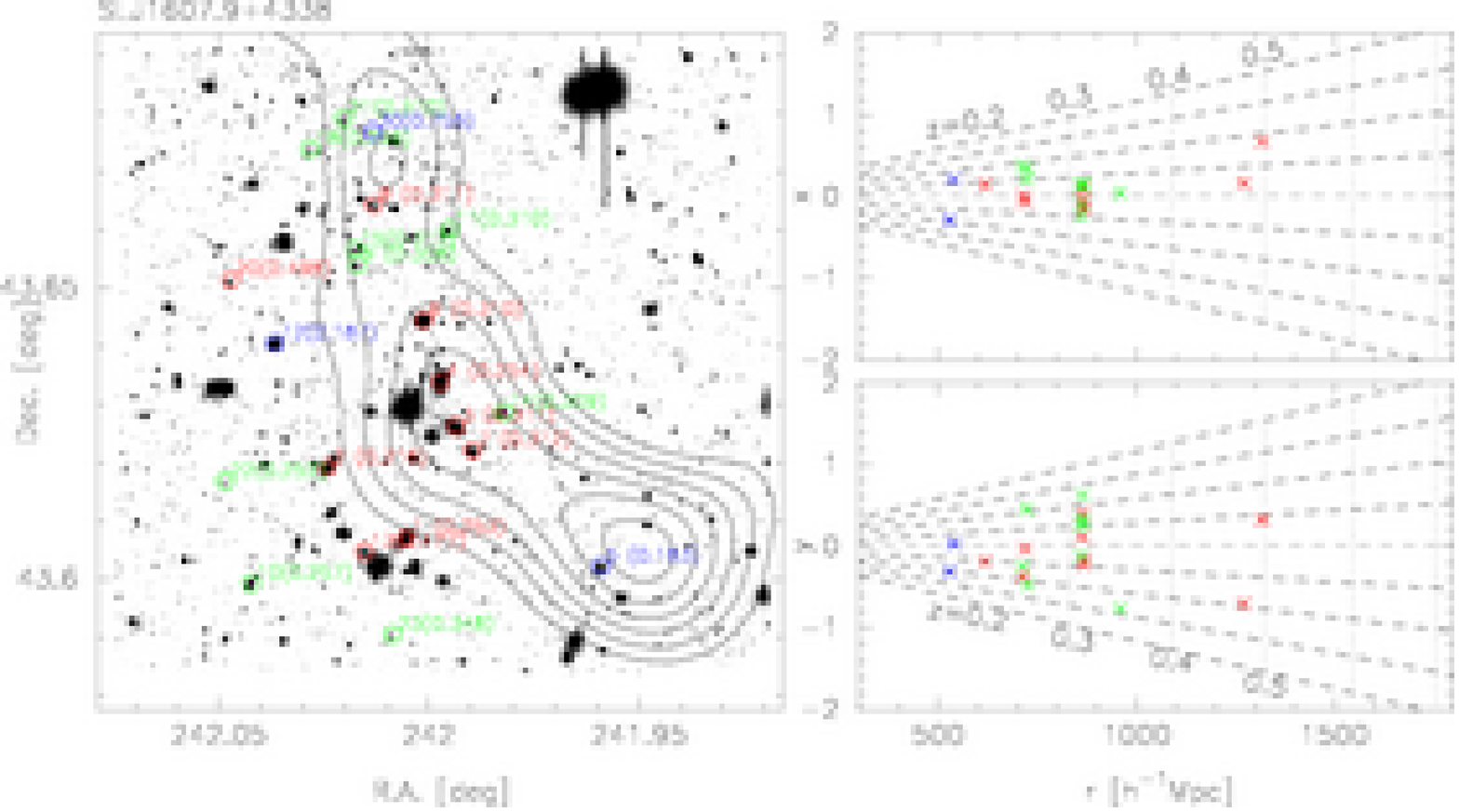}
\vspace{2mm}\\
\includegraphics[width=75mm,clip,angle=-90]{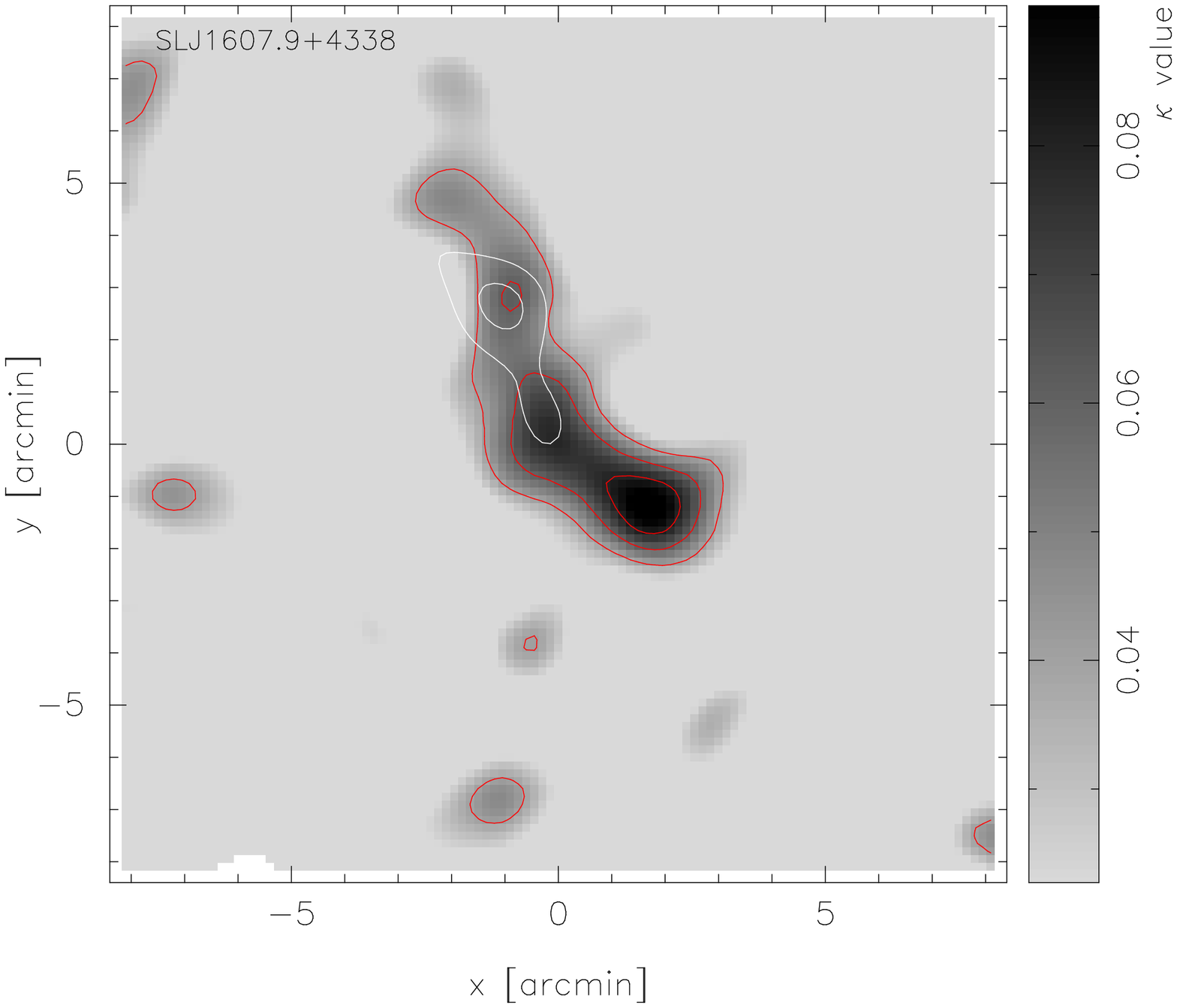}
\hspace{2mm}
\includegraphics[width=75mm,clip,angle=-90]{fig40c.ps}
\caption{Same as Figure \ref{fig:sxds_6} but for SL~J1607.9$+$4338.}
\end{figure*}

\clearpage
\begin{figure*}
\includegraphics[height=160mm,clip,angle=-90]{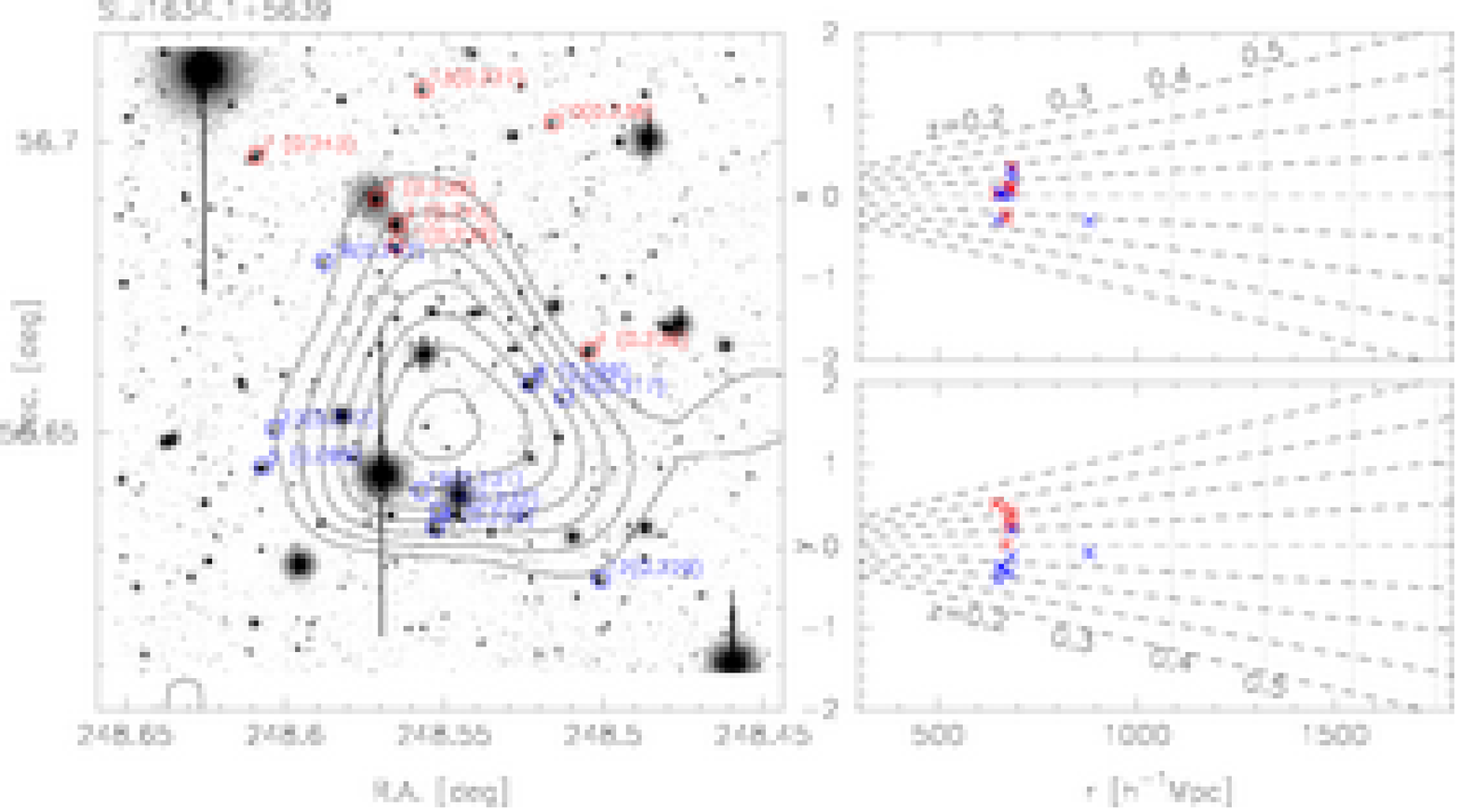}
\vspace{2mm}\\
\includegraphics[width=75mm,clip,angle=-90]{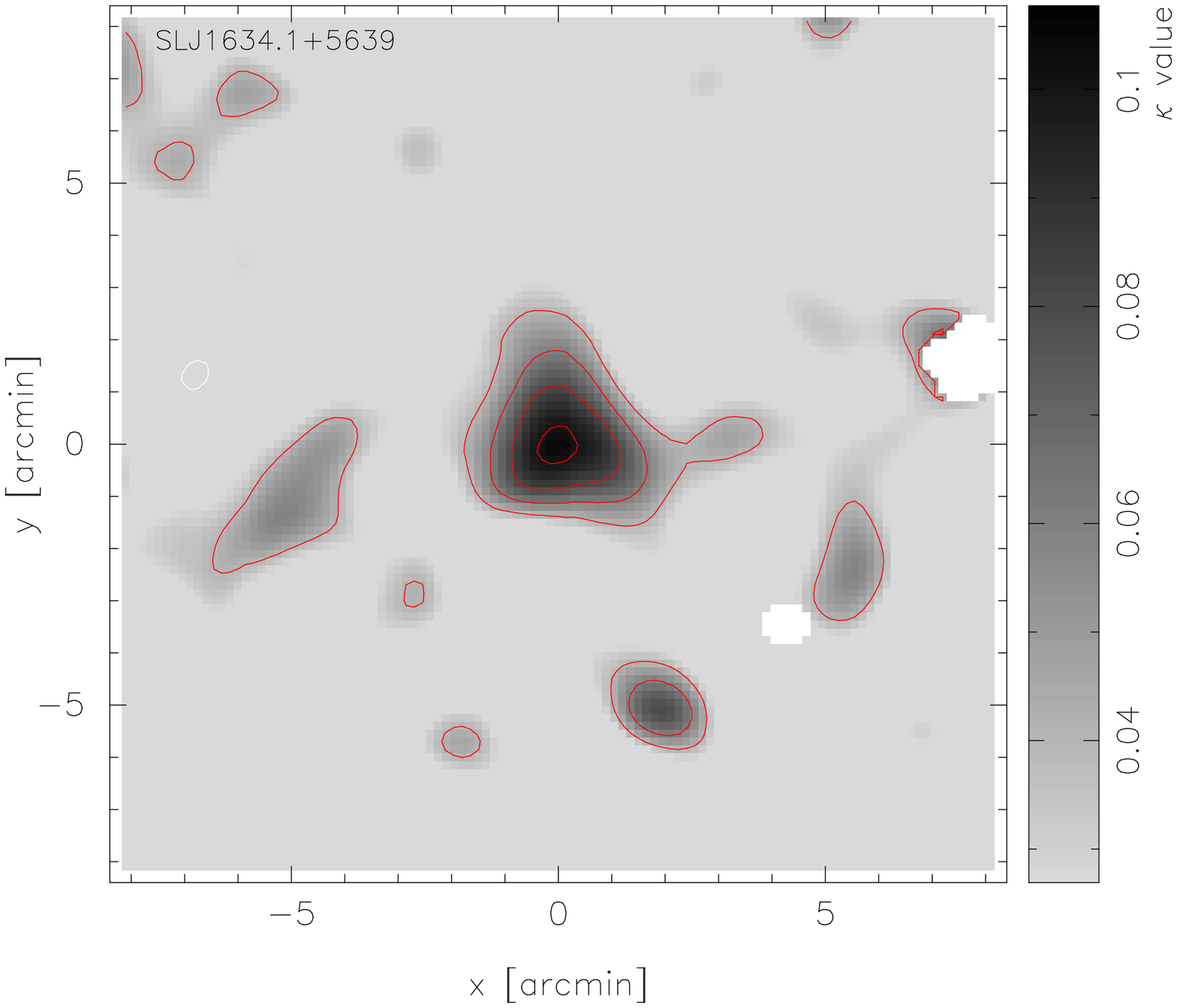}
\hspace{2mm}
\includegraphics[width=75mm,clip,angle=-90]{fig41c.ps}
\caption{Same as Figure \ref{fig:sxds_6} but for SL~J1634.1$+$5639.}
\end{figure*}

\clearpage
\begin{figure*}
\includegraphics[height=160mm,clip,angle=-90]{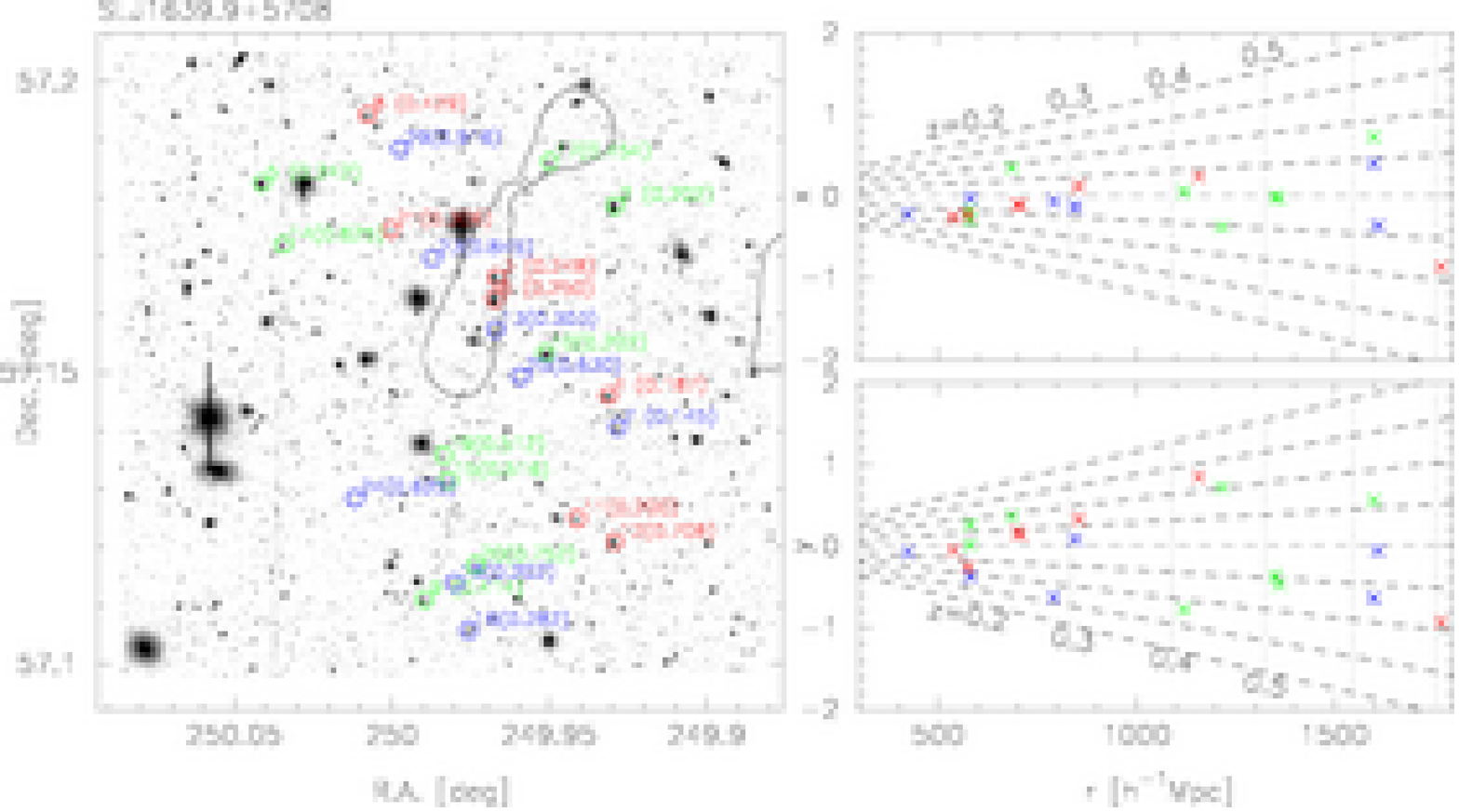}
\vspace{2mm}\\
\includegraphics[width=75mm,clip,angle=-90]{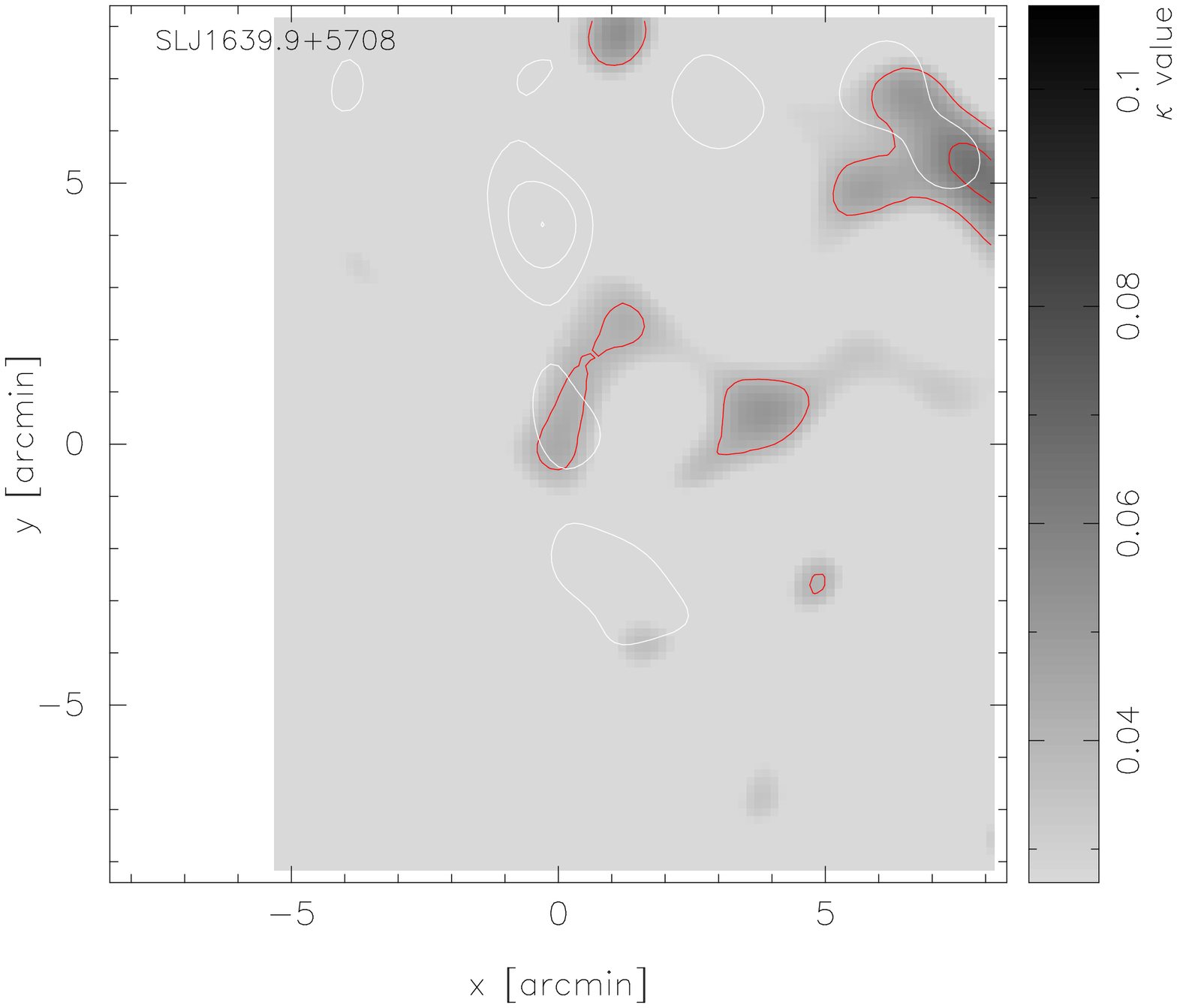}
\caption{Same as Figure \ref{fig:sxds_6} but for SL~J1639.9$+$5708.}
\end{figure*}

\clearpage
\begin{figure*}
\includegraphics[height=160mm,clip,angle=-90]{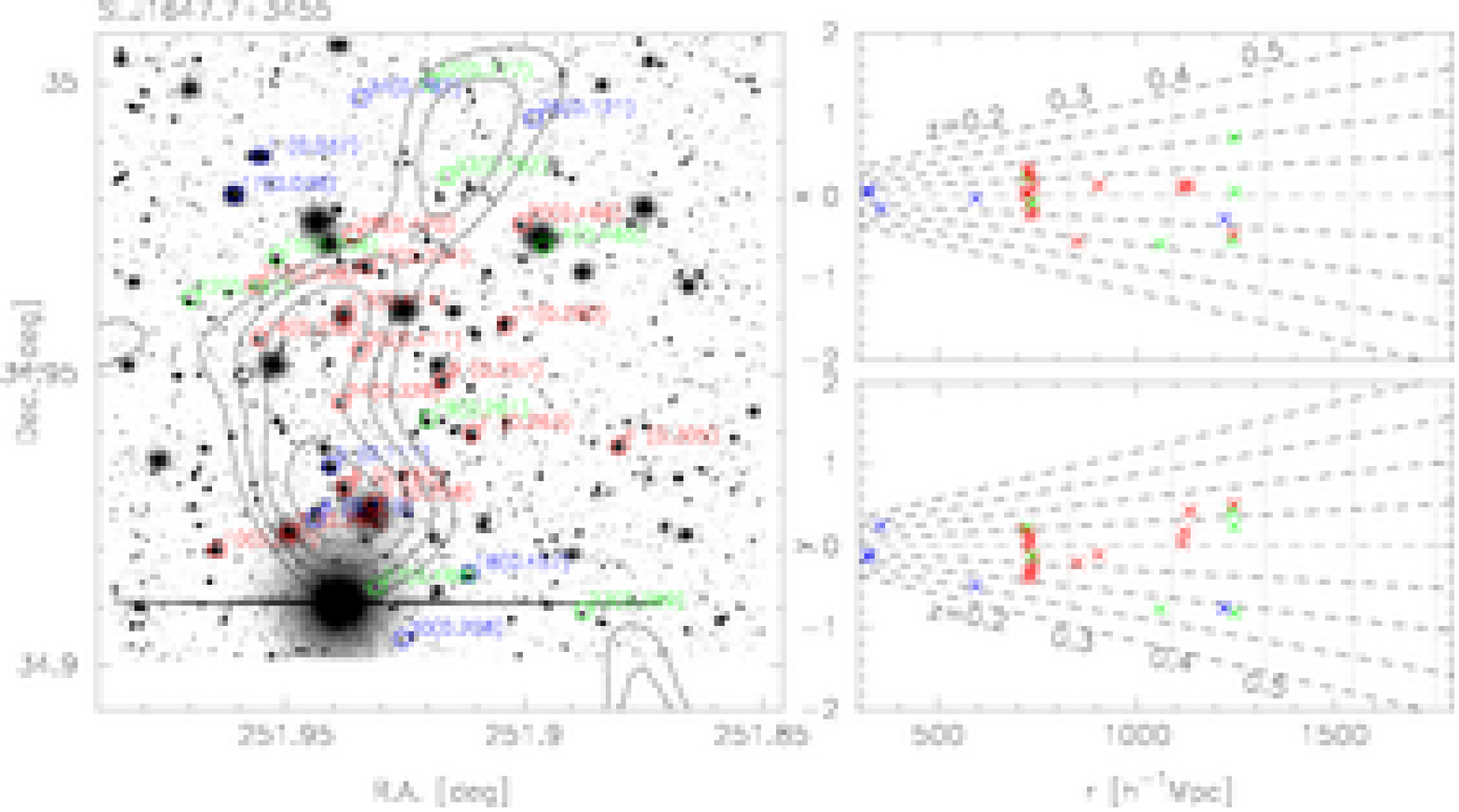}
\vspace{2mm}\\
\includegraphics[width=75mm,clip,angle=-90]{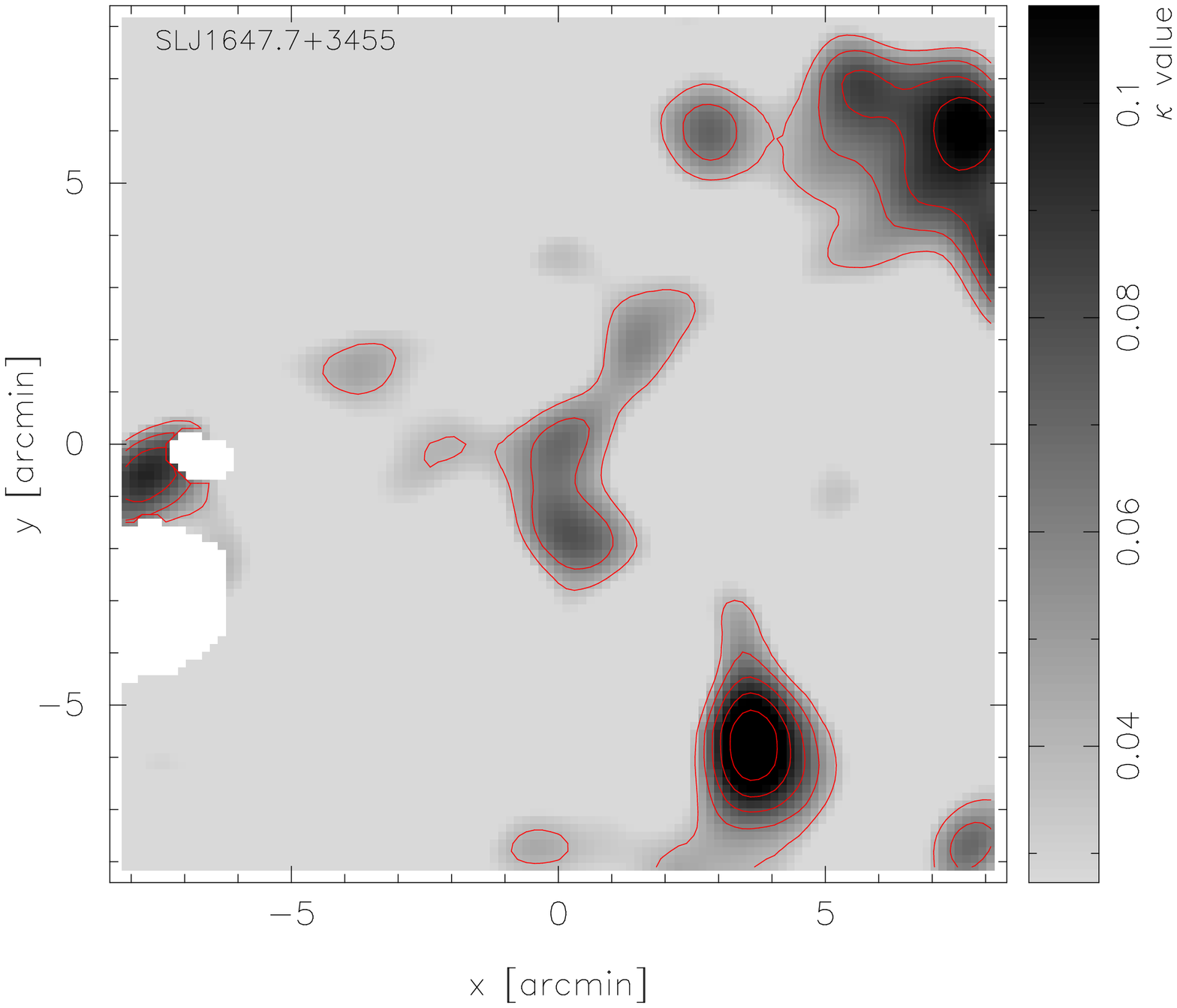}
\hspace{2mm}
\includegraphics[width=75mm,clip,angle=-90]{fig43c.ps}
\caption{Same as Figure \ref{fig:sxds_6} but for SL~J1647.7$+$3455.
\label{fig:deep16_m1_15}}
\end{figure*}

\clearpage
\section{Properties of previously studied clusters}
\label{appendix:xmmlss}

When studying cluster scaling relations in \S \ref{sec:scaling},
we included two cluster candidates from our weak lensing survey that
had already been spectroscopically verified and whose velocity
dispersions were previously known. 
Data from the literature were exposed to the same selection criteria 
used to make our clean sample (see \S \ref{sec:cleansample}).
Observations of clusters SL~J0221.7$-$0345 and SL~J0228.4$-$0425 by 
Willis et al.\ (2005, where they are named XLSSC~006 and XLSSC~012)
both satisfy our conditions. 

These two clusters were originally identified via their X-ray emission.
Later spectroscopic observations by Willis et al.\ (2005) 
revealed galaxy velocity dispersions for 
SL~J0221.7$-$0345 of $\sigma_v=821_{-74}^{+92}$km/s (computed from 39
galaxy redshifts) and for
SL~J0228.4$-$042 of $694_{-91}^{+204}$km/s (from 13 galaxy redshifts).

\begin{figure*}
\includegraphics[width=75mm,clip,angle=-90]{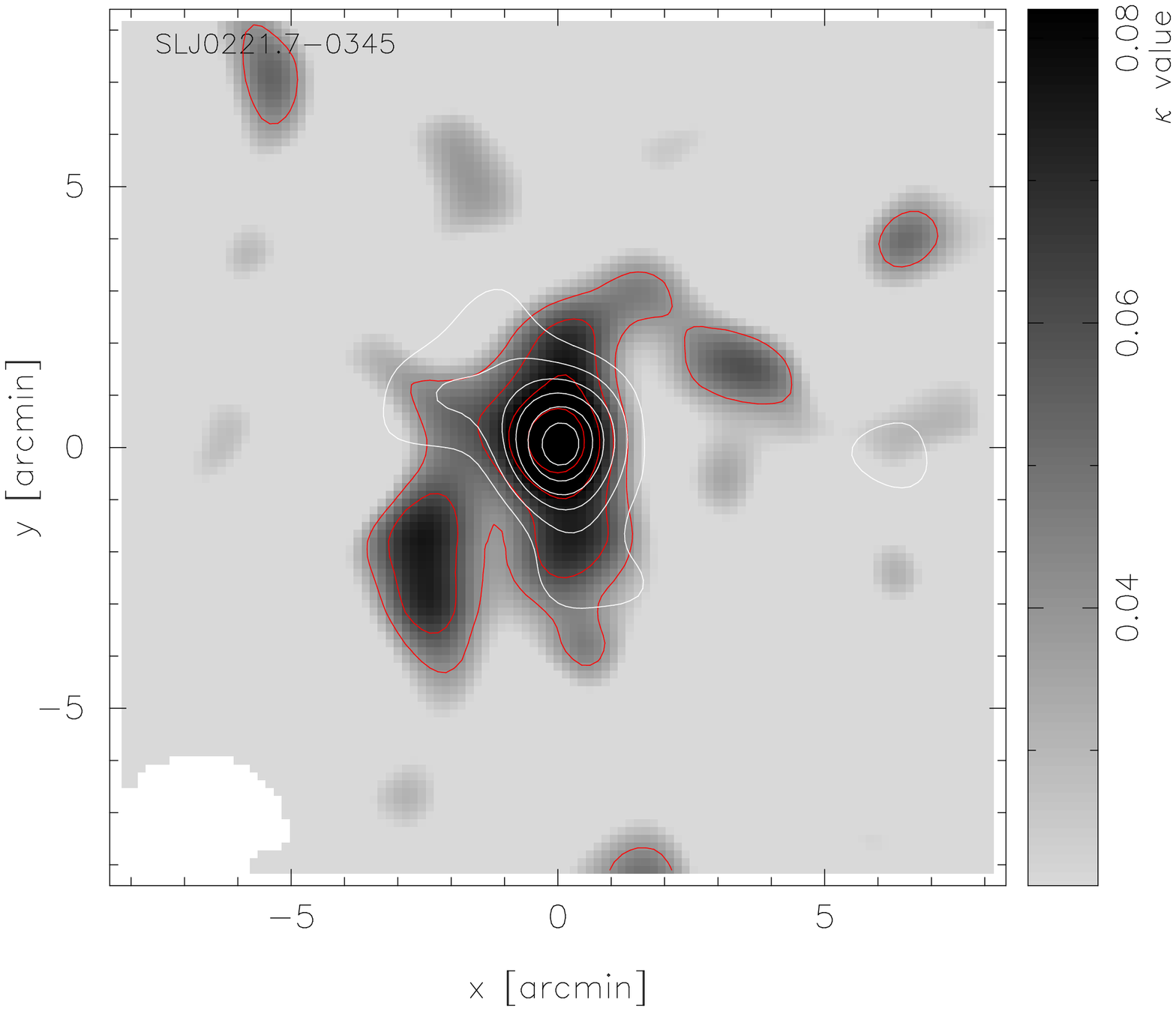}
\hspace{2mm}
\includegraphics[width=75mm,clip,angle=-90]{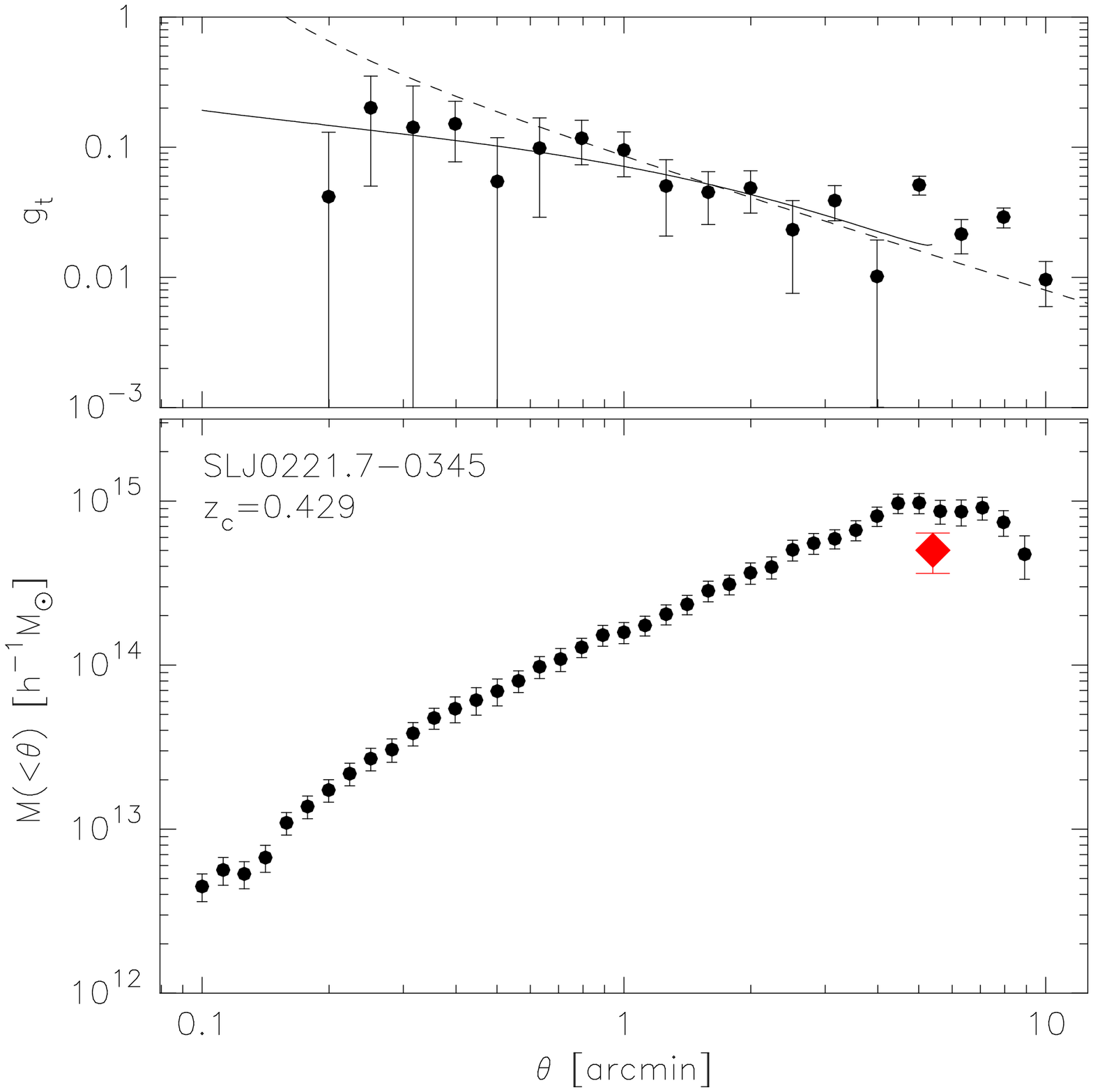}
\caption{Same as the bottom two panels of 
Figure \ref{fig:sxds_6} but for SL~J0222.8$-$0416.
\label{fig:saclay_0}}
\end{figure*}

In P1, the two clusters are listed as XMM-LSS-00 and XMM-LSS-21.
We measured the weak lensing properties of these clusters using the 
method described in \S \ref{sec:wlmass}, and summarize our results
in Table \ref{table:xmmlss}.
Weak lensing mass maps, galaxy density maps, tangential shear profiles and
aperture mass profiles are presented in Figures  \ref{fig:saclay_0} and
\ref{fig:saclay_21}.

\begin{table*}
\caption{Summary of weak lensing analyses: 
(a) the amplitude of the tangential shear profile at 1 arcmin, when fitted
  with an {\it SIS} model (see \S \ref{sec:sis}).
(b) the best-fit SIS velocity dispersion parameter.
(c) the virial mass estimated by fitting the radial shear profile with 
an NFW model.
(d) the $M_{200}$ computed assuming the NFW profile.
(e) the $M_{500}$ computed assuming the NFW profile.
(f) the virial radius computed from the NFW mass using the
relation eq (\ref{stm}).
}
\label{table:xmmlss}
\begin{tabular}{llccccccc}
\hline
Name & XMM-LSS ID & $z_{c}$ & $\gamma_{\rm sis}$$^{(a)}$ & $\sigma_{\rm SIS}$$^{(b)}$  & $M_{\rm
  NFW}$$^{(c)}$ & $M_{200}$$^{(d)}$ & $M_{500}$$^{(e)}$ & $r_{\rm vir}$$^{(f)}$ \\ 
{} & {} & {} & {} & [km/s] &  
\multicolumn{3}{c}{[$\times 10^{14}h^{-1}M_\odot$]} &
[comoving Mpc $h^{-1}$] \\ 
\hline
SL~J0221.7$-$0345 & XLSSC~006 & 0.429 & 0.079 & $ 926_{-438}^{+ 406} $ & $ 5.01_{-1.39}^{+ 1.37}$ & $ 4.35_{-1.21}^{+ 1.19} $ & $ 2.97_{-0.82}^{+ 0.81} $ & 1.7 \\
SL~J0228.4$-$0425 & XLSSC~012 & 0.433 & 0.065 & $ 839_{-448}^{+ 449} $ & $ 3.24_{-1.19}^{+ 1.46} $ & $ 2.82_{-1.04}^{+ 1.27} $ & $ 1.95_{-0.72}^{+ 0.88} $ & 1.6 \\
\hline
\end{tabular}
\end{table*}

\begin{figure*}
\includegraphics[width=75mm,clip,angle=-90]{fig45b.ps}
\hspace{2mm}
\includegraphics[width=75mm,clip,angle=-90]{fig45c.ps}
\caption{Same as  the bottom two panels of 
Figure \ref{fig:sxds_6} but for SL~J0222.8$-$0416.
\label{fig:saclay_21}}
\end{figure*}

\end{document}